\documentclass{emulateapj}
\usepackage{natbib}
\usepackage{amsmath}
\usepackage{color}
\usepackage{xcolor}
\usepackage[colorlinks=true,linkcolor=red,citecolor=blue]{hyperref}
\usepackage{tgcursor}
\singlespace
\usepackage{hypcap}
\usepackage[autostyle]{csquotes}
\usepackage{physics}
\DeclareMathOperator{\sinc}{sinc}
\usepackage{bbold}
\usepackage{array}
\usepackage{multirow}
\newcolumntype{C}[1]{>{\centering\arraybackslash}m{#1}}

\usepackage{orcidlink}

\begin{document}

\title{The Atacama Cosmology Telescope: measurement and analysis of 1D beams for DR4}

\author{
Marius~Lungu\altaffilmark{1},
Emilie~R.~Storer\,\orcidlink{0000-0003-1592-9659}\hspace{-0.3em}\altaffilmark{1},
Matthew~Hasselfield\,\orcidlink{0000-0002-2408-9201}\hspace{-0.3em}\altaffilmark{2},
Adriaan~J.~Duivenvoorden\,\orcidlink{0000-0003-2856-2382}\hspace{-0.3em}\altaffilmark{1},
Erminia~Calabrese\,\orcidlink{0000-0003-0837-0068}\hspace{-0.3em}\altaffilmark{3},
Grace~E.~Chesmore\,\orcidlink{0000-0001-6702-0450}\hspace{-0.3em}\altaffilmark{4},
Steve~K.~Choi\,\orcidlink{0000-0002-9113-7058}\hspace{-0.3em}\altaffilmark{5,6},
Jo~Dunkley\,\orcidlink{0000-0002-7450-2586}\hspace{-0.3em}\altaffilmark{1,7},
Rolando~D\"{u}nner\altaffilmark{8},
Patricio~A.~Gallardo\,\orcidlink{0000-0001-9731-3617}\hspace{-0.3em}\altaffilmark{6},
Joseph~E.~Golec\,\orcidlink{0000-0002-4421-0267}\hspace{-0.3em}\altaffilmark{4},
Yilun~Guan\altaffilmark{9}
J.~Colin~Hill\,\orcidlink{0000-0002-9539-0835}\hspace{-0.3em}\altaffilmark{2,10},
Adam~D.~Hincks\,\orcidlink{0000-0003-1690-6678}\hspace{-0.3em}\altaffilmark{11},
Johannes~Hubmayr\,\orcidlink{0000-0002-2781-9302}\hspace{-0.3em}\altaffilmark{12},
Mathew~S.~Madhavacheril\,\orcidlink{0000-0001-6740-5350}\hspace{-0.3em}\altaffilmark{13,14},
Maya~Mallaby-Kay\,\orcidlink{0000-0002-2018-3807}\hspace{-0.3em}\altaffilmark{15},
Jeff~McMahon\altaffilmark{4,15,16,17},
Kavilan~Moodley\,\orcidlink{0000-0001-6606-7142}\hspace{-0.3em}\altaffilmark{18,19},
Sigurd~Naess\,\orcidlink{0000-0002-4478-7111}\hspace{-0.3em}\altaffilmark{20},
Federico~Nati\,\orcidlink{0000-0002-8307-5088}\hspace{-0.3em}\altaffilmark{21},
Michael~D.~Niemack\,\orcidlink{0000-0001-7125-3580}\hspace{-0.3em}\altaffilmark{5,6,22},
Lyman~A.~Page\,\orcidlink{0000-0002-9828-3525}\hspace{-0.3em}\altaffilmark{1},
Bruce~Partridge\altaffilmark{23},
Roberto~Puddu\altaffilmark{8},
Alessandro~Schillaci\altaffilmark{24},
Crist\'obal~Sif\'on\,\orcidlink{0000-0002-8149-1352}\hspace{-0.3em}\altaffilmark{25},
Suzanne~Staggs\,\orcidlink{0000-0002-7020-7301}\hspace{-0.3em}\altaffilmark{1},
Dhaneshwar~D.~Sunder\altaffilmark{18,19},
Edward~J.~Wollack\,\orcidlink{0000-0002-7567-4451}\hspace{-0.3em}\altaffilmark{26},
Zhilei~Xu\,\orcidlink{0000-0001-5112-2567}\hspace{-0.3em}\altaffilmark{27,28}
}
\altaffiltext{1}{Joseph Henry Laboratories of Physics, Jadwin Hall,
Princeton University, Princeton, NJ 08544, USA}
\altaffiltext{2}{Center for Computational Astrophysics, Flatiron Institute, 162 5th Avenue, New York, NY 10010 USA}
\altaffiltext{3}{School of Physics and Astronomy, Cardiff University, The Parade, Cardiff, Wales CF24 3AA, UK}
\altaffiltext{4}{Department of Physics, University of Chicago, Chicago, IL 60637, USA}
\altaffiltext{5}{Department of Physics, Cornell University, Ithaca, NY 14853, USA}
\altaffiltext{6}{Department of Astronomy, Cornell University, Ithaca, NY 14853, USA}
\altaffiltext{7}{Department of Astrophysical Sciences, Peyton Hall, 
Princeton University, Princeton, NJ 08544, USA}
\altaffiltext{8}{Instituto de Astrof\'isica and Centro de Astro-Ingenier\'ia, Facultad de F\`isica, Pontificia Universidad Cat\'olica de Chile, Av. Vicu\~na Mackenna 4860, 7820436 Macul, Santiago, Chile}
\altaffiltext{9}{Dunlap Institute for Astronomy and Astrophysics, University of Toronto, 50 St. George St., Toronto, ON M5S 3H4, Canada}
\altaffiltext{10}{Department of Physics, Columbia University, New York, New York 10027, USA}
\altaffiltext{11}{David A. Dunlap Department of Astronomy \& Astrophysics, University of Toronto, 50 St. George St., Toronto ON M5S 3H4, Canada}
\altaffiltext{12}{NIST, Quantum Sensors Group, 325 Broadway, Boulder, CO 80305, USA}
\altaffiltext{13}{Perimeter Institute for Theoretical Physics, 31 Caroline Street N, Waterloo ON N2L 2Y5, Canada}
\altaffiltext{14}{Department of Physics and Astronomy, University of Southern California, Los Angeles, CA, 90007, USA}
\altaffiltext{15}{Department of Astronomy and Astrophysics, University of Chicago, 5640 S. Ellis Ave., Chicago, IL 60637, USA}
\altaffiltext{16}{Kavli Institute for Cosmological Physics, University of Chicago, 5640 S. Ellis Ave., Chicago, IL 60637, USA}
\altaffiltext{17}{Enrico Fermi Institute, University of Chicago, Chicago, IL 60637, USA}
\altaffiltext{18}{Astrophysics Research Centre, University of KwaZulu-Natal, Westville Campus, Durban 4041, South Africa}
\altaffiltext{19}{School of Mathematics, Statistics \& Computer Science, University of KwaZulu-Natal, Westville Campus, Durban 4041, South Africa}
\altaffiltext{20}{Institute of Theoretical Astrophysics, University of Oslo, P.O. Box 1029, Blindern, NO-0315 Oslo, Norway}
\altaffiltext{21}{Department of Physics, University of Milano - Bicocca, Piazza della Scienza, 3 - 20126, Milano (MI), Italy}
\altaffiltext{22}{Kavli Institute at Cornell for Nanoscale Science, Cornell University, Ithaca, NY 14853, USA}
\altaffiltext{23}{Department of Physics and Astronomy, Haverford College,
Haverford, PA 19041, USA}
\altaffiltext{24}{Department of Physics, California Institute of Technology, Pasadena, CA 91125, USA}
\altaffiltext{25}{Instituto de F{\'{i}}sica, Pontificia Universidad Cat{\'{o}}lica de Valpara{\'{i}}so, Casilla 4059, Valpara{\'{i}}so, Chile}
\altaffiltext{26}{NASA/Goddard Space Flight Center, Greenbelt, MD 20771, USA}
\altaffiltext{27}{Department of Physics and Astronomy, University of Pennsylvania, 209 South 33rd Street, Philadelphia, PA 19104, USA}
\altaffiltext{28}{MIT Kavli Institute, Massachusetts Institute of Technology, 77 Massachusetts Avenue, Cambridge, MA 02139, USA}

\begin{abstract}
We describe the measurement and treatment of the telescope beams for the Atacama Cosmology Telescope's fourth data release, DR4. Observations of Uranus are used to measure the central portion ($<12^{\prime}$) of the beams to roughly $-$40 dB of the peak. Such planet maps in intensity are used to construct azimuthally averaged beam profiles, which are fit with a physically motivated model before being transformed into Fourier space. We investigate and quantify a number of percent-level corrections to the beams, all of which are important for precision cosmology. Uranus maps in polarization are used to measure the temperature-to-polarization leakage in the main part of the beams, which is $\lesssim 1\%$ ($2.5 \%$) at 150 GHz (98 GHz). The beams also have polarized sidelobes, which are measured with observations of Saturn and deprojected from the ACT time-ordered data. Notable changes relative to past ACT beam analyses include an improved subtraction of the atmospheric effects from Uranus calibration maps, incorporation of a scattering term in the beam profile model, and refinements to the beam model uncertainties and the main temperature-to-polarization leakage terms in the ACT power spectrum analysis.
\end{abstract}

\maketitle

\section{\label{sec:intro}Introduction}
\setcounter{footnote}{0}

The Atacama Cosmology Telescope (ACT) is a 6\,m off-axis Gregorian telescope located at an altitude of 5190\,m in the Atacama Desert of northern Chile. It is designed for millimeter wavelength observations of the cosmic microwave background (CMB) at arcminute resolution. The telescope and receiver are described in \cite{fowler_2007} and \cite{thornton_2016} respectively. This paper describes the treatment of the ACT beams for its fourth data release, referred to as DR4. This data release includes temperature and polarization data collected by ACT between 2013 and 2016, covering seven regions of the sky spanning roughly 18,000 deg$^2$, in frequency bands centered on 98 and 150 GHz \citep{thornton_2016}. The DR4 data were obtained using two 150\;GHz detector arrays, PA1 and PA2, and one dichroic detector array, PA3, operating at 98 and 150 GHz. The three array positions are not identical optically. Each year of data is referred to as a season (S13--S16) and was analyzed separately. The DR4 data products and some of their analyses are presented in  \cite{choi_2020}, \cite{aiola_2020}, \cite{darwish_2020}, \cite{han_2021}, \cite{madhavacheril_2020}, \cite{namikawa_2020}, and \cite{mallabykay_2021}. 

The beams of the telescope determine the instrument's response to different scales on the sky. Quantifying these beams and the uncertainties on the beam measurements is one of the most challenging aspects of characterizing the instrument. An incorrect calibration of the beams would bias virtually all of the ACT science analyses, including the precision measurement of the CMB power spectrum.

The previous paper that focused on ACT beams, \cite{hasselfield_atacama_2013}, dates from DR2, which included data through 2010. The DR3 analysis papers \citep{louis_2017,naess_2014} referenced \cite{hasselfield_atacama_2013} and described the small changes in methodology since then. In recent years, there have been several modifications to our beam pipeline that merit further discussion. This paper provides a comprehensive guide to the analysis of the ACT beams for DR4.

The paper is organized as follows. In \S\ref{sec:obs} we describe the planet observations used to measure the main portion of the ACT beams. In \S\ref{sec:maps} we explain how these observations are made into maps. In \S\ref{sec:pipe} we describe the steps of the beam pipeline from planet maps to a model of the ACT beams and their covariance. \S\ref{sec:pol} accounts for the polarization component of the beams. Finally, in \S\ref{sec:disc} we discuss the publicly released beam products, assumptions made in the analysis, and future directions for ACT beam analyses.
Part of the information in this paper overlaps with that in previous ACT papers, including \cite{aiola_2020}, \cite{choi_2020}, and \cite{hasselfield_atacama_2013}, and is included here for clarity and completeness.

\section{Observations}
\label{sec:obs}

Observations of sufficiently bright point-like objects are required in order to properly characterize the shape of the beams out to large angles. For ACT, planets are the best candidates for this purpose, though not all equally so. The proximity of Mercury and Venus to the Sun make them difficult to observe at night; Mars' temperature is not sufficiently constant \citep[see, for example,][]{wright_1976}, so it intermittently becomes too bright; Jupiter's apparent size is too large, compared to the instrument's angular resolution, to be considered a point source (and it is too bright, as is Saturn); Saturn, though suitable for sidelobe analysis, is bright enough to induce a non-linear detector response near peak amplitude, resulting in signal suppression within the main lobe of the beam;\footnote{Such detector non-linearity may have contributed to the $\sim$7\% discrepancy between the \textit{Planck} and ACT DR2 measurements of Saturn's temperature \citep{planck_planets, hasselfield_atacama_2013}.} Neptune's small angular diameter relative to the beamwidth dilutes its brightness, reducing signal-to-noise. We have thus chosen to base the majority of the ACT beam analysis on observations of Uranus, which achieves adequate signal-to-noise without exceeding the dynamic range of the instrument and can be approximately treated as a point source.\footnote{While not implemented for the current beam analysis, another possibility being studied involves using observations of Uranus to measure the central portion of the beam (where Saturn would induce a non-linear response) and observations of Saturn to measure the parts of the beam further from the peak. This way, Saturn's brightness allows us to measure the outer parts of the beam to greater signal-to-noise than when using observations of Uranus alone.} As described in \S\ref{subsec:sidel}, Saturn is used to study the beam's polarized sidelobes.

Planet observations are done by briefly interrupting the CMB survey scans, changing the azimuth at which the telescope is pointing, and scanning again at a fixed elevation. Uranus is observed roughly once per night during the observing season, but only the higher quality observations are used to characterize the beam. Before making maps of the observations, we apply the same data selection criteria as for the main maps used for cosmological analysis. For example, we discard data taken during the daytime, data taken when the weather was particularly bad, or data from detectors that do not meet criteria summarized in \cite{aiola_2020} and detailed in \cite{dunner_2013}. Once maps of the observations are made, further selection criteria are applied, as described in \S\ref{subsec:mapsel}.

\section{Planet Map-Making}
\label{sec:maps}

\subsection{Filtering And Noise}
\label{subsec:filt}

Uranus maps were made using a special filter-and-bin map-maker built on the same
framework used for building our normal maximum-likelihood maps (\verb|enki|).\footnote{GitHub repository: \url{https://github.com/amaurea/enlib}} The reason we do not use maximum-likelihood maps for planets is that these are not robust to \enquote{model errors}, as described in e.g. \cite{naess_2019}. The difference between the true continuous sky and the pixelated model is interpreted as noise, leading to a bias which propagates along the scan direction for a noise correlation length. This is a significant effect for a bright, localized signal such as a planet.

It is possible to make maximum-likelihood models robust to these effects at significant computational and implementational cost \citep{xgls_paper}, but we here choose a simpler method that exploits the compactness of the planet signal relative to the atmospheric noise correlation length to separate the two.
The idea is to split the data into two regions: a target region (including the planet) which we want to map, and a reference region (the rest of the data) which is used to estimate the behavior of the atmospheric noise.\footnote{The atmospheric fluctuations seen by ACT are described in more detail in \cite{morris_2021}.} The atmosphere in the target region is then estimated by interpolating the atmospheric modes from the reference region, and it is then subtracted from the target region of the map.\footnote{For a previous implementation of a map-making method designed to model and subtract the atmosphere from maps of planets, see \cite{hincks_2010}.} To the extent that the signal in the target region is uncorrelated with any signal in the reference region, then this subtraction will not introduce bias on any scale in the target region. A point source like Uranus is very close to fulfilling these criteria, but in practice the outermost parts of the beam extend into the reference region, introducing a slight bias that we address in \S\ref{subsec:tf}.

The target region must be large enough to capture all of the main beam, but small enough that the atmosphere's behavior in the target region can be adequately predicted using only data from the reference region. We choose a 12$^{\prime}$ radius area centered on the planet as the target region, and estimate the atmosphere's behavior here by exploiting its spatial correlation structure. Nearby detectors see similar parts of the atmosphere, so if at one moment in time some detectors are seeing the target region and some are seeing the reference region, the data from the latter are used to help determine the atmospheric modes in the former.\footnote{Ideally one would also use the temporal correlations of the atmosphere, and solve the most probable value of the atmosphere in the target region given the data in the reference region. This technique may be employed in the future. In preliminary tests, this only results in a mild improvement.} This new interpolation technique significantly improves the noise modelling compared to the technique used in \cite{louis_2017}, which used a simpler variant of filter-and-bin where the detector common mode was subtracted.

The resulting planet maps are cleaner and dominated mostly by white noise. In addition, we now inverse variance weight the detectors when making the planet maps, which was not done previously. This improved mapping has contributed to a more detailed understanding of the temperature-to-polarization leakage, which will be described in \S\ref{subsec:pol_main}.

\subsection{Bias From Map-Making}
\label{subsec:tf}

While it is approximately true that all the planet signal is contained inside the target region, a small part of it extends into the reference region due to the faint wings of the beams. This introduces a small but non-negligible bias in the target region when the atmosphere is subtracted. 

To study this bias, a set of simulated beam-convolved Uranus observations are processed through the map-maker. The beam-convolved planet signal is simulated using a 2D beam model\footnote{This model was developed as part of a larger, ongoing project to fit the beams in two dimensions in order to capture their asymmetry.} which uses the inverse Fourier transforms of two-dimensional Zernike polynomials as its basis set, similar to the 1D model of the beam described in \S\ref{subsubsec:core_mod}.
This simulated planet signal is then injected into several subsets of real time-domain data taken from CMB observations. The resulting data are then processed through the planet map-making pipeline, which includes estimating and subtracting the atmospheric noise. Both the CMB observations alone and the CMB observations with the planet signal injected are mapped separately, and then the former are subtracted from the latter. The result is a set of noise-free simulated maps of Uranus observations.\footnote{In the end, it was discovered that including the atmospheric noise in the simulations does not affect the results, since the atmosphere removal is approximately linear.}

\begin{figure*}
    \centering
    \includegraphics[width=\linewidth]{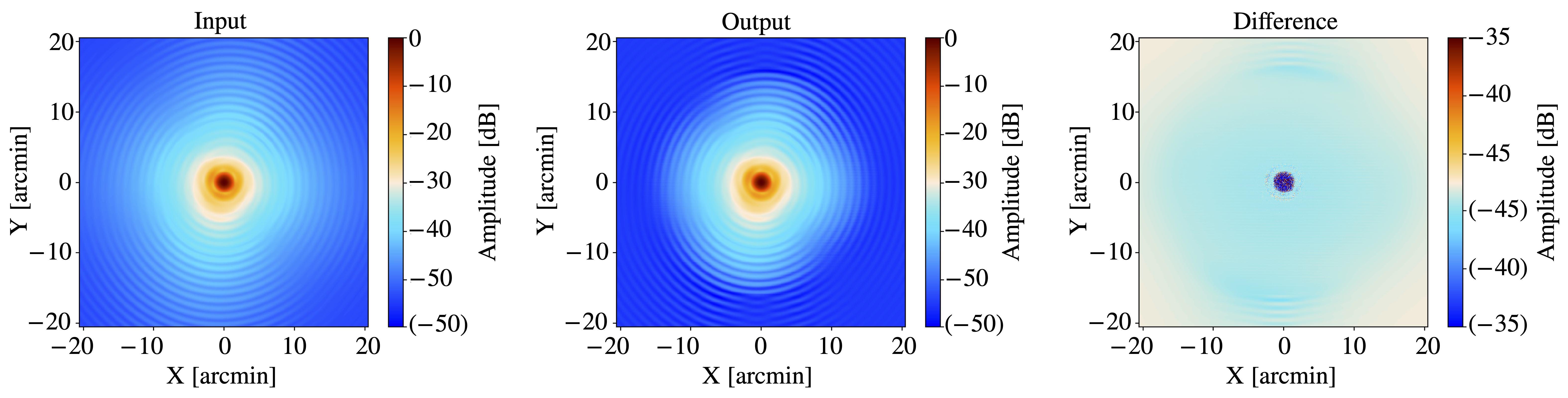}
    \caption{Example of a simulated planet observation used to characterize the mapping transfer function. This example is for S16 PA2 at 150 GHz. (Left) The input 2D beam model. (Center) One simulated observation. (Right) The difference between the \enquote{input} beam model (left) and the map of the \enquote{output} simulated observation (center).
    Note that in each case the color scale is a symmetrical log scale\footnote{A symmetrical log scale is logarithmic in both the positive and negative directions, with a small linear range around zero to avoid having values tend to infinity (\url{https://matplotlib.org/stable/api/scale_api.html\#matplotlib.scale.SymmetricalLogScale}).} in dB (with a linear threshold of $-$50 dB for the input and output maps and $-$45 dB for the difference map) and negative values are enclosed by parentheses. Here the target region where the atmosphere was estimated and removed during the map-making process was chosen to have a radius of $16^{\prime}$. The large-scale residuals seen are mostly constant within this target region of the map. The slight asymmetry in the beam (which may be safely ignored for DR4, as explained in \S\ref{subsec:asym}) is due primarily to the positions of the telescope's optics tubes in the focal plane.
    }
    \label{fig:bias_maps}
    \vspace{1em}
\end{figure*}

\begin{figure}
\vspace{1em}
    \centering
    \includegraphics[width=\linewidth]{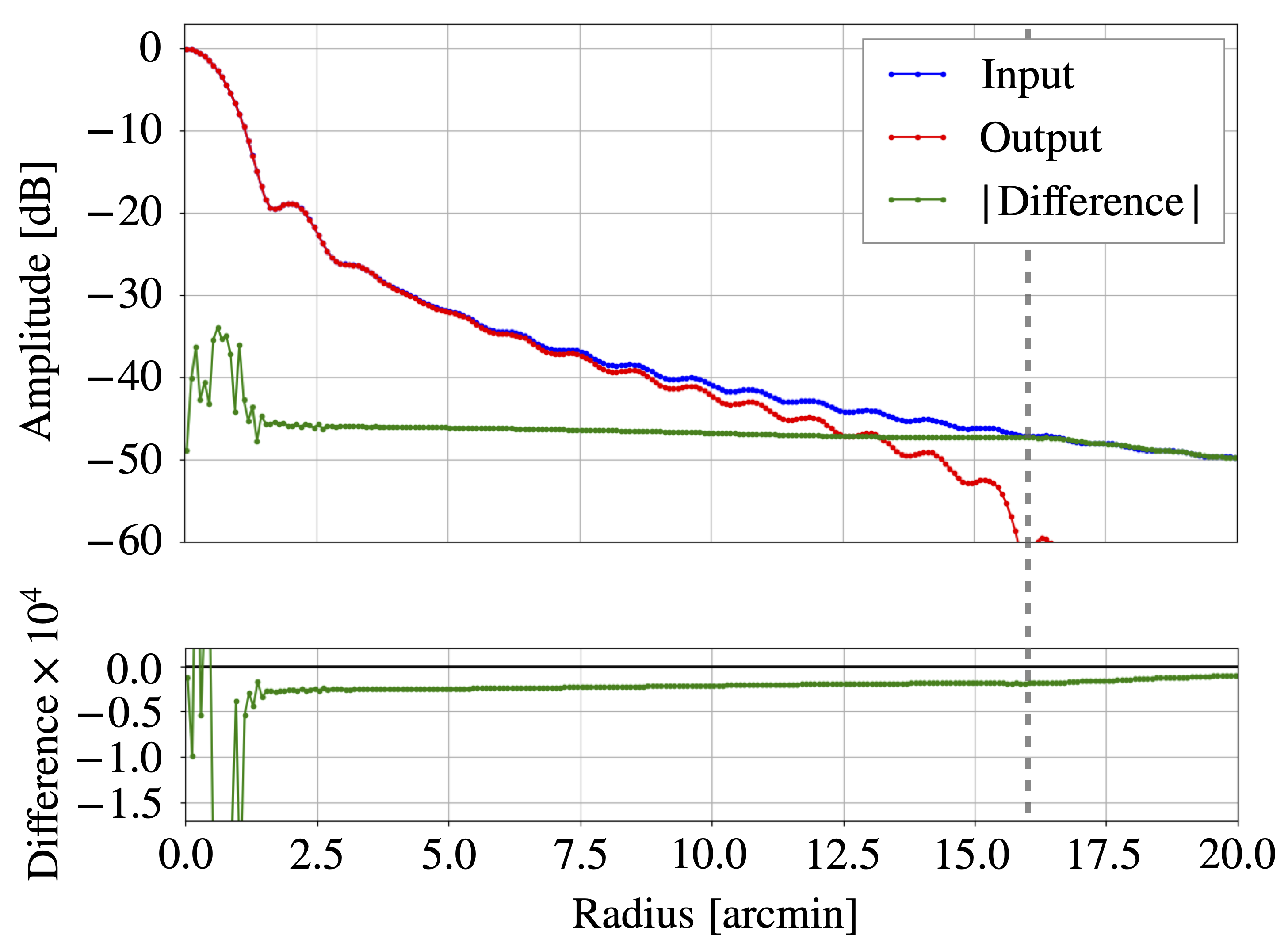}
    \caption{The average radial profile of the simulated Uranus observations for S16 PA2 at 150 GHz, comparing the azimuthal averages of the \enquote{input} 2D beam model and the mean map of the \enquote{output} simulated observations. The vertical dashed line indicates the $16^{\prime}$ radius target region of the map within which the atmosphere was estimated and subtracted. (For the Uranus maps used to characterize the beam, a $12^{\prime}$ radius target region is used, but the qualitative conclusions we draw here still apply.) Other than the small (roughly $-$40 to $-$30 dB) variations near the beam center, the difference is well approximated by a constant offset in the region from 3.5$^{\prime}$ to 10.0$^{\prime}$ where we fit for it. As described in \S\ref{subsec:trans_covmat}, we adjust our beam covariance matrices to account for possible uncertainty due to variations in the range over which we fit the offsets, exploring the three independent ranges of 3.5$^{\prime}$--5.0$^{\prime}$, 5.0$^{\prime}$--7.0$^{\prime}$, and 7.0$^{\prime}$--10.0$^{\prime}$.}
    \label{fig:bias_plot}
    \vspace{1em}
\end{figure}

An example of a map of a simulated Uranus observation is shown in Figure~\ref{fig:bias_maps}. Radial profiles are constructed by taking the azimuthal average of the input beam model and the mean output of the simulated observations. The difference between these two radial profiles represents the bias due to the atmospheric subtraction. As seen in Figure~\ref{fig:bias_plot}, other than small residual fluctuations near the center of the beam, the bias can be well approximated by a constant offset.\footnote{These small residual features only appear in the $T$ maps when solving for the maps in $T$, $Q$, $U$ simultaneously. When we solve only for $T$ and set $Q=U=0$, the fluctuations disappear. This seems to be due to an issue with the conditioning of the covariance matrix in the simulations, and is not a physical effect.} In these simulations the target region for atmosphere subtraction was chosen to have a 16$^{\prime}$ radius, but the qualitative conclusions we draw are applicable to the final planet maps used for the beam analysis which have a target region with a 12$^{\prime}$ radius.\footnote{During preliminary beam studies, maps of both Uranus and Saturn were made, with target regions both of radius 12$^{\prime}$ and 16$^{\prime}$. Saturn was found to be too bright. In the Uranus maps with a target region of radius 16$^{\prime}$, the signal-to-noise in the radial profiles past 12$^{\prime}$ was insufficient to be of use in the analysis. Hence, the Uranus maps with a target region of radius 12$^{\prime}$ were used, in order to make the best possible prediction of the atmosphere in the region of interest.} As shown in Figure~\ref{fig:bias_plot}, the effect of the atmosphere removal is almost exactly as expected: the value of the input beam at the target radius (16$^{\prime}$ in this case) is uniformly subtracted within that radius to produce the output beam.

So the planet map-making bias is well approximated by a constant offset in each map, determined by the value of the Uranus signal at the edge of the target radius used for noise estimation. This offset is estimated and subtracted prior to analysis, as described in \S\ref{subsubsec:prof}.

\section{Beam Pipeline}
\label{sec:pipe}

\subsection{Map Selection And Processing}
\label{subsec:mapsel}

A map of Uranus is made for each detector array, frequency, and observation.\footnote{So we do not make any per-detector maps. For each detector array, frequency, and observation of Uranus, we combine the data from all the detectors that meet the quality criteria mentioned in \S\ref{sec:obs} into one map.} Only a fraction of these maps are used in the final beam analysis. 
We select maps made from observations done at night (23--11 UTC) taken with the secondary mirror in its final focus position for the season in question. We pick the maps where there are not too many zero-weight pixels (which are indicative of poor scan coverage), where the signal-to-noise is sufficiently high ($>10$), and where there are not too many horizontal stripes (determined by visual inspection of the maps). 
This striping in the maps is caused by large-scale variations due to residual atmosphere.
The rejection of maps with too much striping is new, and is done to simplify the analysis.
It results in fewer maps being selected than in \cite{louis_2017}. For each season, array, and frequency, the number of maps discarded due to striping ranges from 3 to 42 (on average 19), which represents between 4\% and 36\% of the maps (on average 16\%).

\begin{figure}
\vspace{1em}
    \centering
    \includegraphics[width=\linewidth]{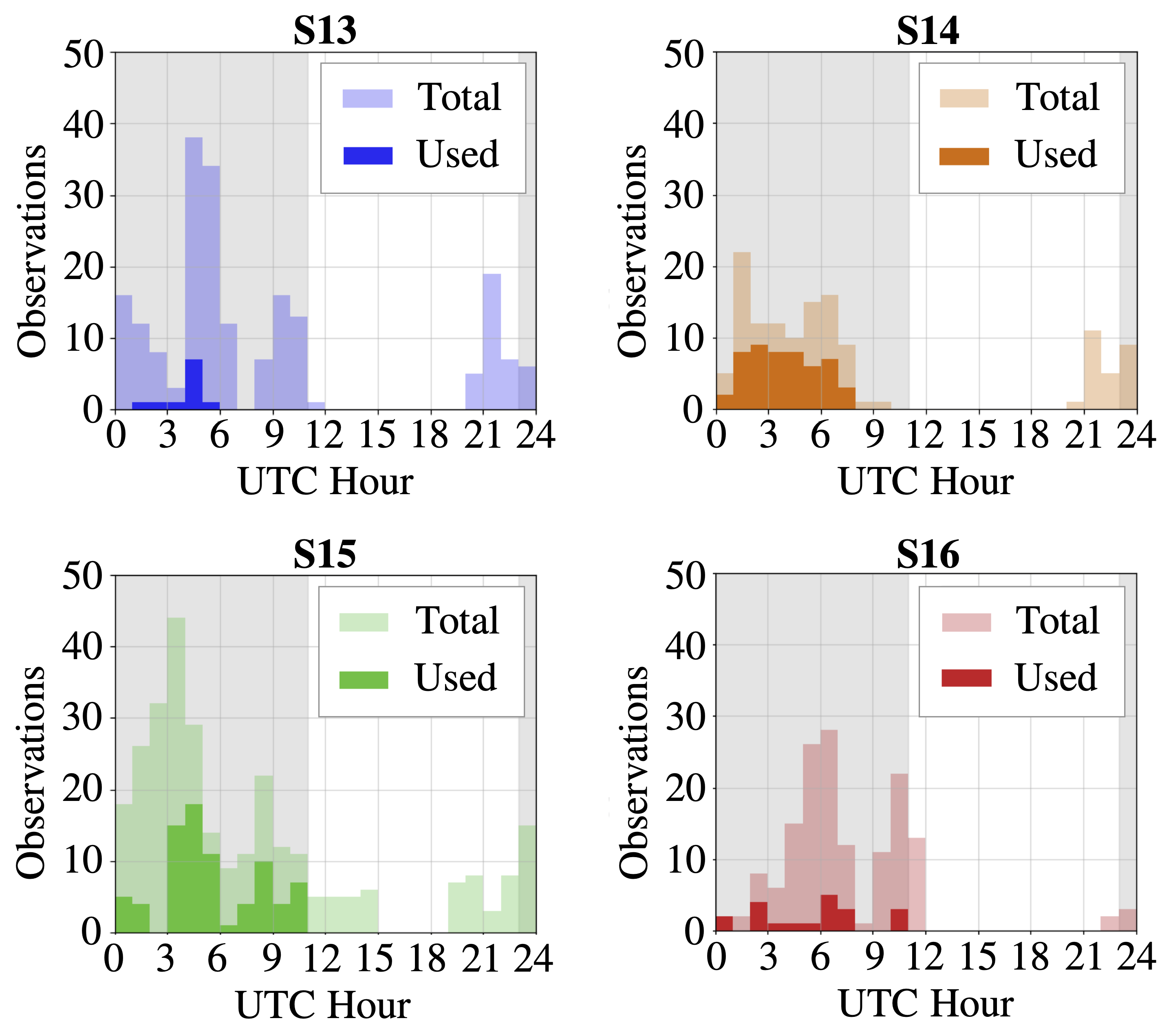}
    \caption{Distribution of the total number of Uranus observations that were made (faded colors) and ultimately became part of the final beam analysis (solid colors) for all arrays combined, shown by observing seasons from 2013--16. The shaded regions between 0 UTC and 11 UTC, as well as 23 and 24 UTC, demarcate the ACT nighttime dataset (note that local time at the observing site fluctuates between UTC-3 and UTC-4). See Table~\ref{tab:mapsel} for a summary of the number of observations used vs total for each detector array and season.}
    \label{fig:obs_summary}
    \vspace{1em}
\end{figure}

\begin{figure}
    \centering
    \includegraphics[width=6.5cm]{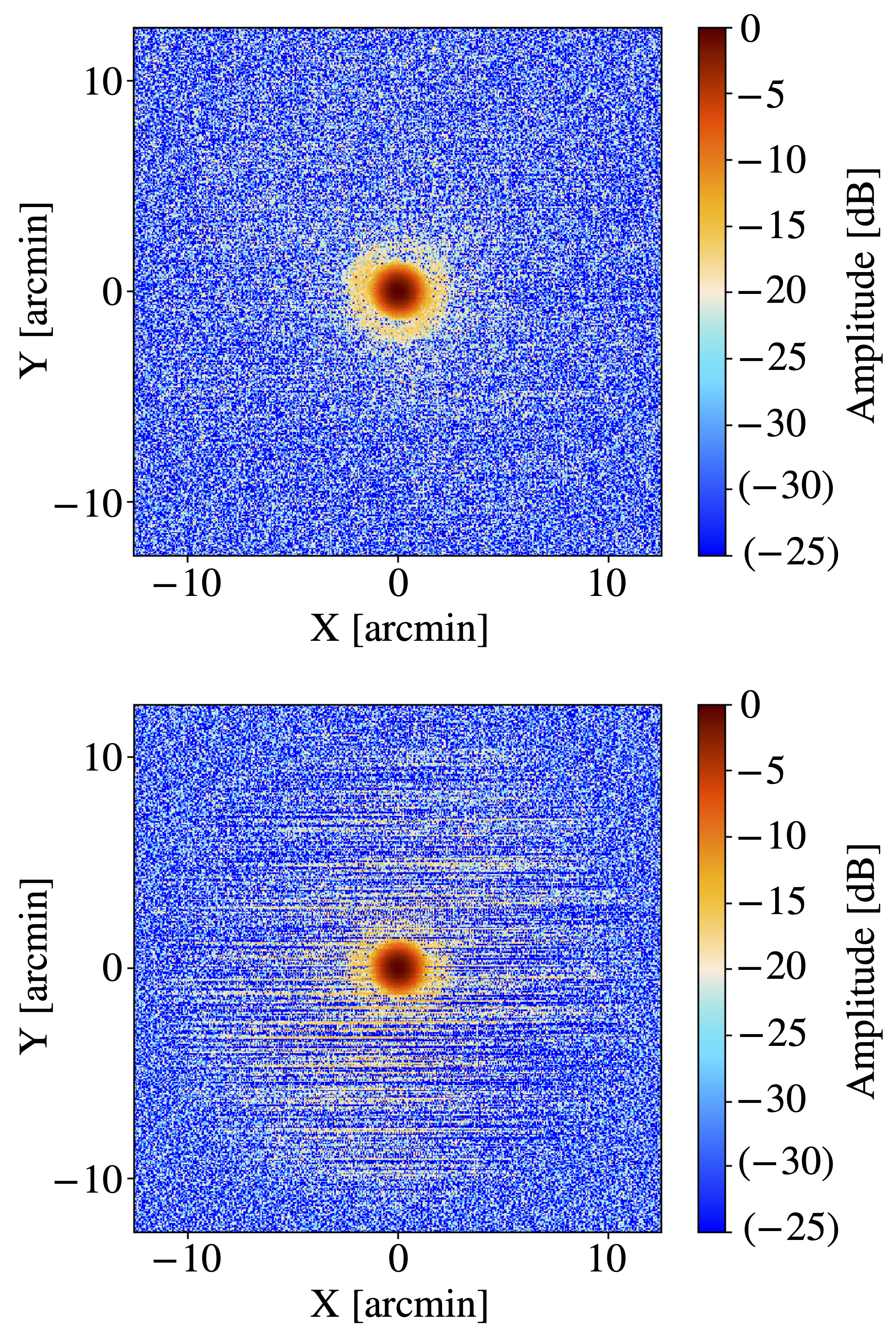}
    \caption{Two maps of individual Uranus observations for S15 PA2 at 150 GHz. (Top) A \enquote{good} map which was selected for the beam analysis. (Bottom) A \enquote{bad} map which was thrown out due to having too many horizontal stripes (caused by large-scale variations due to residual atmosphere remaining after the atmosphere subtraction described in \S\ref{subsec:filt}). In each case, the color scale is a symmetrical log scale in dB, with a linear threshold of $-$30~dB and negative values enclosed by parentheses. Any slight asymmetry visible here may be safely ignored for DR4, as explained in \S\ref{subsec:asym}, and is due primarily to the positions of the telescope's optics tubes in the focal plane.}
    \label{fig:example_maps}
    \vspace{1em}
\end{figure}

The number of Uranus maps used for the beam analysis versus the total number of observations is shown in Figure~\ref{fig:obs_summary} and Table \ref{tab:mapsel}. For example, in 2014 (S14) we made 129 observations of Uranus. In the case of PA2, we discarded 34 observations because at the time the telescope pointing was optimized for PA1 (so Uranus was not well measured by PA2), 36 observations because the resulting maps had too much striping, and 8 because the signal-to-noise was too low. This leaves 51 Uranus observations to measure the beam. An example \enquote{good} map which was selected for the beam analysis and a \enquote{bad} map which was discarded due to too much striping are shown in Figure \ref{fig:example_maps}. In S13, a large fraction of the Uranus maps were thrown out because the observations were made in the early commissioning phase of the telescope, before it had achieved its final focus.

\renewcommand{\arraystretch}{1.3}
\begin{table}
\caption{Summary of Uranus observations - number used/total.}
\centering
\begin{tabular}{ | C{2.5em} | C{2.8em} | C{3.8em} | C{2.7em} | C{2.7em} | }
\hline
Array\vspace{0.1em} & Band (GHz) & Season\vspace{0.1em} & Used\vspace{0.1em} & Total\vspace{0.1em}\\
\hline
\multirow{3}{*}{PA1} & \multirow{3}{*}{150} & S13 & 11 & 197\\
 &  & S14 &  45 & 129 \\
 &  & S15 &  17 & 133 \\
\hline
\multirow{3}{*}{PA2} & \multirow{3}{*}{150} & S14 & 51 & 129\\
 & & S15 & 38 & 164\\
 & & S16 & 11 & 86\\
\hline
\multirow{2}{*}{PA3} & \multirow{2}{*}{150} & S15 & 8 & 117\\
 & & S16 & 6 & 78\\
\hline
\multirow{2}{*}{PA3} & \multirow{2}{*}{98} & S15 & 33 & 117 \\
 & & S16 & 9 & 78 \\
\hline
\end{tabular}
\label{tab:mapsel}
\vspace{1em}
\end{table}

For the purposes of the beam measurements, we re-center each Uranus map by fitting a 2D Gaussian in the vicinity of the planet. The amplitudes from these fits are used to normalize the beam profiles to have peak values of unity. We also estimate the noise level in each map by computing the standard deviation of the pixel values outside the target region (the 12$^{\prime}$ radius area centered on the planet, described in \S\ref{subsec:filt}).

\pagebreak
\subsection{Radial Profile Fitting}
\label{subsec:prof_fit}

\subsubsection{Radial Profiles}
\label{subsubsec:prof}

The instantaneous ACT beams are slightly elliptical. However, when fitting the beam we work with azimuthally averaged (\enquote{symmetrized}) radial beam profiles, treating the beams as if they were circular. In our case, the cross-linking of the scans \citep[visible in Figure 2 of][]{choi_2020} means that the telescope beams contribute to the maps with roughly equal weight at two different orientations that are approximately 90 degrees apart for a large part of the ACT sky coverage. The effective beams are thus circularized.\footnote{In reality, anisotropic noise weighting (due to ACT's noise being correlated along the scan direction) makes the beam symmetrization from the map-maker is different from what one gets by simply averaging. The overall effect is to either make the circularized beam smaller or larger than the naive prediction, depending on the direction the raw beam is elongated relative to the scanning direction. This effect is absorbed into the jitter correction described in \S\ref{subsec:jitter}.}

This symmetrizing effect works well in temperature, but it does not help with the temperature-to-polarization leakage caused by beam asymmetry. However, this effect has been quantified and may be safely ignored for DR4, as explained in \S\ref{subsec:asym}.

The ACT beams are small enough that we use a flat sky approximation for modeling them. We denote each beam map by $B(\theta,\phi)$, where $\theta$ is the radial distance from the beam center and $\phi$ is the polar angle. We set $B(0,\phi)=1$. 
The symmetrized radial beam profile is then
\begin{equation}
    B(\theta) = \frac{1}{2\pi}\int d\phi'\; B(\theta,\phi')\; .
\end{equation}

Each map, with a target region of radius 12$^{\prime}$, is binned into a symmetrized radial profile with bins of width 11$^{\prime\prime}$ out to a radius of 10$^{\prime}$, for a total of 55 bins. The dominant source of noise in the Uranus maps comes from variations in the atmosphere which occur at relatively large angular scales, so there is significant correlation between the bins in each radial profile. 

\begin{figure}
    \centering
    \includegraphics[width=7.5cm]{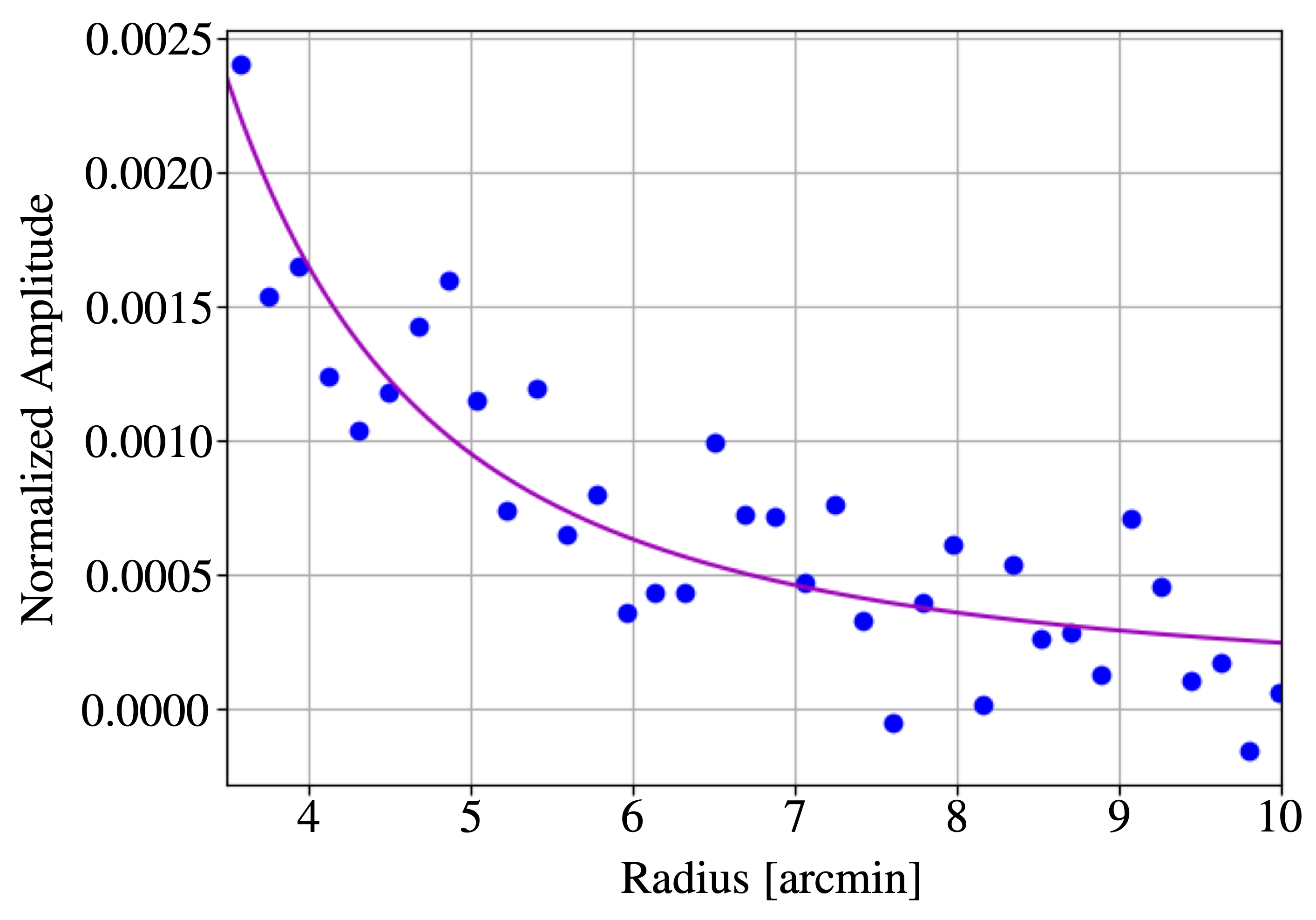}
    \caption{Example of the offset fit for the radial profile of one Uranus observation for S14 PA2 at 150 GHz. The points are found by taking the azimuthal average of the map for each radial bin. Uncertainties on these points are not computed.
    The curve is the best-fit model $\alpha/\theta^3+\beta$, where in this case $\alpha=0.08934$ and ${\beta=0.00012}$.
    }
    \label{fig:offset_fit}
    \vspace{1em}
\end{figure}

Before combining the radial profiles into one mean profile for each season, array, and frequency, each map's profile must be corrected for the offset due to the bias induced by the planet map-maker that was described in \S\ref{subsec:tf}.
This is done by fitting the model $\alpha /\theta^3 + \beta$ (motivated in \S\ref{subsubsec:asympt}) plus a scattering term (described in \S\ref{subsubsec:scatt}) to each profile in the range 3.5$^{\prime}$--10.0$^{\prime}$ and then subtracting the measured offset, $\beta$, from each. An example of such a fit is shown in Figure \ref{fig:offset_fit}.

We then compute an average radial profile independently for each season, detector array, and frequency. When averaging radial profiles, each profile is weighted by $1/N$ where $N$ is the white noise variance estimated from the corresponding map. The averaged radial profiles extend to roughly $-$40 dB of the peak, or 35 dBi (40\;dBi) at 150 GHz (98 GHz).

An advantage of computing each radial profile then taking the average, compared to making one average Uranus map then computing its radial profile (as done in \cite{hincks_2010}, for example), is that it more easily allows for the estimation of the full covariance matrix for the averaged radial profile. 
The off-diagonal elements of this matrix are important since there is significant noise correlation between the radial bins, which propagates into the low-$\ell$ uncertainty of the beam window function described in \S\ref{subsec:window}.

\subsubsection{Radial Profile Covariance Matrix}
\label{subsubsec:prof_covmat}

As the averaged radial profile is computed for each season, array, and frequency, the covariance matrix between the amplitudes of the 55 radial bins is also computed in order to account for the covariant uncertainty on large angular scales. Given the small sample size, with only between 6 and 51 profile measurements used to estimate each matrix, a shrinking algorithm \citep{schafer_2005} is used to down-weight the off-diagonal terms of the matrix. This method is described in detail in Appendix \ref{appen:shrink}. In our case this shrinking procedure is necessary as a standard estimate of the covariance matrix is often so noisy it becomes ill-conditioned and cannot be inverted.

The shrinkage technique works by combining an empirical estimate of the covariance matrix $\mathbf{S}$ (a high-dimensional estimate of the underlying covariance with little or no bias) with a model $\mathbf{T}$ (a low-dimensional estimate which may be biased but has much smaller variance) to minimize the total mean squared error (sum of bias squared and variance) with respect to the true underlying covariance: 
\begin{equation}
\label{eq:shrink_cov}
    \mathbf{C} = \widehat{\lambda}^{*}\mathbf{T} + (1-\widehat{\lambda}^{*})\mathbf{S} \; .
\end{equation}
Here $\widehat{\lambda}^{*}$ is the parameter (often referred to as shrinkage intensity) that determines the contribution of each matrix. In our case, the covariance matrix $\mathbf{S}$ is an unbiased empirical estimate of the covariance, the sample covariance matrix, and the model matrix $\mathbf{T}$ is given by the diagonal of $\mathbf{S}$:
\begin{equation}
\label{eq:target}
    T_{ij} = 
    \begin{cases}
    S_{ii} & \text{if $i=j$}\\
    0 & \text{if $i \neq j$} \; ,
    \end{cases}
\end{equation}
a common choice.
We analytically calculate the optimal combination of the low and high dimensional estimates by determining $\widehat{\lambda}^{*}$ from $\mathbf{S}$:
\begin{equation}
    \hat{\lambda}^* = \frac{\sum_{i\neq j}\widehat{\mathrm{Var}}(S_{ij})}{\sum_{i\neq j} S_{ij}^2}\; ,
\end{equation}
where $\widehat{\mathrm{Var}}(S_{ij})$ is an estimate of the variance of each covariance matrix element. The derivation of $\widehat{\lambda}^{*}$ can be found in Appendix~\ref{appen:shrink}. For the analysis presented here, $\widehat{\lambda}^{*}$ ranges from 0.24 to 1.

\subsubsection{Core Model}
\label{subsubsec:core_mod}

Since ACT's primary mirror has a diameter of ${D = 6}$\,m, at a given wavelength $\lambda$ the diffraction limited optical response of the telescope is restricted to spatial frequencies below $\ell_{\mathrm{max}} \simeq 2\pi D/\lambda$. The Fourier transform of each beam is therefore compact on a disk. Thus, a natural choice of basis functions to model the beams in harmonic space is the Zernike polynomials, an orthonormal set of basis functions that is complete on the unit disk.\footnote{The Zernike polynomials are usually used to fit effects in the electric field, not in intensity, but since they are a basis set, we can use them to fit the intensity beams. So the asymptotic behavior of $f_n$ in Equation \ref{eq:f_n} is $1/\theta^{1.5}$, not $1/\theta^3$ as we expect for the beams (as described in \S\ref{subsubsec:asympt}).} 

Expressed in polar coordinates $\rho$ and $\phi$ in the aperture plane (where $\rho$ is the radial distance $0\leq \rho \leq 1$, and $\phi$ is the azimuthal angle $0\leq \phi \leq 2\pi$), the Zernike polynomials may be written as 
\begin{equation}
V^m_n(\rho,\phi)=R^m_n(\rho)e^{i m\phi}\; ,
\end{equation}
where $m$ and $n$ are integers such that $n \geq \abs{m}$ and $R^m_n (\rho)$ are a set of radial functions defined as 
\begin{equation}
\label{eq:model}
    R^{\pm m}_n(\rho) \hspace{-0.25em} = \hspace{-0.35em}
    \begin{cases}
    \displaystyle 
    \sum_{k=0}^{\frac{n-m}{2}} \frac{(-1)^k(n-k)!}{k!(\frac{n+m}{2}-k)!(\frac{n-m}{2}-k)!}\rho^{n-2k} \hspace{1em} \parbox[t]{1.0cm}{for\hspace{1em} $n - \abs{m}$\\ even}\\
    \\
    0 \hspace{1em}\text{for $n - \abs{m}$ odd}\; .
    \end{cases}
\end{equation}

For an azimuthally symmetric beam, it is only necessary to consider the polynomials for which $m=0$, which may be expressed as 
\begin{equation}
R^0_{2n}(\rho)=P_n(2\rho^2-1)\; ,
\end{equation}
where $P_n(x)$ are Legendre polynomials (\cite{born_1980}). The Fourier transform of these polynomials is
\begin{equation}
\tilde{R}^0_{2n}(\theta) = \int_0^{1} \rho\; d\rho\hspace{0.2em} e^{-i\rho \theta}R^{0}_{2n}(\rho) = \frac{(-1)^n J_{2n+1}(\theta)}{\theta}\;,
\end{equation}
where $J_{2n+1}$ is a Bessel function of the first kind.

Motivated by this, and following previous analyses \citep{hasselfield_atacama_2013}, we adopt the basis functions 
\begin{equation}
    f_n(\theta \ell_{\mathrm{max}}) = \frac{J_{2n+1}(\theta \ell_{\mathrm{max}})}{\theta \ell_{\mathrm{max}}}
    \label{eq:f_n}
\end{equation}
to fit the core of the beams in real space, where the parameter $\ell_{\mathrm{max}}$ varies the scale of the function's argument and $n$ is a non-negative integer.

\subsubsection{Asymptotic Behavior}
\label{subsubsec:asympt}
On scales smaller than a few arcminutes, the noise is subdominant to the planetary signal, and so the beam profiles are well measured even by a single observation of Uranus. However, at larger angular scales, the noise quickly becomes non-negligible. Thus, the asymptotic behavior of the beams cannot be separated from the background without making some assumptions about the shape of the beams far from the core.

The illumination of the ACT optics is controlled by a cryogenic Lyot stop. The beam's Fraunhofer diffraction pattern for monochromatic radiation is described by the squared modulus of the Fourier transform of this circular aperture \citep{born_1980}. The resulting intensity response with angle, or Airy pattern, is
\begin{equation}
A(\theta) = \Big[\frac{2J_1[k(D/2)\sin\theta]}{k(D/2)\sin\theta} \Big]^2 \; ,
\end{equation}
where $k=2\pi/\lambda$ is the wavenumber.\footnote{Note that this corresponds to the $f_0$ term from the core model described in \S\ref{subsubsec:core_mod}.}
For ACT, $k(D/2)=\pi D/\lambda$ is large, since the primary is several thousand wavelengths across. 
When the argument of the Bessel function is $\gg 1$ and real, as is the case for $\theta > 2^{\prime}$, we can make the approximation

\begin{equation}
J_{1}(x) = \sqrt{\frac{2}{\pi x}}\cos(x -\frac{3\pi}{4})
\end{equation}
\citep{abramowitz_stegun}. In this regime, the Airy pattern asymptotically approaches

\begin{equation}
A(\theta) \simeq  \frac{8\cos^2[k(D/2)\sin\theta-3\pi/4]}{\pi [k(D/2)\sin\theta]^3} \; ,
\end{equation}
such that the envelope of $A(\theta)$ falls as $1/\sin^3\theta$. This implies that at angles larger than the \enquote{near sidelobes} (described in \S\ref{subsec:sidel}) and neglecting the effects of scattering, the beams should fall as $1/\sin^3\theta \simeq 1/\theta^3$, since $\theta$ is small in the region considered for the beam models. 

A different, more general, way to derive this asymptotic behavior of the beams comes from the geometric theory of diffraction from Keller \citep{keller1, keller2, keller3}. As detailed in \cite{page_2003}, for an illuminated disk, the diffraction pattern from the rim, regardless of the interior, leads to a gain (response) of the shape
\begin{subequations}
\begin{align}
        G(\theta) & \propto \frac{1}{\sin\theta}\Big(\frac{1}{\sin\theta/2}\pm \frac{1}{\cos \theta/2} \Big)^2\\
        & \propto \frac{1}{\sin^3 \theta}(1\pm \sin\theta),
\end{align}
\end{subequations}
for an electric field parallel (+) or perpendicular ($-$) to the edge, in the far field-limit. For unpolarized light, one can simply average over both polarizations.

The beam behavior derived here assumes perfectly smooth and uniform surfaces within the telescope (i.e. perfect coherence). Nonuniformity, including mirror surface roughness, leads to diffuse scattering (i.e. decoherence of the fields), and thus a reduction in the telescope gain \citep{ruze_1966}. As described in the next section, an additional term is included in our beam model to account for this effect.

\subsubsection{Scattering}
\label{subsubsec:scatt}

The primary telescope surface can be characterized by an rms surface roughness, $\epsilon$, and a correlation length, $c$. We measured these properties for ACT using photogrammetry,\footnote{VSTARS by GSI, \url{https://www.geodetic.com/v-stars/}} placing 284 targets on the corners of the primary's 71 panels where the panels join. The rms of the raw measurements was found to be approximately $30$\,$\mu$m. However, these measurements are not a representative sample of the rms of the surface, since the targets are at the most extreme points on the corners of the panels. By fitting a polynomial surface to the measurements and looking at the residuals over the entire surface, we find an rms of $\epsilon = 20$\,$\mu$m. Using the photogrammetry software, the uncertainty on the rms measurements is estimated to be $10$\,$\mu$m. This does not include any misalignments due to macroscopic deformations of the telescope.

From these measurements, we can also compute the correlation length, $c$. The correlation function is 
\begin{equation}
C(d) = \sum_{ij}z(\vec{r_i})z(\vec{r_j}) = C_0 e^{-d^2/c^2} \; ,
\label{eq:corr_func}
\end{equation}
where $\vec{r}$ is the position on the surface, $z$ is the mean-subtracted departure from the best-fit designed surface \citep[described in Section 3 of][]{fowler_2007}, $d = |\vec{r}_i - \vec{r}_j|$ is the separation, and the sum is over all measurement pairs with a separation $d$. We find that the shape of the correlation function follows the above form reasonably well. By fitting the photogrammetry measurements with Equation \ref{eq:corr_func}, for the ACT primary, we find $c = 280$\,mm. This scattering leads to another angular scale (or shoulder) in the beam response, not governed by the diameter of the Lyot stop.

For a beam $B(\theta)$ normalized to unity at $\theta=0$, the corresponding solid angle $\Omega$ is 
\begin{equation}
\Omega = 2\pi \int_{0}^{\pi}B(\theta)\sin\theta\hspace{0.2em} d\theta \; .
\label{eq:solid_angle}
\end{equation}
The expression for the gain due to scattering off a rough surface, $G(\theta)$, given by Equation 8 in \cite{ruze_1966}\footnote{There is a typo in this equation in the original paper. Inside the summation, the variance $\delta^2$ should be raised to the $n^{\mathrm{th}}$ power, as written here, instead of being simply squared, as in \cite{ruze_1966} \citep{ruze_typo}.} can be re-written in terms of the unit-normalized beam for an ideal reflector, $B_0(\theta)$, its corresponding solid angle, $\Omega_0$, the total beam, $B(\theta)$, and its corresponding solid angle, $\Omega$, by using the relation $G(\theta) = (4\pi/\Omega)B (\theta)$. We then obtain the equation
\begin{equation}
    B(\theta)=\frac{\Omega}{\Omega_0}B_0(\theta)e^{-\expval{\delta^2}}+S(\theta),
\label{eq:ruze}
\end{equation}
with the $S(\theta)$ term given by 
\begin{equation}
    S(\theta) = \frac{\Omega}{4\pi}\Big(\frac{2\pi c}{\lambda} \Big)^2e^{-\expval{\delta^2}}\sum_{n=1}^{n=\infty}\frac{\expval{\delta^2}^n}{n\cdot n!}e^{-(c\pi\sin(\theta)/\lambda)^2/n} \; ,
\end{equation}
where $\expval{\delta^2}=(4\pi\epsilon/\lambda)^2$ is the variance in the surface phase (and $\epsilon$ and $c$ are the surface rms and correlation length, as described earlier).

The sum converges quickly, with four terms sufficient for our purposes. We have simulated the ACT beam by taking the Fourier transform of a perturbed surface with deformations of a Gaussian shape and verified that we indeed recover the Ruze equation as it is written above.\footnote{More specifically, deformations of a Gaussian shape of width (standard deviation) 250\,mm were placed on a square grid of dimension 580\,mm on a side to create deformations over a surface. This gave 71 squares inside a diameter of 5500\,mm, the effective diameter of ACT's primary. This is a reasonable approximation given that the primary has 71 panels \citep{thornton_2016}. The amplitudes of the deformations were drawn from a normal distribution, and the overall rms of the surface was adjusted to be $\epsilon=100$\,$\mu$m, and the modeled surface had $c=440$\,mm. The resulting beam followed the Ruze equation for these values of $\epsilon$ and $c$.}

The first term in Equation~\ref{eq:ruze} above is expected to decay as $1/\theta^3$, but the same does not hold true for $S(\theta)$, so the inclusion of this term in our fitting procedure is important as it affects how the beam fits are extrapolated. This scattering term $S(\theta)$, which we refer to as the Ruze beam, contains roughly 1.5\% of the solid angle of the main beams at 150 GHz. The addition of this term to the beam model is new to the DR4 analysis. 

\subsubsection{Beam Model Fitting}
\label{subsubsec:fit}

We separate the main beam for each season, array, and frequency into two domains: the core, where we fit the set of basis functions described in \S\ref{subsubsec:core_mod}, and the wing, where the model is composed of a $1/\theta^3$ term in addition to the Ruze beam (where we use the measured values for $\epsilon$ and $c$ described in the previous section). As explained in \S\ref{subsubsec:prof}, a constant offset has already been subtracted from each measured beam profile before combining them into an average radial profile for each season, detector array, and frequency. The full model for each averaged beam profile can be written as 
\begin{equation}
\label{eq:model}
    B(\theta) = 
    \begin{cases}
    \displaystyle 
    \sum_{n=0}^{n_{\mathrm{mode}}-1}a_n f_n(\theta \ell_{\mathrm{max}})\hspace{1.5em} \text{for $\theta \leq \theta_1$}\\
    \\
    {\alpha}/{\theta^3} + S(\theta)\hspace{4.915em}  \text{for $\theta_1 < \theta$} \; .
    \end{cases}
\end{equation}

The initial linear least squares fits are performed for the $a_n$ and $\alpha$ for a range of values of $\ell_{\mathrm{max}}$, $\theta_1$, and $n_{\mathrm{mode}}$ (allowing $n_{\mathrm{mode}}$ to vary up to a maximum value of half the number of data points in the core). We first sample a range of values for the scaling parameter $\ell_{\mathrm{max}}$ and then allow this non-linear parameter to vary later on. Based on the optics of the system, at 150 GHz (98 GHz)  we expect to have $\ell_{\mathrm{max}} \simeq 2\pi D/\lambda \simeq$ 19,000 (12,500). We sample $\ell_{\mathrm{max}}$ in integer steps from 1 to 30,000.

We fit the model out to 10$^{\prime}$, a value which is chosen because this is the radius out to which the beam is measured to sufficient signal-to-noise. Despite the atmosphere subtraction in the target region that extends out to 12$^{\prime}$, residual modes remain, so the signal-to-noise out to 12$^{\prime}$ is not sufficient. By only fitting out to 10$^{\prime}$ we ensure that the fit remains in the region where the measurement is dominated by the beam, not atmosphere.

\begin{figure}
    \centering
     \includegraphics[width=\linewidth]{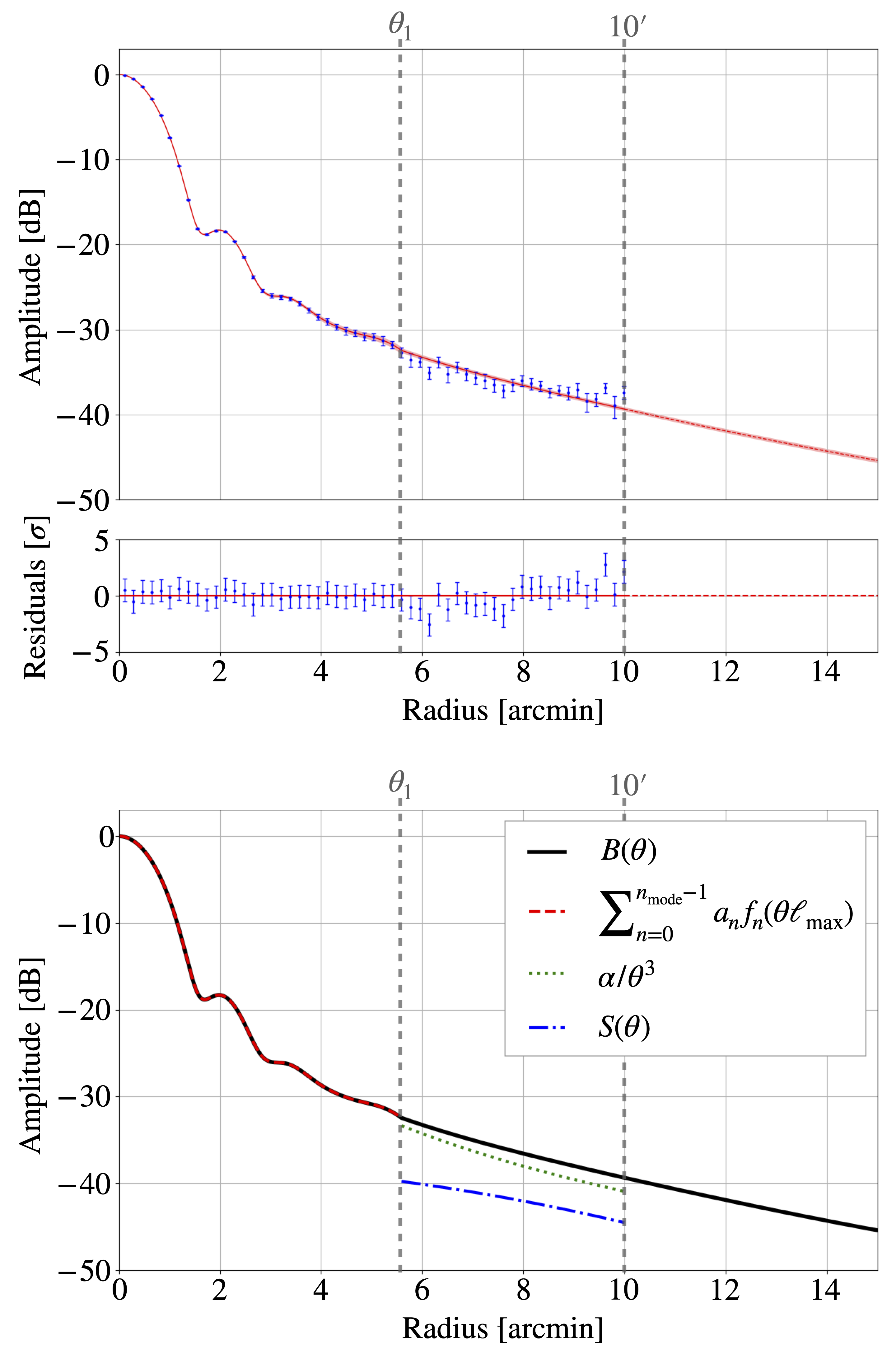}
    \caption{Top: Measured radial beam profile (blue) for S15 PA2 at 150 GHz and the best-fit model (red), with the (narrow) red shaded region indicating the 1$\sigma$ model uncertainty bounds, and residuals shown in the middle panel. Note that the bins are correlated. Bottom: The terms in the model for the beam. Here the final best-fit value of $\theta_1$, where the model for the beam changes, is 5.59$^{\prime}$. The number of modes in the core fit ($n_{\mathrm{mode}}$) is 13.
    }
    \label{fig:profile_t}
    \vspace{1em}
\end{figure}

The model is fit to each averaged radial profile, with its associated covariance matrix. We do not include continuity conditions at $\theta_1$, but any small discontinuities are well below ACT's resolution.\footnote{For example, the fit shown in the top panel of Figure~\ref{fig:profile_t} may have a slight discontinuity in amplitude at $\theta_1$. However, as part of the beam processing we take the harmonic transform of our beam model and then transform back to radial space to obtain the final beam profiles shown in Figure~\ref{fig:inst_beams}. Any small discontinuity that may have been present in the initial fit disappears due to the resolution with which we do the transform.}

In order to select the optimal number $n_{\mathrm{mode}}$ for each beam profile, we must strike the right balance between minimizing the $\chi^2$ and not adding too many parameters. To do this, we use the corrected Akaike information criterion (AICc), which estimates the relative quality of models based on both their goodness of fit and their simplicity. For each choice of $\ell_{\mathrm{max}}$, $\theta_1$ and $n_{\mathrm{mode}}$, the AICc is computed for the best-fitting model. 

The uncorrected AIC is given by
\begin{equation}
\mathrm{AIC} = 2k - 2\ln(\hat{\mathcal{L}}) \; ,
\end{equation}
where $k$ is the number of estimated parameters in the model and $\hat{\mathcal{L}}$ is the maximum value of the likelihood function for the model \citep{akaike_1973, akaike_1974}. For small sample sizes, as is the case here, the AIC can exhibit a large bias. To account for this, we use the corrected criterion AICc, where
\begin{equation}
\mathrm{AICc}=\mathrm{AIC}+ \frac{2k^2+2k}{n-k-1} \; ,
\end{equation}
and $n$ is the sample size \citep{hurvich_1989}. 

We select the values for $\ell_{\mathrm{max}}$, $\theta_1$, and $n_{\mathrm{mode}}$ corresponding to the lowest AICc.\footnote{In reality, this is not always precisely the case. When there exists a set of values with $n_{\mathrm{mode}}$ smaller by one (hence one fewer parameter) and an AICc that is not significantly different (${\Delta\mathrm{AICc}<2}$), we actually select the model with the slightly higher AICc. We do this to avoid over-fitting the data, since in this case the AICc does not justify the addition of the extra parameter.}\textsuperscript{,}\footnote{In earlier analyses \citep{louis_2017, hasselfield_atacama_2013}, we simply increased $n_{\mathrm{mode}}$ until the reduced-$\chi^2$ fell below 1.} 
For the DR4 beams, the best-fit values for $\ell_{\mathrm{max}}$ range from 15,604 to 17,936 (from 10,960 to 11,285) at 150 GHz (98 GHz), the best-fit values for $\theta_1$ range from 3.58$^{\prime}$ to 7.79$^{\prime}$, and the best-fit values for $n_{\mathrm{mode}}$ range from 9 to 13.
We then use MCMC to sample the posterior distribution of the parameters $\ell_{\mathrm{max}}$, the $a_n$, and $\alpha$, assuming uniform priors \citep{metropolis_1953}. This method produces an estimate of the parameter means and the full covariance between all the basis functions and wing parameters, including the non-linear scaling parameter $\ell_{\mathrm{max}}$.

The term $S(\theta)$ in Equation~\ref{eq:model} depends on the beam's solid angle, $\Omega$, but the calculation of $\Omega$ depends on the beam model. To get around this issue, the beam fit is performed iteratively, first using a rough estimate for $\Omega$, and then re-computing $\Omega$ once a beam model is obtained, and then performing the fit again. In total, we fit the beam four times, re-estimating $\Omega$ each time. By the last iteration, the change in $\Omega$ becomes undetectable, and so we have converged to a final value for the beam solid angle.

\begin{figure*}
    \centering
    \includegraphics[width=\linewidth]{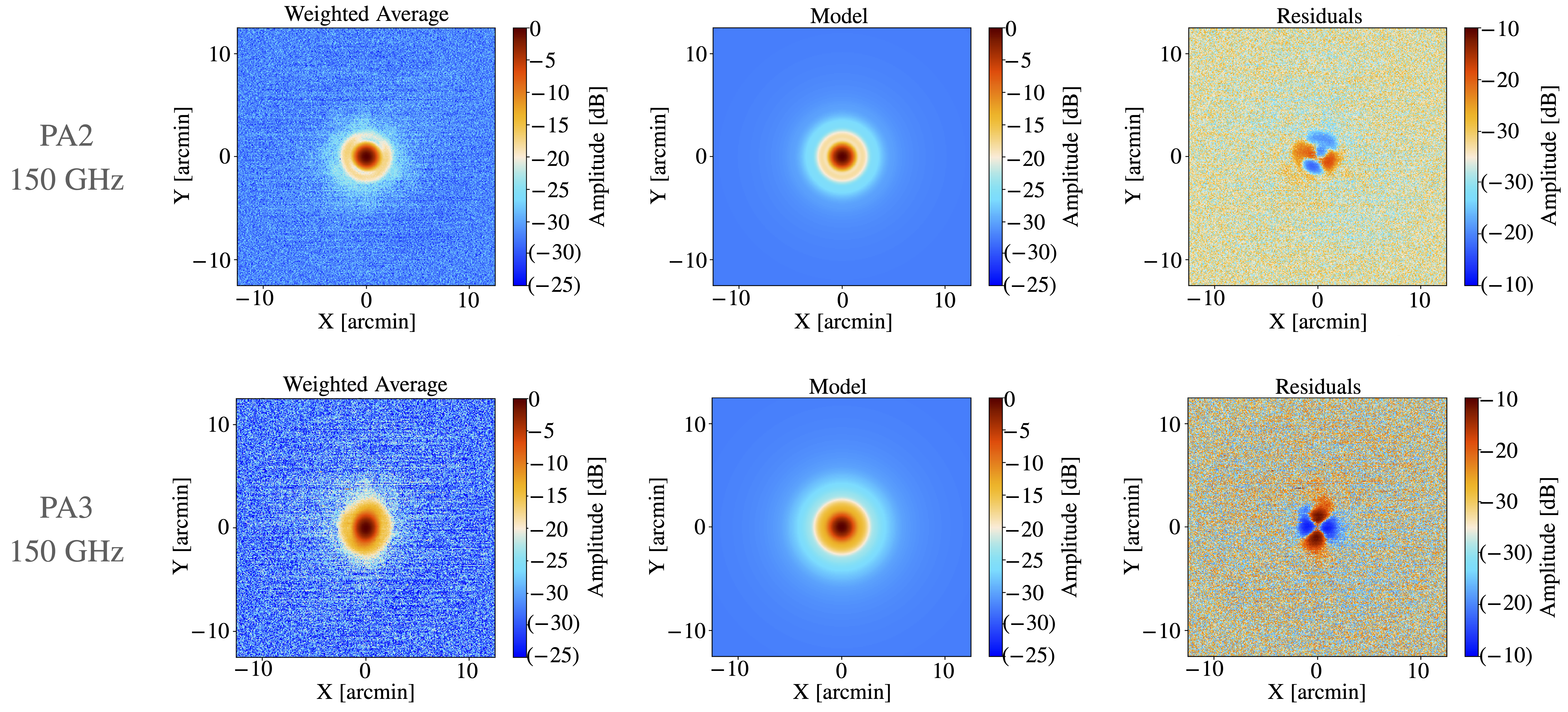}
    \caption{Maps of the main beam for S15 PA2 (top) and PA3 (bottom) at 150 GHz. In each case, the color scale is a symmetrical log scale in dB, with a linear threshold of $-$30 dB and negative values enclosed by parentheses. (Left) The weighted average of the Uranus observations used to characterize the beam. (Center) The model which is fit to the radial profile of the measured beam. (Right) The difference between the measured beam (left) and the beam model (center). Note that the color scale for the residuals is different from the other two maps. 
    Both PA2 and PA3 are shown here, since PA2 serves as an example of a less elliptical beam, and PA3 serves as an example of a more elliptical one.
    The corresponding maps for S15 PA1 at 150 GHz resemble the ones shown for PA2 here, and the maps for S15 PA3 at 98 GHz are similar to the ones shown here for PA3 at 150 GHz, but broader. For each individual detector array, the residuals are fairly constant from one season to the next.
    As shown in Figure \ref{fig:profile_t} for PA2, the azimuthal average of the residuals is consistent with zero, which is why the fit is successful, despite the residuals visible in the maps.
    These residuals are expected to have a quadrupole-like shape, since we know our beams are slightly elliptical, and the quadrupole is the dominant asymmetric azimuthal mode for an elliptical beam.
    }
    \label{fig:map_t}
    \vspace{1em}
\end{figure*}

We did test whether the data prefer an asymptotic wing fit term that differs from $\alpha/\theta^3$, by fitting for $\alpha/\theta^b$ instead. 
While the best-fitting value for $b$ often differs (by 10--20\%) from three, it is not strongly constrained by the data. 
The AICc indicates that the addition of this new parameter is not warranted.

An example of the beam model fit to radial profile data is shown in Figure~\ref{fig:profile_t}, for the S15 PA2 Uranus maps at 150~GHz, together with the residuals. The core functions, $\alpha/\theta^3$ term, and the Ruze beam are indicated in the lower panel, with the core functions used out to a radius of ${\theta_1=5.59'}$, then transitioning to the $\alpha/\theta^3$ plus Ruze beam model at larger radii, and using this model to extrapolate past 10$^{\prime}$. We measure the beam profiles down to $-$40 dB from the peak, which would leave a few percent of the beam's solid angle unaccounted for if we did not use our fit to extrapolate past 10$^{\prime}$.

Figure~\ref{fig:map_t} shows how the symmetrized model compares to the average Uranus map data for this case. The full set of radial profile fits for the S15 data is shown in Figure~\ref{fig:inst_beams} (with some small additional corrections, as described in \S\ref{subsec:add_corr}). 
We also obtained model fits for the S13, S14 and S16 data that make up DR4. The solid angles, gains, and FWHMs for all seasons are reported in Table~\ref{tab:inst_sa} (again, with the small corrections from \S\ref{subsec:add_corr}). For reference, $1\; \mathrm{nsr}\; \simeq \; 0.0118\; \mathrm{arcmin}^2$.

Even though we make a high signal-to-noise measurement of each beam, the uncertainty on the solid angle is limited to $\sim$ 2\% because of the uncertainty in extrapolation, which depends on the model. The fractional uncertainties in the solid angles at 98 GHz may be larger than those at 150 GHz in part because at 98 GHz the beam is broader, so a larger fraction of the beam is affected by the uncertainty in our extrapolation. 

When using the beam solid angle to calculate the flux density of a point source, the uncertainty on the measurement may be reduced by first applying a matched filter to both the beam and the source in order to remove the scales associated with the extrapolation uncertainty.

\begin{figure*}
    \centering
    \includegraphics[width=\linewidth]{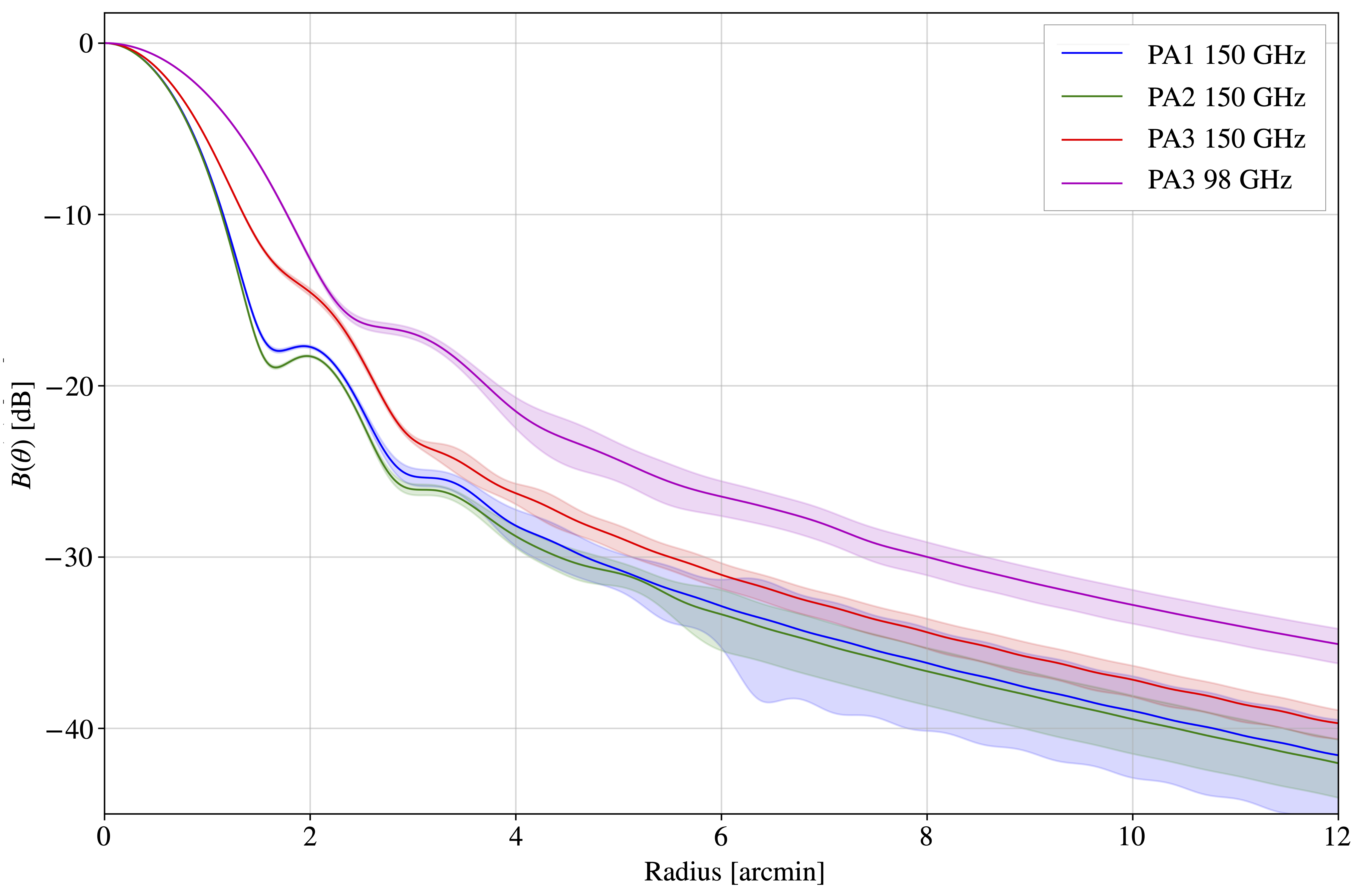}
    \caption{The estimated instantaneous beam profiles for S15, for the three arrays (PA1, PA2 and PA3), and for both 98~GHz and 150~GHz for PA3. The shaded bands indicate the 1$\sigma$ uncertainty bounds. The uncertainties are strongly correlated between radial bins. The S13, S14 and S16 beam profiles are similar, and are included in the public data release. At 150 GHz, the beam for PA3 is quite different from those for PA1 and PA2. This is because PA3 was not focused as well, due to its position, further off-axis, compared to PA1 and PA2.}
    \label{fig:inst_beams}
    \vspace{1em}
\end{figure*}

While much of this fitting method follows the approach used in \cite{hasselfield_atacama_2013} and \cite{louis_2017}, notable improvements are the addition of the scattering beam term in the model, the use of MCMC sampling that includes estimating the non-linear parameter $\ell_{\mathrm{max}}$, the exploration of different radii ($\theta_1$) out to which the core model is fit, and the use of the AICc to choose the final best-fit model.

\renewcommand{\arraystretch}{1.3}
\begin{table}
\caption{Instantaneous beam solid angles, gains, and FWHMs}
\centering
\setlength\tabcolsep{1.5pt}
\begin{tabular}{ | C{2.55em} | C{2.8em} | C{3.0em} | C{5.5em} | C{6.5em} | C{6.1em} | }
\hline
Array \vspace{-0.3cm} & Band (GHz) & Season \vspace{-0.3cm} & Solid Angle (nsr) & Forward Gain (dBi) & FWHM (arcmin)\\
\hline
\multirow{3}{*}{PA1} & \multirow{3}{*}{150} & S13 & 201.5\hspace{-0.2em} $\pm$\hspace{-0.2em} 3.8 & 77.94\hspace{-0.2em} $\pm$\hspace{-0.2em} 0.08 & 1.330\hspace{-0.2em} $\pm$\hspace{-0.2em} 0.001\\
 &  & S14 &  198.5\hspace{-0.2em} $\pm$\hspace{-0.2em} 3.3 & 78.01\hspace{-0.2em} $\pm$\hspace{-0.2em} 0.07 & 1.330\hspace{-0.2em} $\pm$\hspace{-0.2em} 0.002\\
 &  & S15 &  196.5\hspace{-0.2em} $\pm$\hspace{-0.2em} 8.8 & 78.06\hspace{-0.2em} $\pm$ \hspace{-0.2em}0.19 & 1.321\hspace{-0.2em} $\pm$\hspace{-0.2em} 0.002\\
\hline
\multirow{3}{*}{PA2} & \multirow{3}{*}{150} & S14 & 183.1\hspace{-0.2em} $\pm$\hspace{-0.2em} 3.2 & 78.37\hspace{-0.2em} $\pm$\hspace{-0.2em} 0.08 & 1.310\hspace{-0.2em} $\pm$\hspace{-0.2em} 0.001\\
 & & S15 & 187.8\hspace{-0.2em} $\pm$\hspace{-0.2em} 4.7 & 78.26\hspace{-0.2em} $\pm$\hspace{-0.2em} 0.11 & 1.311\hspace{-0.2em} $\pm$\hspace{-0.2em} 0.001\\
 & & S16 & 185.4\hspace{-0.2em} $\pm$\hspace{-0.2em} 4.8 & 78.31\hspace{-0.2em} $\pm$\hspace{-0.2em} 0.11 & 1.316\hspace{-0.2em} $\pm$\hspace{-0.2em} 0.001\\
\hline
\multirow{2}{*}{PA3} & \multirow{2}{*}{150} & S15 & 269.6 \hspace{-0.2em}$\pm$\hspace{-0.2em}5.5 & 76.68\hspace{-0.2em} $\pm$\hspace{-0.2em} 0.09 & 1.461\hspace{-0.2em} $\pm$\hspace{-0.2em} 0.002\\
 & & S16 & 237.5\hspace{-0.2em} $\pm$\hspace{-0.2em} 8.5 & 77.24\hspace{-0.2em} $\pm$\hspace{-0.2em} 0.16 & 1.444\hspace{-0.2em} $\pm$\hspace{-0.2em} 0.003\\
\hline
\multirow{2}{*}{PA3} & \multirow{2}{*}{98} & S15 & 503.9\hspace{-0.2em} $\pm$\hspace{-0.2em} 21.8 & 73.97\hspace{-0.2em} $\pm$\hspace{-0.2em} 0.19 & 2.001\hspace{-0.2em} $\pm$\hspace{-0.2em} 0.004\\
 & & S16 & 476.3\hspace{-0.2em} $\pm$\hspace{-0.2em} 21.8 & 74.21\hspace{-0.2em} $\pm$\hspace{-0.2em} 0.20 & 2.002\hspace{-0.2em} $\pm$\hspace{-0.2em} 0.004\\
\hline
\end{tabular}
\label{tab:inst_sa}
\vspace{1em}
\end{table}

\subsection{Beam Window Functions}
\label{subsec:window}

In spherical harmonic space, the beam information is encoded in the harmonic transform $b_{\ell}$ and the window function $w_{\ell} = b_{\ell}^2$, which describes the instrument's response to different multipoles, $\ell$. This window function is an essential component of the DR4 power spectrum analysis in \cite{choi_2020}.

The harmonic transform $b_{\ell}$ is the Legendre transform, or more accurately the Legendre polynomial transform, of the beam radial profile:
\begin{equation}
b_{\ell} = \frac{2\pi}{\Omega}\int_{-1}^{1} B(\theta)P_{\ell}(\cos\theta)\; d(\cos\theta) \; .
\label{eq:legendre}
\end{equation}
For small beams such as that of ACT, this is effectively a Fourier transform. The derivation of the Legendre transform and details about how the transform is computed are presented in Appendix \ref{appen:trans}.

We use $b_{\ell}$ instead of $B_{\ell}$ to indicate the division by $\Omega$, which normalizes $b_{\ell}$ to unity at $\ell = 0$ (since $P_0 = 1$). $B_{\ell} = \Omega b_{\ell}$ has units of $\mathrm{sr}$, whereas $b_{\ell}$ is dimensionless. We extrapolate the model beyond the fit radius of 10$^{\prime}$ when computing the transform.
This is necessary to capture the low-$\ell$ part of the window function, and to account for the part of the beam solid angle that is beyond the range we fit.

\begin{figure*}
    \centering
    \includegraphics[width=\linewidth]{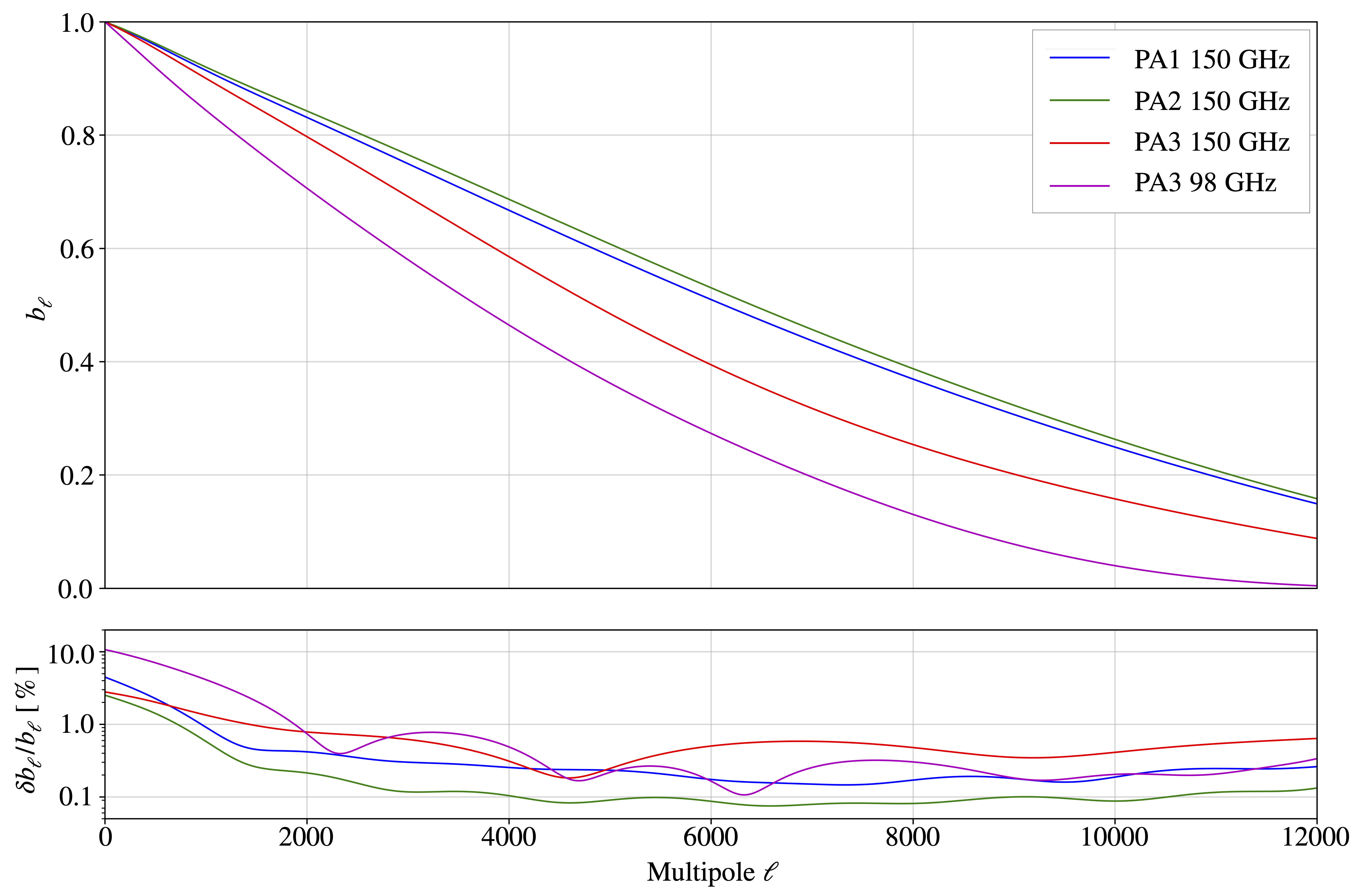}
    \caption{Instantaneous beam transforms and their uncertainties for the S15 data, for all the detector arrays. The uncertainties are strongly correlated between multipoles. For context when looking at this figure along with Figure~\ref{fig:inst_beams}, in the mapping from angle to multipole ($\ell \simeq \pi/\theta$), 1$^{\prime}$ corresponds to $\ell \simeq 10800$, 1.8$^{\prime}$ corresponds to $\ell \simeq 6000$, and 10.8$^{\prime}$ corresponds to $\ell \simeq 1000$.}
    \label{fig:inst_tforms}
    \vspace{1em}
\end{figure*}

A subset of the beam transforms from DR4 is shown in Figure~\ref{fig:inst_tforms}. A similar figure in \cite{aiola_2020} shows window functions, $b_{\ell}^2$, which are used to correct the power spectra. For a given array and season, if the beam transforms for PA1 and PA2 are, respectively, $b_{\ell}^{\mathrm{PA1}}$ and $b_{\ell}^{\mathrm{PA2}}$, then for the auto-power spectrum of the PA1 or PA2 maps the window functions are $(b_{\ell}^{\mathrm{PA1}})^2$ and $(b_{\ell}^{\mathrm{PA2}})^2$, and for the cross-spectrum of the PA1 and PA2 maps the window function is $b_{\ell}^{\mathrm{PA1}}b_{\ell}^{\mathrm{PA2}}$.

\subsection{Additional Corrections}
\label{subsec:add_corr}

\begin{figure}
\vspace{1em}
    \centering
    \includegraphics[width=\linewidth]{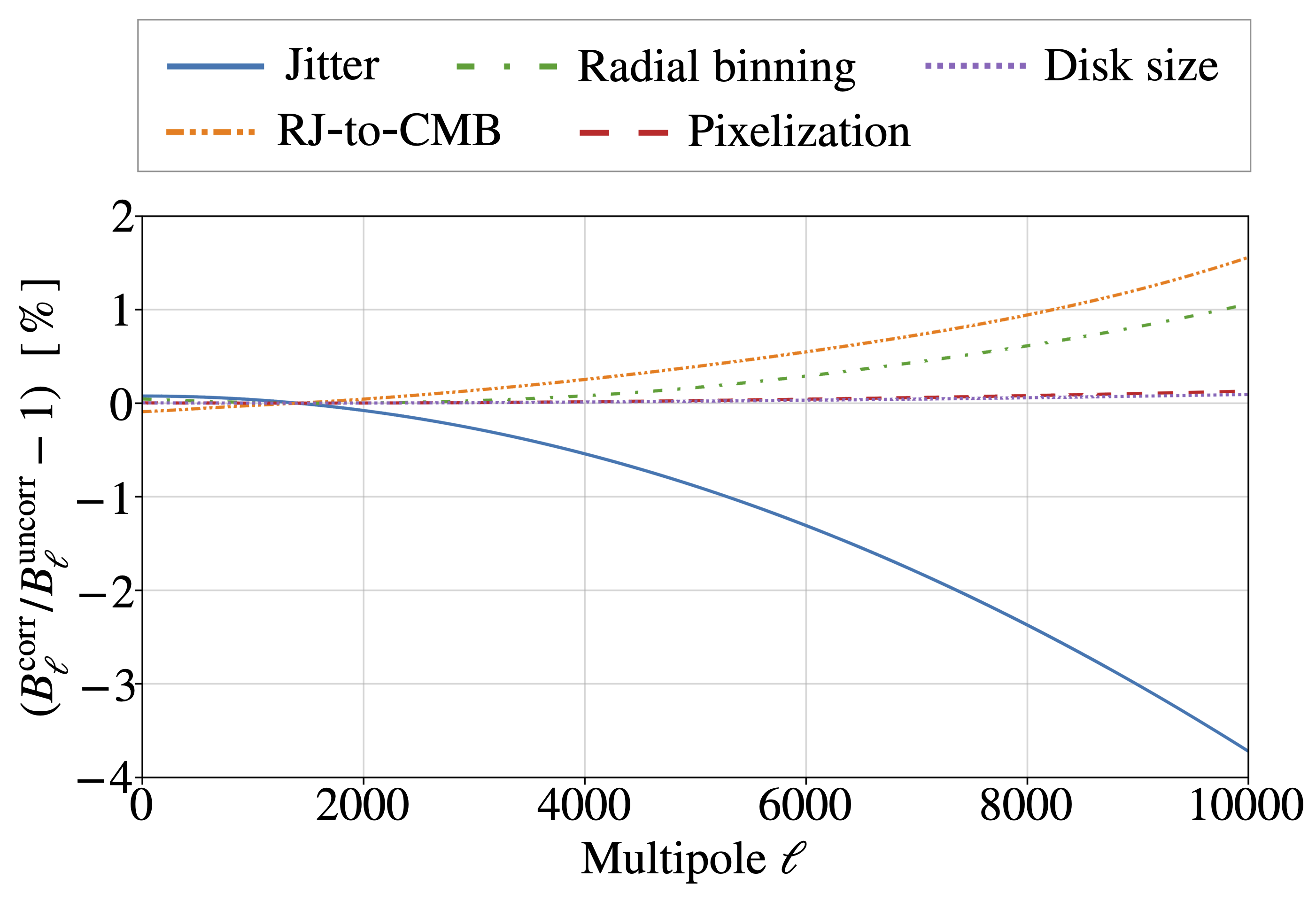}
    \caption{Change in the beam transform (in \%) due to applying different corrections for S15 PA2 at 150 GHz. For this plot, the beam transforms have all been normalized at $\ell = 1400$. This corresponds roughly to our effective calibration scale, so any change in our beam at this multipole would be corrected by our subsequent calibration to \textit{Planck}.}
    \label{fig:effect_allcorr}
    \vspace{1em}
\end{figure}

The resulting beam models and covariance matrices are an accurate description of the binned radial beam profiles, but they must be corrected for some systematic effects. Following the same approach as in \cite{hasselfield_atacama_2013}, corrections are applied to account for the pixelization of the planet maps, the binning of the maps into radial annuli, Uranus' angular diameter, and the planet's effective frequency. The effect of each of these corrections is shown in Figure \ref{fig:effect_allcorr} for S15 PA2 at 150\;GHz, as a typical example.

To correct for the pixelization of the planet maps, we divide the beam transform by the azimuthal average of the pixel window function, $\sinc ({p k_x}/{2\pi}) \sinc ({p k_y}/{2\pi})$, where $\{k_x, k_y\}$ are the spatial frequencies and $p$ is the pixel size in radians. This is a $\lesssim 0.1\%$ effect for ${\ell < 10,000}$.

Then, we estimate and correct for the transfer function induced due to binning the planet maps into radial annuli.
This is done by simulating planet maps, using the best-fitting beam profile model as the input, and then estimating their radial profiles following the same procedure as with the data. Comparing the input model with the output radial profile gives an estimate for the transfer function resulting from the radial binning. This is a $\lesssim 1\%$ effect for $\ell < 10,000$.

We also correct for Uranus' angular diameter, since Uranus is large enough that it cannot be treated as a point source given the precision to which we measure the beam. For each season, array, and frequency, we assume Uranus is a disk with radius equal to the weighted mean of the radii for all the Uranus observations contributing to the beam measurement. We then deconvolve this shape in harmonic space using the function $2{J_1(\ell r)}/{\ell r}$, where $r$ is the radius of Uranus' disk in radians. The factor of 2 normalizes the function to unity as $\ell \rightarrow 0$. This is a $\lesssim 0.1\%$ effect for $\ell < 10,000$.

The beam at this point describes the telescope's response to a point-like source with an approximately Rayleigh-Jeans (RJ) spectrum. Near our frequencies, the temperature spectrum of Uranus goes roughly as $\nu^{-0.25}$ \citep{planck_planets}. It is sufficiently close to the RJ limit ($\nu^0$) for our purposes. We apply a simple, first-order correction to obtain the relevant beam for the CMB blackbody spectrum using the band effective frequencies from \cite{thornton_2016}.\footnote{For the foreground modeling in \cite{choi_2020}, these effective frequencies were re-computed with improved passband data and upgraded code (as described in Appendix \ref{appen:sas}). Since the beam analysis for DR4 was done before this new work on the effective frequencies, the values from \cite{thornton_2016} were used. Considering the uncertainties on the passbands, the two sets of effective frequencies are consistent. In addition, given that the uncertainty on the beams is subdominant in the power spectrum analysis, whether one uses the effective frequencies from \cite{thornton_2016} or \cite{choi_2020} for the beam spectral correction does not have a significant effect on the results.} For this radiation with a band effective frequency $\nu_{\mathrm{CMB}}$, the beam is taken to be $B'(\ell)=B(\ell \nu_{RJ}/\nu_{\mathrm{CMB}})$, where $\nu_{RJ}$ is the effective frequency for radiation with a Rayleigh-Jeans spectrum.
This is a $\lesssim 1.5\%$ effect for $\ell < 10,000$.

The resulting beam models are referred to as the \enquote{instantaneous} beams, and it is in fact these corrected beams that are shown in Figures~\ref{fig:inst_beams} and \ref{fig:inst_tforms}.

\subsection{Jitter Beams}
\label{subsec:jitter}

In practice, the effective beam for a given sky region, season, detector array, and frequency is broader than the instantaneous beam. This is due to combining observations taken on multiple different nights throughout each season, so the resulting effective beam for each map is not as sharp as the beam inferred from planet observations which have been carefully recentered before co-adding. Broadening can be caused by variations in the pointing and global alignment, as well as possible small changes in the beam over the course of a season.

As described in \cite{aiola_2020}, ACT's blind pointing accuracy for DR4 is comparable to the average beam FWHM. If left uncorrected, this would significantly broaden the effective beams. Instead, as was done in \cite{louis_2017}, we correct the pointing by comparing the observed positions of bright point sources to their known catalog positions. This is done for each 10-minute section of the time-ordered data from each detector array, as described in \cite{aiola_2020}. Instead of performing the fit in the time domain as in \cite{louis_2017}, due to the larger data volume the fit is now performed in map-space. The resulting fit is obtained more quickly, but is slightly less accurate, leaving a larger residual variation in the beam due to pointing uncertainty.

This residual variation is captured by the effective beam, $B_{\ell}^{\rm eff}$, which is parametrized in terms of the instantaneous beam, $B_{\ell}$, and a correction term that is Gaussian in $\ell$, as
\begin{equation}
\label{eq:jitter}
B_{\ell}^{\rm eff} = B_{\ell}\times e^{-\ell (\ell+1)V/2}\; .
\end{equation}
If we were to interpret the Gaussian correction term as arising purely from residual pointing errors, then $V$ would be the residual pointing variance in square radians, which is why it is also referred to as \enquote{pointing jitter.} This variance, which in practice also includes the additional errors due to alignment and seasonal changes in the beam, is expected to be different for each sky region used in the power spectrum analysis. These different regions (Deep1, Deep5, Deep6, Deep56, Deep8, BOSS, and AdvACT) are shown in Figure 2 of \cite{choi_2020}.

\begin{figure}
    \centering
    \includegraphics[width=\linewidth]{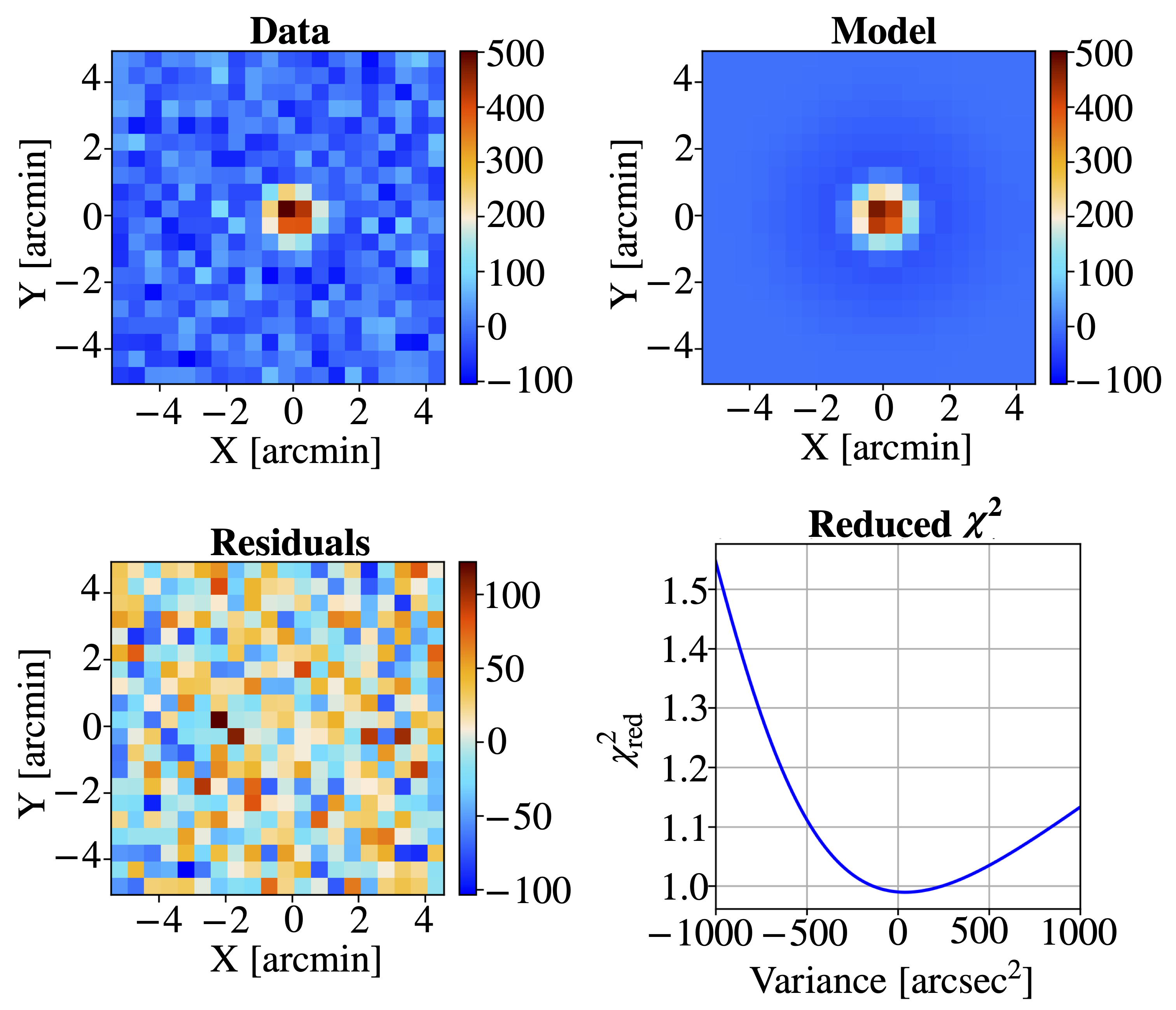}
    \caption{Example of the residual pointing variance, estimated by fitting an effective beam to a point source. The top panels compare the flux from the best-fitting model (right) to the data (left), with the residuals (bottom left), and the reduced $\chi^2$ as a function of pointing variance (bottom right). In each case, the data is high-pass filtered, and the colorbars are in units of $\mu$K. The best-fit variance for this source is 41 $\pm$ 93 arcsec$^2$.}
    \label{fig:jitter_source}
    \vspace{1em}
\end{figure}

To estimate $V$, in each region we create a catalog of the brightest point sources that are found in the maps using a matched filter, that have a signal-to-noise of at least 10 in each map, and that are matched to catalogs of known sources. For each of these sources, we then obtain an estimate of the variance $V$ and its associated uncertainty for each season, array, and frequency. To do this, we crop a section ($10^{\prime}\times10^{\prime}$) of each map around the source and remove large-scale variations by high-pass filtering the data (by multiplying the data by $1-G$, where $G$ is a Gaussian filter with a FWHM of 8$^{\prime}$). We then compute the effective beam using Equation~\ref{eq:jitter}, multiply this model by the local pixel-window of the cropped map, transform it into map space, high-pass filter it, and then compute the $\chi^2$ using an estimate of the local white noise amplitude. An example of such an individual source fit is shown in Figure~\ref{fig:jitter_source}. 

This procedure was validated by verifying that when simulated point sources convolved with Equation~\ref{eq:jitter} (with a known value for $V$) are injected into real CMB data and run through the pipeline, the input value for $V$ is recovered.

For convenience, we exclude sources identified as galaxies, pairs of galaxies, and planetary nebulae in the known source catalogs from this fitting procedure, as these types of sources sometimes appear to ACT to be extended. This leaves approximately 20 sources (for the smaller, deep regions) to 520 (for the largest region) that are included in the fits. They are primarily quasars and radio sources.

\begin{figure}
    \centering
    \includegraphics[width=\linewidth]{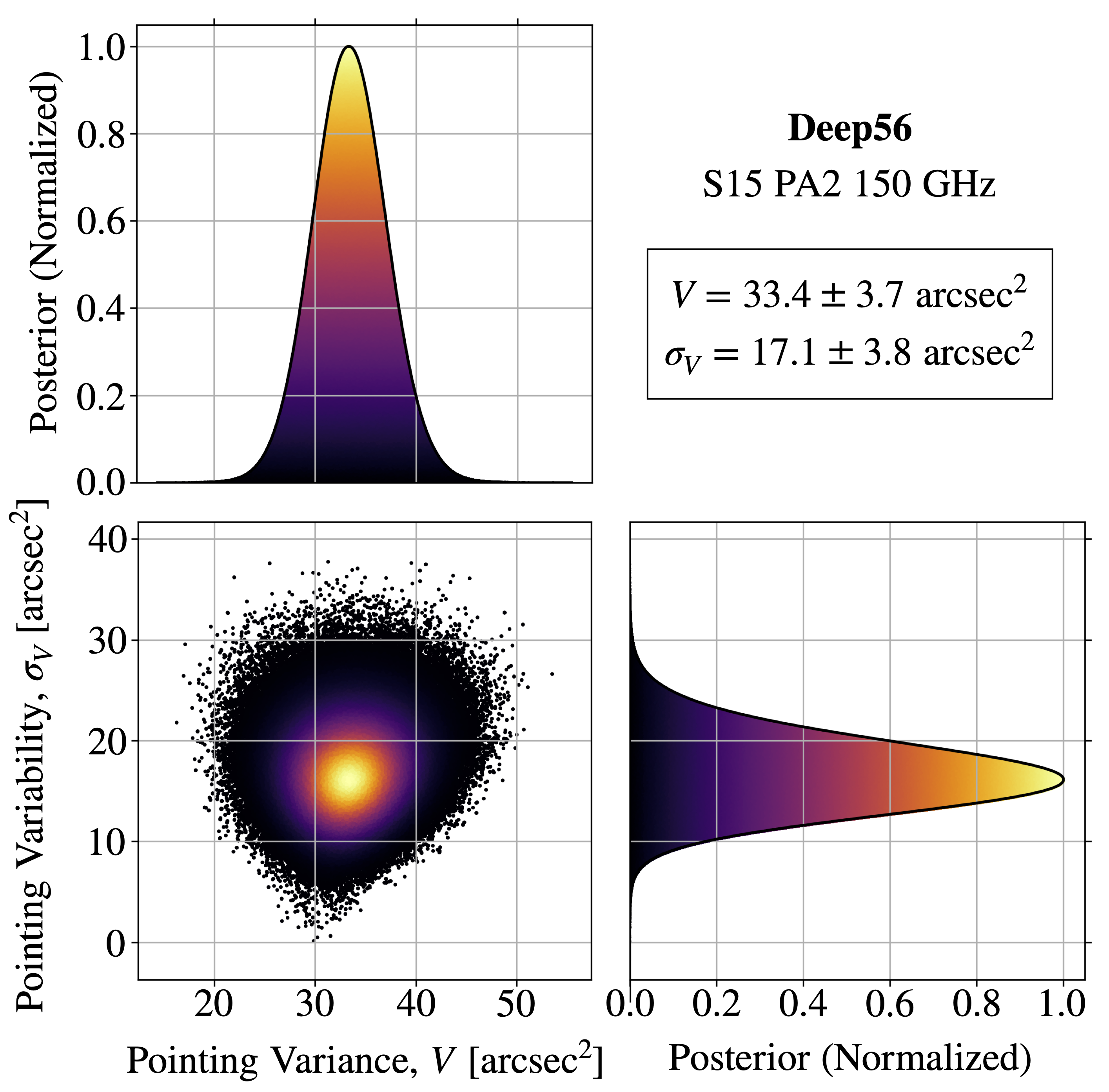}
    \caption{Distribution of the residual pointing variance, $V$, and pointing variability, $\sigma_V$, for the S15 PA2 beam at 150 GHz in the Deep56 region, resulting from fitting a pointing variance model to point sources. Notice here that the intrinsic scatter (measured with $\sigma_V$) is significantly larger than the measurement uncertainty on $V$. }
    \label{fig:jitter_dist}
    \vspace{1em}
\end{figure}

Once all $N$ sources have been fit using this method, we use the resulting set of pointing jitter estimates $\mathbf{d}$ and their uncertainties $\boldsymbol{\sigma}$ to obtain an estimate for the mean and intrinsic scatter, $V$ and $\sigma_V$, of the
effective pointing jitter in each map.
The likelihood for this is written as
\begin{equation}
     \mathcal{L}(\mathbf{d},\boldsymbol{\sigma}\,|\,V,\sigma_V) = A(\boldsymbol{\sigma},\sigma_V)\,e^{-\sum_{i=1}^N (d_i - V)^2/2(\sigma_i^2+\sigma_V^2)} \; ,
\end{equation}
where $A(\boldsymbol{\sigma},\sigma_V) = \Pi_i^N[2(\sigma_i^2+\sigma_V^2)]^{-1/2}$ is a normalization factor. We estimate the posterior distribution for $V$  and $\sigma_V$ using MCMC with the Metropolis-Hastings algorithm \citep{hastings_1970}, assuming a uniform prior on $V$. An example of these posterior distributions for one of the maps is shown in Figure~\ref{fig:jitter_dist}.

This likelihood, which was not used for previous ACT analyses, better takes into account the intrinsic scatter in the residual pointing variance, resulting in an improved estimate.

After computing the pointing variance independently in each map, we use Equation~\ref{eq:jitter} to obtain an effective beam, and associated covariance matrix, for each season, region, detector array, and frequency. An example of the effect of the pointing jitter correction on the beam transform is shown in Figure~\ref{fig:effect_allcorr}. The jitter values and uncertainties are given in Table~\ref{tab:jitter} and the resulting solid angles and uncertainties for the effective beams are given in Table~\ref{tab:eff_sa}.

The jitter values for PA3 at 98 GHz are significantly greater than those at 150 GHz. This empirical finding serves as a useful reminder that this \enquote{jitter} encompasses not only pointing errors, but also any changes in the beam throughout each season. These changes may be greater at 98 GHz since that is where the beam is broader.

Some of the best-fit jitter values for the Deep8 region are negative, but considering their uncertainties, they are consistent with positive values. We allow for negative jitter values in the fits in order to account for all possible effects on the beams, not just those due to pointing. While we have not investigated whether these negative values are related to a data quality issue, it should be noted that the data from Deep8 were not used in the cosmological analysis due to the poor cross-linking in the region. 

In Appendix~\ref{appen:sas} we provide the effective beam solid angles for beams at different frequencies. We applied the first-order spectral correction from \S\ref{subsec:add_corr} with effective central frequencies corresponding to synchrotron emission, dust, and the thermal Sunyaev-Zel'dovich effect. These beam solid angles differ from the CMB estimates by 0.3 to 5.3\%.

\begin{table*}
\caption{Jitter values and uncertainties (arcsec$^2$).}
\centering
\begin{tabular}{|c|c|c|C{5em}|C{5em}|C{5em}|C{5.5em}|C{5.5em}|C{5.5em}|C{5.5em}|}
\hline
Array & Band & Season  & Deep1 & Deep5 & Deep6 & Deep56 & Deep8 & BOSS & AdvACT \\
\hline
\multirow{3}{*}{PA1} & \multirow{3}{*}{150 GHz} & S13 & 47.6 $\pm$ 18.5 & 26.6 $\pm$ 8.9 & 30.3 $\pm$ 10.2 & - & - & - & - \\
 & & S14 & - & - & - & 12.1 $\pm$ 4.1 & - & - & - \\
 & & S15 & - & - & - & 34.9 $\pm$ 4.4 & -2.0 $\pm$ 15.4 & 30.8 $\pm$ 3.3 & - \\
\hline
\multirow{3}{*}{PA2} & \multirow{3}{*}{150 GHz} & S14 & - & - & - & 22.5 $\pm$ 3.2 & - & - & - \\
 & & S15 & - & - & - & 33.4 $\pm$ 3.7 & 4.2 $\pm$ 6.9 & 13.3 $\pm$ 3.3 & - \\
 & & S16  & - & - & - & - & - & - & 35.7 $\pm$ 2.4 \\
\hline
\multirow{2}{*}{PA3} & \multirow{2}{*}{150 GHz} & S15 &  - & - & - & 6.6 $\pm$ 5.8 & -21.3 $\pm$ 24.1 & 65.1 $\pm$ 6.8 & - \\
 & & S16 &  - & - & - & - & - & - & 12.2 $\pm$ 4.5 \\
\hline
\multirow{2}{*}{PA3} & \multirow{2}{*}{98 GHz} & S15 &  - & - & - & 190.9 $\pm$ 7.0 & 163.2 $\pm$ 18.5 & 246.0 $\pm$ 7.2 & - \\
 & & S16 & - & - & - & - & - & - & 238.2 $\pm$ 5.8 \\
\hline
\end{tabular}
\label{tab:jitter}
\end{table*}

\begin{table*}
\caption{Effective beam solid angles and uncertainties (nsr).}
\centering
\begin{tabular}{|c|c|c|C{5em}|C{5em}|C{5em}|C{5.5em}|C{5.5em}|C{5.5em}|C{5.5em}|}
\hline
Array & Band & Season  & Deep1 & Deep5 & Deep6 & Deep56 & Deep8 & BOSS & AdvACT \\
\hline
\multirow{3}{*}{PA1} & \multirow{3}{*}{150 GHz} & S13 & 206.1 $\pm$ 5.0 & 202.7 $\pm$ 4.1 & 203.3 $\pm$ 4.2 & - & - & - & - \\
 & & S14 & - & - & - & 197.2 $\pm$ 3.4 & - & - & - \\
 & & S15 & - & - & - & 199.0 $\pm$ 8.9 & 193.0 $\pm$ 9.0 & 198.3 $\pm$ 8.9 & - \\
\hline
\multirow{3}{*}{PA2} & \multirow{3}{*}{150 GHz} & S14 & - & - & - & 183.9 $\pm$ 3.3 & - & - & - \\
 & & S15 & - & - & - & 190.3 $\pm$ 4.8 & 185.7 $\pm$ 4.8 & 187.1 $\pm$ 4.7 & - \\
 & & S16  & - & - & - & - & - & - & 188.2 $\pm$ 4.8 \\
\hline
\multirow{2}{*}{PA3} & \multirow{2}{*}{150 GHz} & S15 &  - & - & - & 267.1 $\pm$ 5.5 & 262.0 $\pm$ 6.9 & 278.0 $\pm$ 5.8 & - \\
 & & S16 &  - & - & - & - & - & - & 236.3 $\pm$ 8.5 \\
\hline
\multirow{2}{*}{PA3} & \multirow{2}{*}{98 GHz} & S15 &  - & - & - & 510.7 $\pm$ 22.0 & 506.0 $\pm$ 21.0 & 520.2 $\pm$ 22.4 & - \\
 & & S16 & - & - & - & - & - & - & 490.3 $\pm$ 22.4 \\
\hline
\end{tabular}
\label{tab:eff_sa}
\vspace{2em}
\end{table*}

\subsection{Beam Transform Covariance Matrix}
\label{subsec:trans_covmat}

There have been several modifications since \cite{hasselfield_atacama_2013} in how we estimate uncertainties in the beam model. Previously, the non-linear scaling parameter $\ell_{\mathrm{max}}$ was varied in the fits, but the uncertainty in this parameter was not propagated to the covariance matrices.
We now estimate the covariance of each beam transform, $b_\ell$, by computing the Legendre transform of the sampled posterior distribution for the parameters describing the radial profile, as in Equation~\ref{eq:legendre}, and then applying the small-order corrections from \S\ref{subsec:add_corr}.
We then estimate the covariance matrix from this suite of $\ell$-space beam transforms and the added uncertainty associated with the jitter correction. These matrices are large, since the beam transforms are computed at each integer $\ell$ from 0 to 30,000, so we do not store them in their entirety. Rather, we decompose each matrix into independent modes (via singular value decomposition) and store the largest 10 modes. We find that this is sufficient to capture the majority of the covariance (we do not discard any modes with a singular value larger than $10^{-3}$ of the maximum value).

We include additional modes to account for possible uncertainty due to variations in the surface rms $\epsilon$ (which enters into the calculation of the Ruze beam, as described in \S\ref{subsubsec:scatt}, and was fixed to our estimate of 20\,$\mu$m),\footnote{We do not account for possible variations in the other measured parameter for the Ruze beam, the correlation length $c$, since it is well constrained by our measurements.} as well as the range over which the model for each beam offset is fit (which initially was 3.5$^{\prime}$--10.0$^{\prime}$).
For the surface rms, the values explored are $20$ $\mu$m and $30$ $\mu$m. For the region over which the offsets are fit, the three independent ranges explored are 3.5$^{\prime}$--5.0$^{\prime}$, 5.0$^{\prime}$--7.0$^{\prime}$, and 7.0$^{\prime}$--10.0$^{\prime}$. We store the top 3 modes associated with these model variations. 
The final uncertainty for each beam is thus composed of a total of 13 modes. These beam uncertainties are later added to the data covariance matrix as part of the power spectrum analysis pipeline.

\begin{figure}
    \centering
    \includegraphics[width=\linewidth]{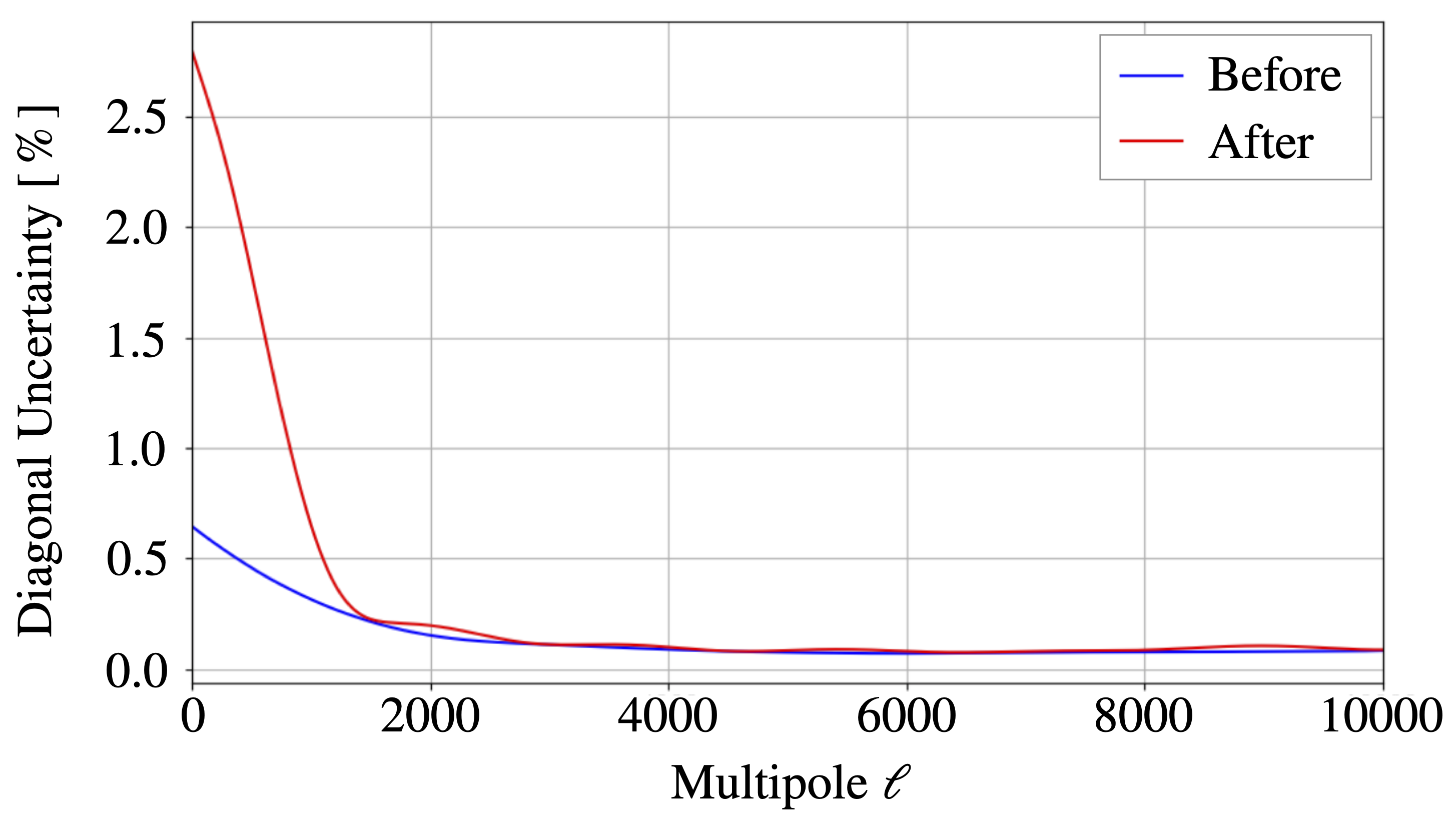}
    \caption{Uncertainties on the beam transform (in \%) for S15 PA2 at 150 GHz, before (blue) and after (red) adding the additional modes to account for different surface rms values and different fit ranges for the beam offset levels. These uncertainties are included in Figure \ref{fig:inst_tforms}.}
    \label{fig:add_err}
    \vspace{1em}
\end{figure}

An example of the effect of these additional
modes on the beam transform uncertainties is shown in Figure~\ref{fig:add_err}. The significant increase in the uncertainty at low multipoles is mainly due to the inclusion of the different fit ranges for the offsets. Previously, as described in \cite{hasselfield_atacama_2013}, the uncertainties were simply doubled from their formal values to account for potential systematic variations due to different fitting ranges. Even though this earlier method lead to a smaller estimate of the beam uncertainties at low multipoles, this did not have a significant effect on our results for DR3, since the uncertainty on the beams is subdominant in the power spectrum analysis, and a low-$\ell$ cutoff of 500 (350) was applied to the $TT$ ($TE$ and $EE$) data.

Another difference compared to the DR3 analysis in \cite{louis_2017} is the treatment of calibration uncertainty. As described in \cite{choi_2020}, for each season, region, array, and frequency the angular power spectra from ACT are calibrated to the \textit{Planck} temperature maps in the range $600< \ell <1800$. The \cite{louis_2017} analysis factored out the beam amplitude and uncertainty at an \enquote{effective} calibration scale (chosen in that case to be $\ell = 1400$).\footnote{This procedure is described by Equation 11 of \cite{hasselfield_atacama_2013}.}
However, we have changed this procedure for DR4 to reflect the fact that we are not calibrating the data at a single $\ell$, but over a range of $\ell$ values. We now simply normalize the beam to unity at $\theta=0$ and treat the calibration, and its uncertainty, separately in the ACT analysis. This is why the fractional beam uncertainties in Figure~\ref{fig:inst_tforms} are no longer smallest at $\ell=1400$, differing from those reported for ACT DR3.

\section{Polarization}
\label{sec:pol}

\subsection{Main Beam}
\label{subsec:pol_main}

Measurements of Uranus in polarization at visible and near-infrared wavelengths have shown that its disk-integrated polarization is less than $0.05\%$ \citep{schmid}. While we are not aware of any similarly precise data at  millimeter wavelengths, it is expected that the relevant scattering effects in the planetary atmosphere would be much weaker, resulting in net polarization levels considerably lower than the measurement cited above. Measurements by \textit{Planck} at 100 and 143 GHz place 68\% (95\%) confidence upper limits on the polarization fraction of Uranus at 2.6\% (3.6\%) and 1.5\% (2.0\%), respectively \citep{planck_planets}.

Since we do not expect Uranus to be significantly polarized in the bands observed by ACT, we interpret any polarization response measured from Uranus as being due to temperature-to-polarization (T-to-P) leakage. Although this leakage is relatively small in magnitude, the ACT DR4 data are now sensitive enough that we must account for it in our analysis. To do this, we use observations of Uranus to build an $\ell$-space T-to-P leakage function for each season, array, and frequency. This measurement and correction for the leakage in the main part of our beams is new to DR4 \citep[see also][]{aiola_2020, choi_2020}.

We begin by making maps of the $Q$ and $U$ Stokes parameters for the same set of Uranus observations chosen in \S\ref{subsec:mapsel} to fit the main beam. We then convert each set of $\{Q,U\}$ polarization maps to the radial Stokes parameters $\{Q_r, U_r\}$ \citep[following the third definition of cross polarization in][]{ludwig_1973} using the flat-sky approximation:
\begin{subequations}
\begin{align}
    Q_r(\boldsymbol{\theta}) &= Q(\boldsymbol{\theta}) \cos 2\phi_{\theta} + U(\boldsymbol{\theta}) \sin 2\phi_{\theta}\\
    U_r(\boldsymbol{\theta}) &= U(\boldsymbol{\theta}) \cos 2\phi_{\theta} - Q(\boldsymbol{\theta}) \sin 2\phi_{\theta} \;,
\end{align}
\label{eq:qr_ur}%
\end{subequations}
where $\boldsymbol{\theta} \equiv (\theta,\phi_\theta)$ are standard polar coordinates with the beam centroid as their origin and $\phi_{\theta}$ increases clockwise from the positive $y$-axis (assuming that one uses the convention in which $+x$ points to the right and $+y$ points upward). Here $Q$ and $U$ follow the \texttt{COSMO} convention \citep{healpix_2005}, whereas for the ACT maps released as part of DR4, the polarization components are defined by the \texttt{IAU} convention \citep{hamaker_bregman_1996}. The polarization convention initially used for $\{Q, U\}$ does not matter once we have transformed to $\{Q_r, U_r\}$.

While it is not obvious how the polarized $E$ or $B$ beams ought to behave at larger radii, we do have some intuition about $Q$ and $U$. For an unpolarized point source such as Uranus, any polarized signal will be the result of beam differences between the two axes of a polarimeter. Although the beams may be slightly different, due to e.g., differential ellipticity, we still expect each of them to decay radially as $1/\theta^{3}$ as they are part of the same diffraction-limited optical system. The $Q$ and $U$ maps are essentially just radially independent linear combinations of the various detector axes in an array, appropriately weighted by the map-maker, so the same asymptotic behavior should apply. This asymptotic behavior would then hold true for the $Q_r$ and $U_r$ maps as well. This was confirmed with simulations where the beams for individual polarimeter axes were constructed using an \enquote{elliptified} version of the azimuthally averaged intensity beam for a given season and array. We thus use the same basis functions in the core and the $\alpha/\theta^{3}$ term in the wing to fit the beam in polarization as we did for the main beam in \S\ref{subsubsec:fit}.

Since we are ultimately interested in how leakage manifests itself in the angular power spectra, we need to translate any polarized beam models of the azimuthally averaged radial profiles of $Q_r$ and $U_r$, $\tilde{Q}_r$ and $\tilde{U}_r$, to an $\ell$-space representation of $E$ and $B$. Conveniently, $\tilde{Q}_r$ and $\tilde{U}_r$ have a direct correspondence to the azimuthally averaged $\ell$-space $E$ and $B$ transforms, $\tilde{E}(\ell)$ and $\tilde{B}(\ell)$. As shown in Appendix~\ref{appen:qr_ur}, there exists a simple relation between these components, which in the flat-sky approximation takes the form of a second-order Hankel transform:
\begin{equation}
\label{eq:trans_e_b}
    \{\tilde{E}(\ell), \tilde{B}(\ell)\} = -2\pi\int \{\tilde{Q}_r(\theta),\tilde{U}_r(\theta) \} J_2(\ell\theta)\;\theta\;d\theta \; .
\end{equation}

\begin{figure*}
    \centering
    \includegraphics[width=\linewidth]{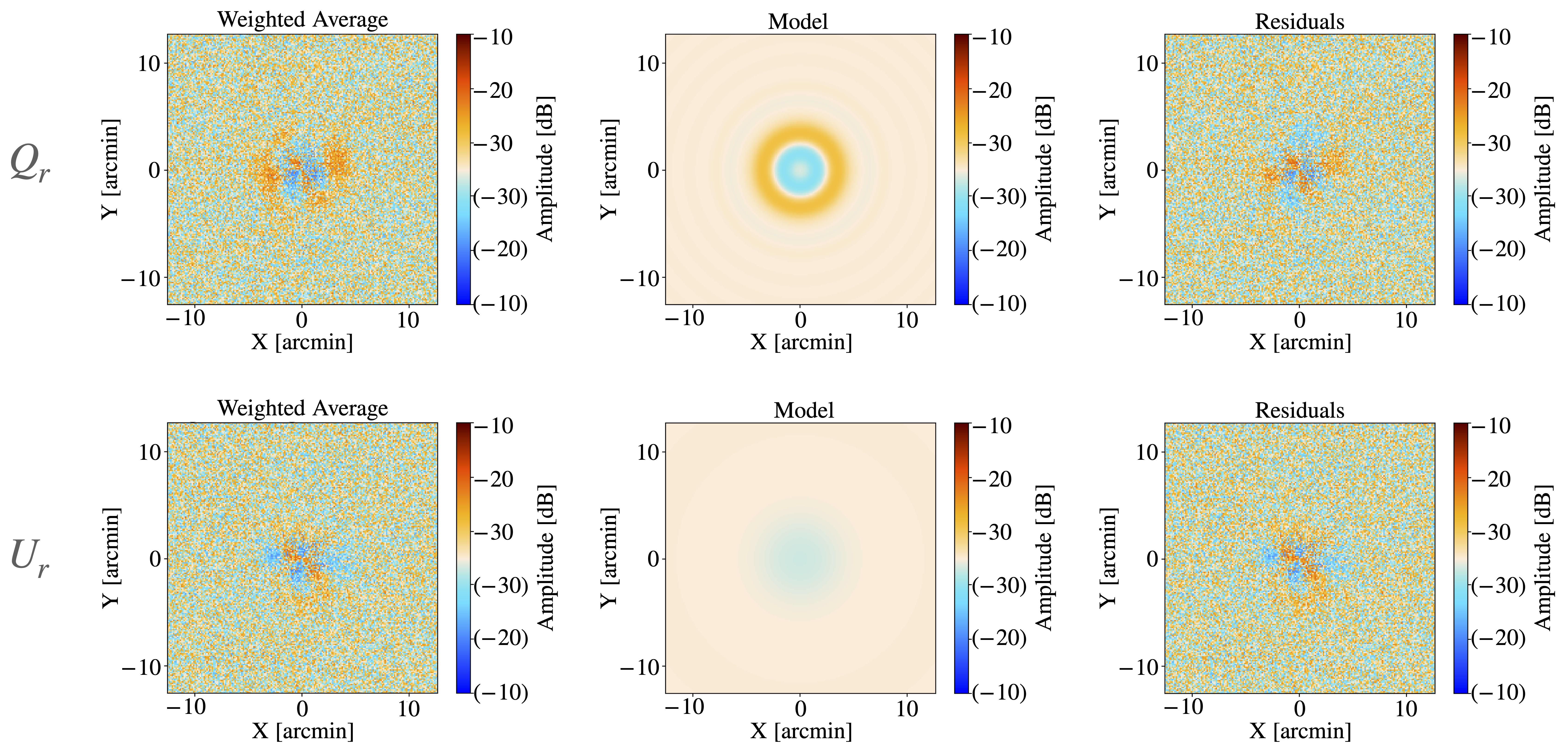}
    \caption{Maps of the leakage beam in $Q_r$ (top) and $U_r$ (bottom) for S15 PA2 at 150 GHz. In each case, the color scale is a symmetrical log scale in dB, with a linear threshold of $-$30 dB and negative values enclosed by parentheses. (Left) The weighted average of the Uranus observations used to characterize the beam. (Center) The model that is fit to the radial profile of the measured beam. (Right) The difference between the measured beam (left) and the beam model (center). 
    Features in the residuals could arise due to physical effects, such as differential beam ellipticity between the two axes of a polarimeter or a polarization angle offset \citep{hu_2003}. The level of residuals seen in the last column has an insignificant effect on the results of the power spectrum analysis. As shown in Figure \ref{fig:profile_pol}, the azimuthal averages of the residuals are close to zero, which is why the fits are successful.}
    \label{fig:map_qur}
    \vspace{1em}
\end{figure*}

\begin{figure}
    \centering
    \includegraphics[width=\linewidth]{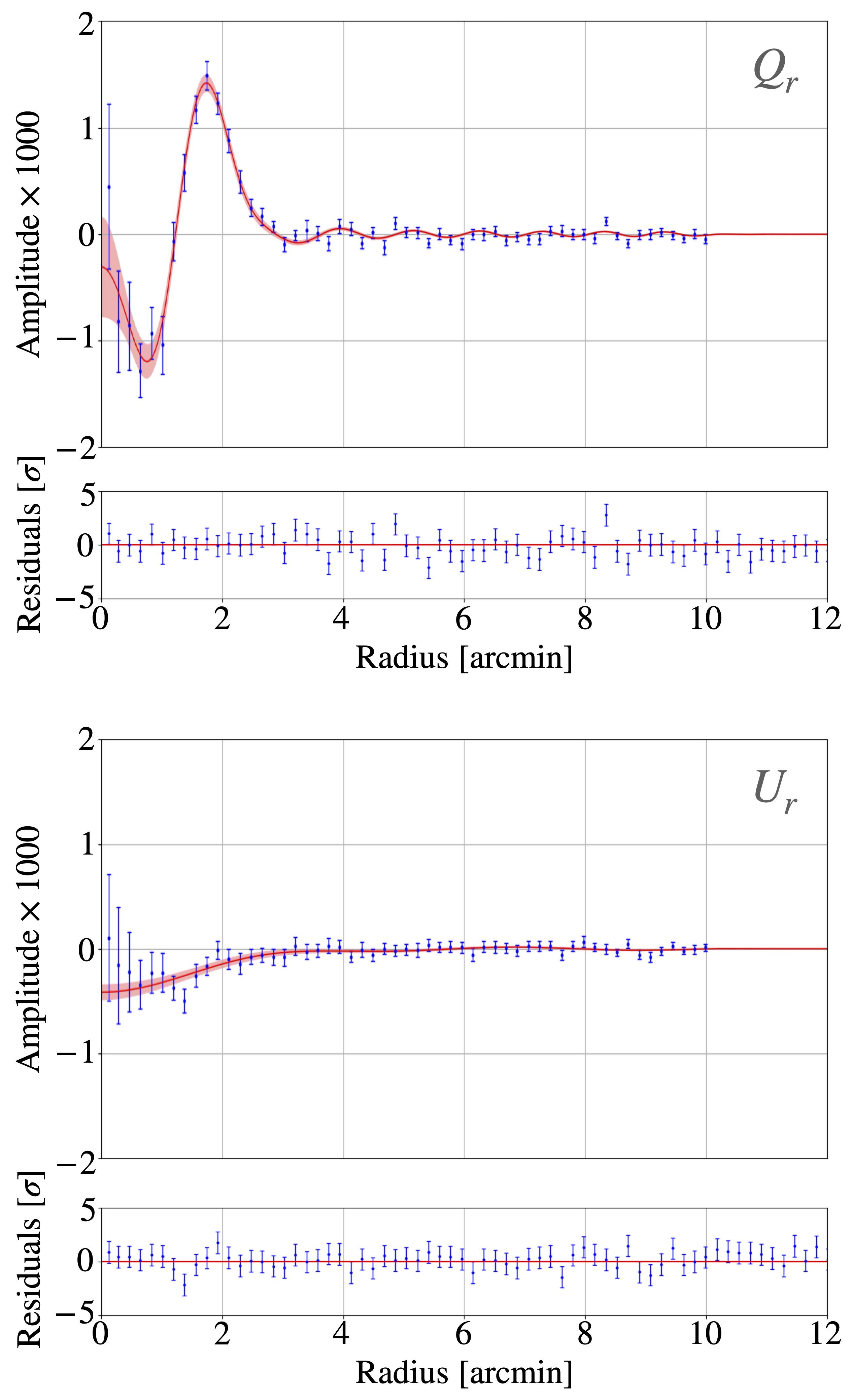}
    \caption{Measured radial polarization profiles (blue) in $Q_r$ (top) and $U_r$ (bottom) for S15 PA2 at 150 GHz and the model we fit to each profile (red), with the red shaded region indicating the 1$\sigma$ model uncertainty bounds. The profiles have been normalized by the peak amplitude of the corresponding $T$ beam (at the beam centroid, $\theta=0$). Note that the bins are correlated.}
    \label{fig:profile_pol}
    \vspace{1em}
\end{figure}

From the set of Uranus maps in the $\{Q_r, U_r\}$ basis, examples of which are shown in Figure~\ref{fig:map_qur}, we thus construct average radial profiles for each season, array, and frequency, and we fit them in a similar way as the main beam is fit in \S\ref{sec:pipe}. While for the usual temperature beam fitting pipeline we fit an offset to each individual Uranus profile before taking an average, we do not do this in polarization since we expect the mapping transfer function in that case to be zero.\footnote{Even if the mapping transfer function were non-zero in polarization, it would be a sub-percent-level effect in the measurement of the leakage beam, which itself is percent-level in terms of our power spectrum analysis and results, so the effect would be insignificant.} The only difference in the fitting procedure here is that we do not include a scattering term in the polarized beam model, since it is not expected to matter to first order, and it is unclear how it would vary for the different detector polarizations.\footnote{Again, this would be a sub-percent-level effect in the measurement of the percent-level leakage beams, so there would be no significant impact on our analysis and results.} Examples of the radial profile model fits in polarization are shown in Figure~\ref{fig:profile_pol}.

As can be seen in Figures \ref{fig:map_qur} and \ref{fig:profile_pol}, while our model is a good fit to the $Q_r$ and $U_r$ radial profiles, there are significant, quadrupole-like residuals in the maps when our model is subtracted. It turns out that the leakage beams are far less azimuthally symmetric than the temperature beams (as can be seen by comparing Figures \ref{fig:map_t} and \ref{fig:map_qur}). The treatment of this leakage will be revisited for upcoming ACT beam analyses and may be improved upon, for example, by fitting a 2D model to the polarized beam profiles in order to properly capture the asymmetry.\footnote{Work on fitting the beams in 2D is in progress.} In the meantime, the fitting done here is sufficient. The level of residuals seen in Figure \ref{fig:map_qur} has an insignificant effect on the results of the power spectrum analysis for DR4 (see \S\ref{subsec:asym}). The features visible in the residual maps can be expected to occur due to physical effects. For example, we simulated the T-to-P leakage for a point source due to both differential beam ellipticity between the two axes of a polarimeter and a polarization angle offset. The resulting maps of the simulated leakage beam in $Q_r$ and $U_r$ had strong quadrupole-like features. As shown in Figure \ref{fig:profile_pol}, the azimuthal averages of the residuals in $Q_r$ and $U_r$ are close to zero, which is why the fits works well, despite the appearance of the residuals in the maps.

\begin{figure}
    \centering
    \includegraphics[width=\linewidth]{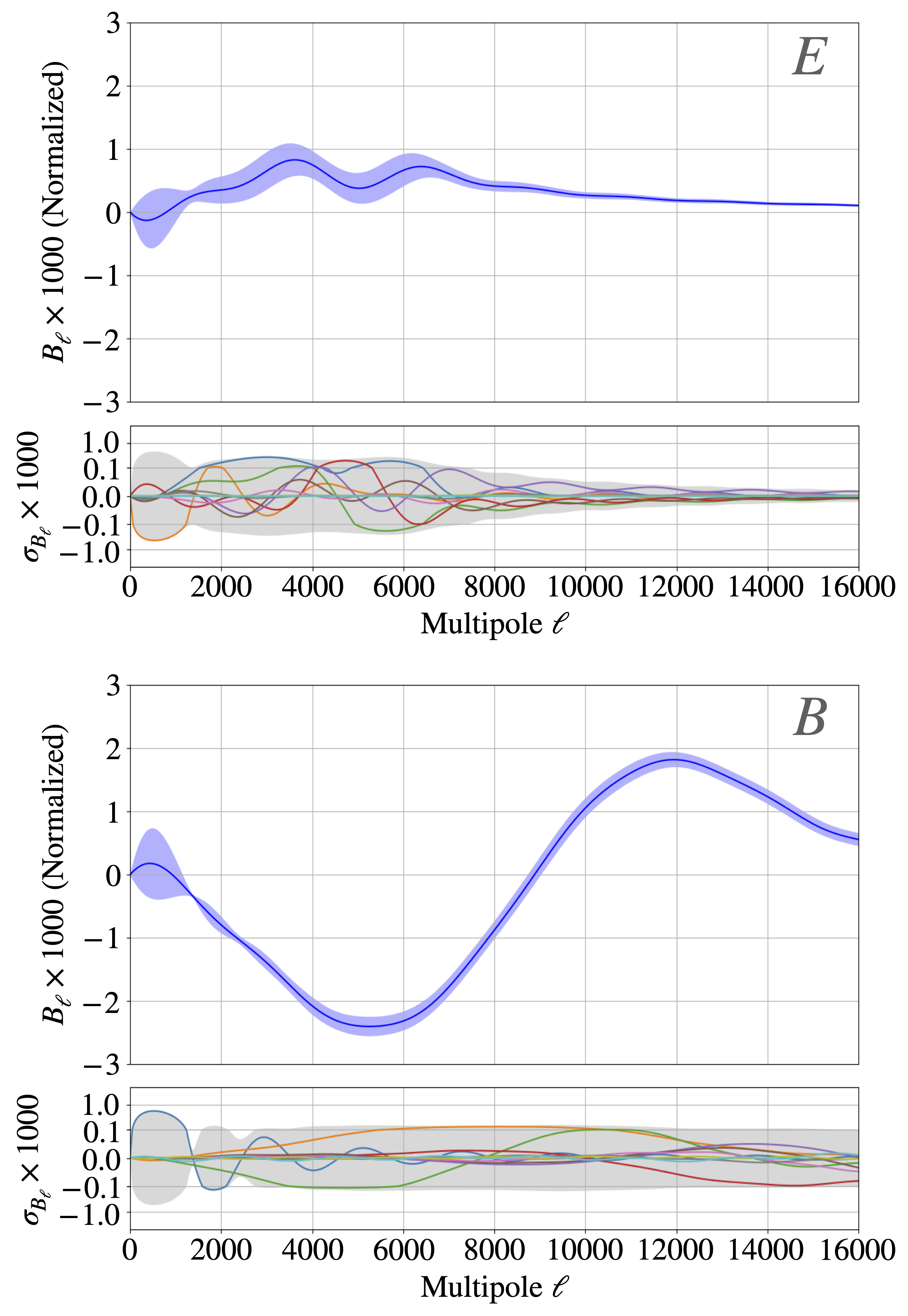}
    \caption{The transform of the radial beam profiles in $E$ (top) and $B$ (bottom) for S15 PA2 at 150 GHz. These transforms are normalized by the amplitude of the corresponding $T$ beam transform at $\ell =0$. The lower panels show the dominant independent modes of each transform's covariance matrix (in color) and the magnitude of the diagonals (the gray shaded regions). }
    \label{fig:transform_pol}
    \vspace{1em}
\end{figure}

\begin{figure*}
    \centering
    \includegraphics[width=\textwidth]{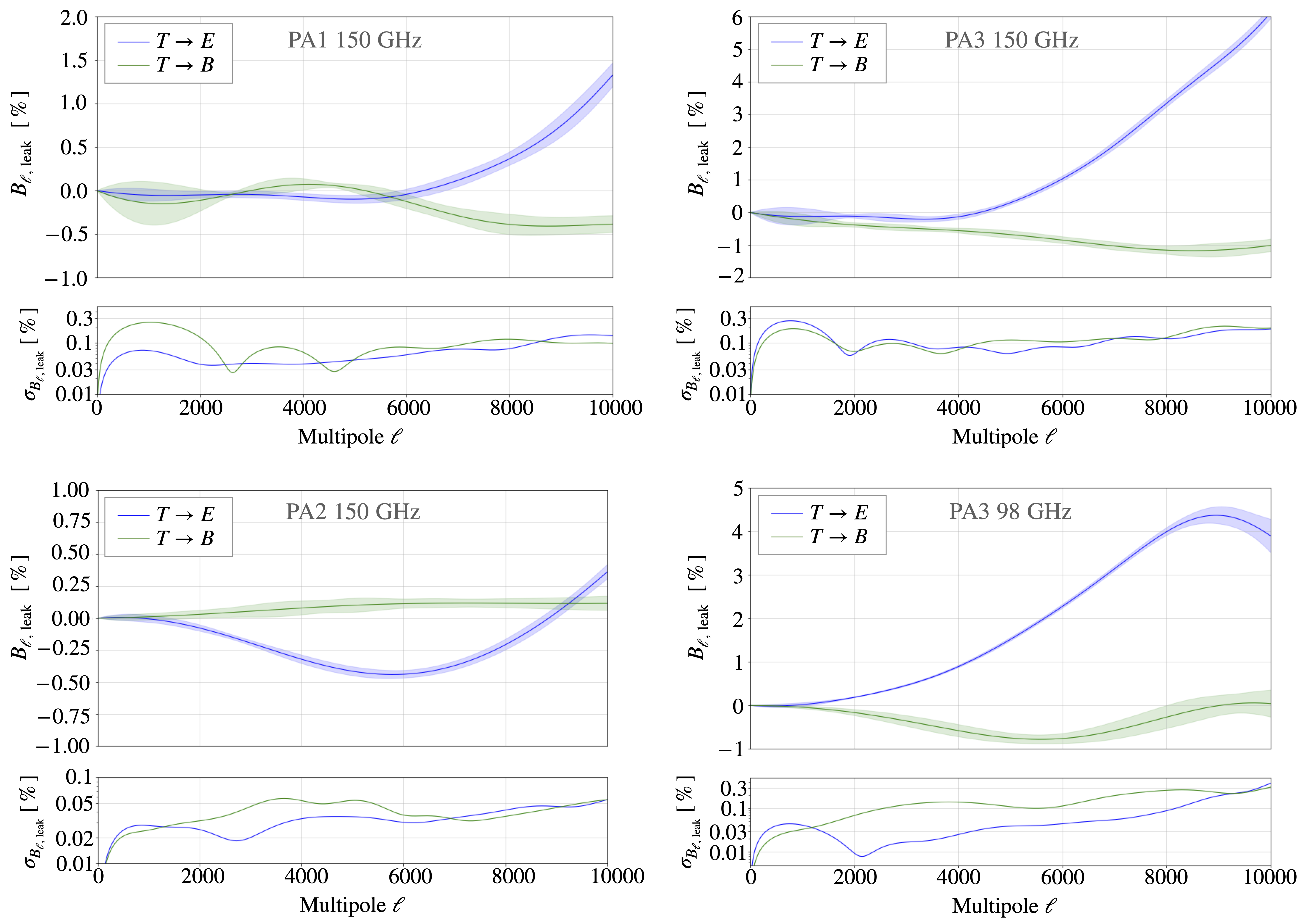}
    \caption{Measured temperature-to-polarization leakage in the main beams for S15 for all arrays, showing the temperature leaking into $E$-mode and $B$-mode polarization. Note the different $y$-axis ranges. For PA1 and PA2 the leakage values are within 1.5\%, whereas for PA3, the leakage increases around $\ell \simeq$ 4000--6000, and reaches over 6\% (4\%) at $\ell$ $\sim$10,000 at 150 GHz (98 GHz). We attribute the higher leakage for PA3 to an imperfectly optimized horn design in this first generation of multichroic polarimeters. The uncertainties (lower panels) include the adjustments for model variability.}
    \label{fig:leak}
    \vspace{1em}
\end{figure*}

The $Q_r$ and $U_r$ radial profiles fits and their uncertainties are transformed to $\ell$-space using a similar approach as for the temperature data, except using Equation~\ref{eq:trans_e_b} instead of the usual Legendre transform. The transforms for our example case are shown in Figure~\ref{fig:transform_pol}. These transforms do not include corrections for small systematic effects, as was done for the main beam fitting pipeline in \S\ref{subsec:add_corr}. This is because we are ultimately interested in the ratios of transfer functions for the leakage beams, in which these $\ell$-space corrections cancel out. To determine the polarization leakage beams, an example of which is shown in Figure~\ref{fig:leak}, we divide the $E$ and $B$ transforms by their corresponding (uncorrected) $T$ transform. For PA1 and PA2 the leakage values are within 1.5\%  \citep[comparable to the leakage values for SPT-3G,][]{dutcher_2021}, whereas for PA3, the leakage increases around $\ell \simeq$ 4000--6000, and reaches over 6\% (4\%) at $\ell$ $\sim$10,000 at 150 GHz (98 GHz). We attribute the higher leakage for PA3 to an imperfectly optimized horn design in this first generation of multichroic polarimeters.

The top 10 modes from the leakage beam covariance matrices are stored. In addition, similar to the main beam analysis, two modes are added to account for variations in the model for the temperature beam.\footnote{We only consider variations in the range over which the beam offsets are fit (3.5$^{\prime}$--5.0$^{\prime}$, 5.0$^{\prime}$--7.0$^{\prime}$, 7.0$^{\prime}$--10.0$^{\prime}$). The Ruze beam is a smaller effect, so it is (safely) ignored here.} The final leakage beam uncertainties are thus comprised of 12 modes.

The leakage beams were used in the DR4 power spectrum likelihood, as explained in \S\ref{subsec:pol_corr}. 

\subsection{Polarized Sidelobes}
\label{subsec:sidel}

As described in \cite{louis_2017} and \cite{aiola_2020}, we detect polarized sidelobes of the main ACT beams. Although weak in amplitude, these sidelobes cause noticeable T-to-P leakage. The sidelobes for PA1 and PA2 were shown in \cite{louis_2017}. While PA3 was not used in that analysis, it was mentioned at the time that polarized sidelobes were not detected for PA3. However, using additional observations of Saturn to conduct a more thorough analysis, we have detected sidelobes in PA3 at 150 GHz, with an amplitude roughly 10\% that of the sidelobes in PA1 and PA2. 
The general features of the sidelobes are common to all detector arrays; they consist of a group of compact lobes, each resembling a slightly elongated image of the main beam, with approximate four-fold symmetry, strongly polarized perpendicular to the radius from the beam center (which corresponds to $-Q_r$). This results in most of the leakage due to the sidelobes being from temperature into $E$-mode polarization rather than into $B$-mode polarization. These sidelobes are stable in time. We also observe that sidelobes from Saturn are only seen if Saturn lies in a focal plane's field of view. As explained in Section 3.8 of \cite{aiola_2020}, this is consistent with the sidelobes being due to an optical effect inside the receiver. 

The sidelobes for PA1 and PA2 are at a distance of roughly 15$^{\prime}$ from the beam centroid, with an additional set visible at 30$^{\prime}$ for PA1. The sidelobes for PA3 are located approximately 30$^{\prime}$ to 40$^{\prime}$ from the beam centroid. This larger angular separation means the strongest T-to-P leakage occurs at $\ell \simeq 300$ for PA3 compared to $\ell \simeq 500$ for PA1 and PA2. Also, since the sidelobes only map to the sky when the main beam is also in the field of view, fewer detectors are affected by each sidelobe for PA3. We do not detect any sidelobes in PA3 at 98 GHz, which implies they must either be of significantly lower amplitude than those at 150 GHz, or in a different position. For PA1, PA2, and PA3, the amount of solid angle contained in these sidelobes is roughly 1.2\%, 3.4\%, and 0.15\%, respectively, of that in the main beam. 

Preliminary studies suggest that these sidelobes are due to diffraction caused by the arrays of metal elements that make up ACT's optical filters. We note that similar filters \citep{ade_2006} are also used for the South Pole Telescope (SPT), POLARBEAR, and BICEP/Keck \citep{padin_2008, arnold_2010, keating_2003} at varying locations in the optical path.\footnote{This issue with ACT is related to the location of the filters in the optical path and the diffraction angle produced by the filters relative to the effective view passed by the system at this location. So other experiments may or may not see such an effect, depending on the details of their optical implementation.}

If the sidelobes were indeed due to diffraction from the filters, we would expect them to occur at a larger radius for lower frequencies. Based on our calculations, at 98 GHz, the sidelobes would appear starting at a distance of approximately 47$^{\prime}$ from the beam centroid. Since this is roughly the size of our field of view, the sidelobes would then map to the sky for only a small fraction of the detectors, the ones at the edges of the focal plane. This is consistent with the lack of detectable sidelobes at 98\;GHz. We do not apply any sidelobe corrections to the PA3 data at 98 GHz.

\begin{figure}
    \centering
    \includegraphics[width=\linewidth]{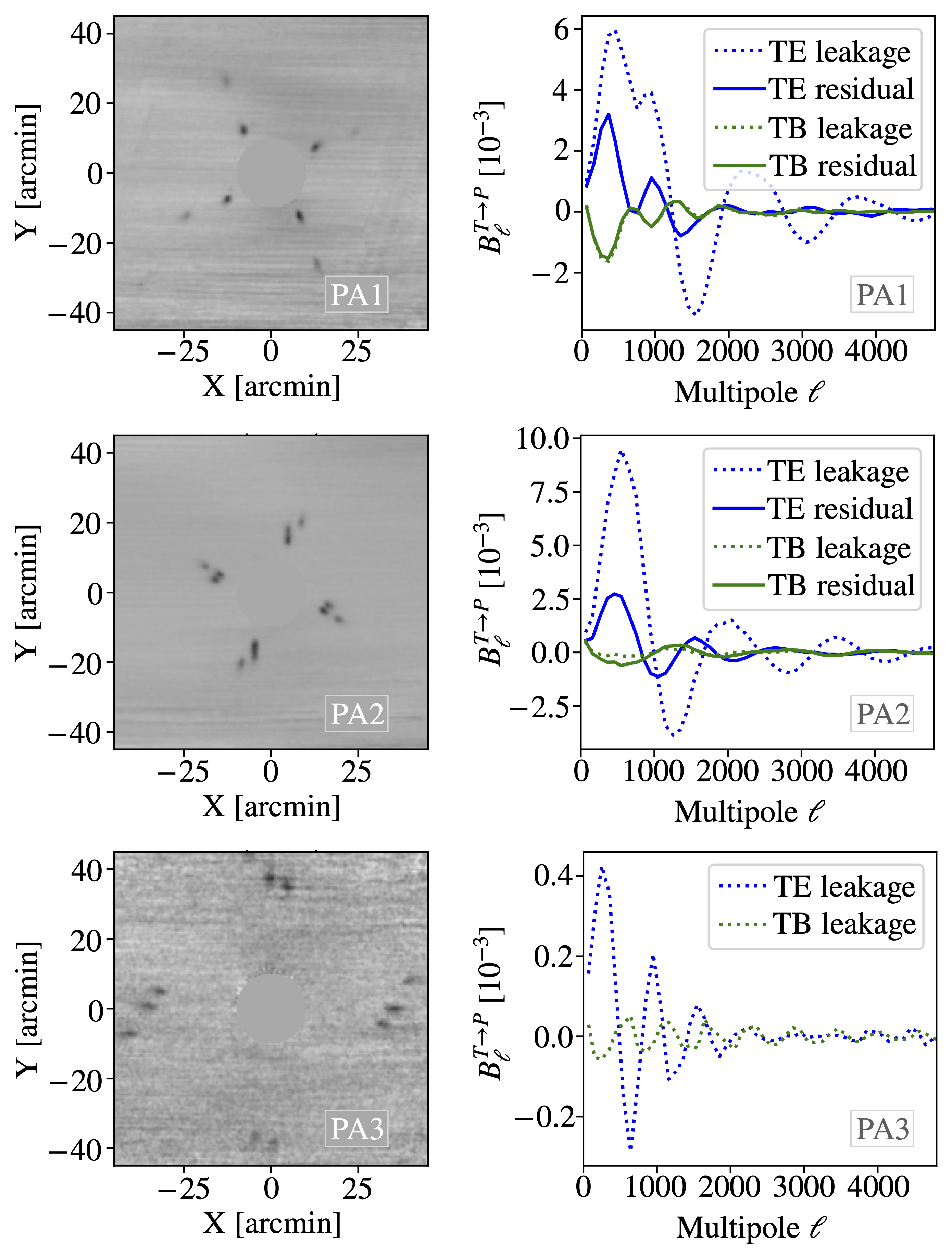}
    \caption{Sidelobes in the PA1, PA2, and PA3 detectors at 150 GHz. (Left) Maps of the polarized sidelobes, obtained by stacking observations of Saturn. The main part of the beam at the center of each image is masked. In each the grayscale is linear, indicating the sidelobe amplitude in the range from $-0.002$ (black) to $+0.001$ (white) relative to the main beam peak,\footnote{Saturn's brightness appears to compress the detector gain by a few percent. The effect can be ignored here because it is a small correction on the sidelobe leakage, which itself is a small effect.} with positive (negative) numbers corresponding to polarization parallel (perpendicular) to the radial direction. The complementary polarization leakage (corresponding to $TB$ leakage) is smaller and not shown in the maps, but it is also estimated. (Right) The effect of the sidelobes on our measurements, expressed as beam transfer functions $B_{\ell}^{T\rightarrow E}$ and $B_{\ell}^{T\rightarrow B}$. Note that the scales on the right differ for all three arrays. The sidelobes are projected out of the time-ordered data prior to mapping, so the effective (residual) leakage is what affects our spectra. Since the sidelobes for PA3 are significantly weaker than PA1 and PA2, we do not compute residuals for PA3. No sidelobes are seen for PA3 at 98 GHz.}
    \label{fig:buddies}
    \vspace{1em}
\end{figure}

As mentioned in \cite{aiola_2020}, to study the sidelobes we use observations of Saturn. While Saturn's brightness appears to induce a non-linear response near peak amplitude, it is useful for studying the relatively weak sidelobes. There are no issues due to non-linearity when observing the sidelobes. This is confirmed by the fact that the amplitude of the sidelobes seen by a detector is consistent, whether they are seen before or after the detector sees Saturn's peak.

We apply the same treatment to the sidelobes for all detectors at 150 GHz. In short, we model the sidelobes as a sum of polarized, spatially shifted copies of the main beam and fit the amplitudes of the $T$, $Q$, and $U$ components of each beam instance using maps of Saturn. We then use this model to deproject the sidelobes from the time-ordered data prior to map-making. The idea is to subtract the total flux in the sidelobes, even if their shapes are not exactly zeroed out in the maps.
It can be shown that this removes the low-$\ell$ T-to-P leakage. Maps of the sidelobes are shown in Figure~\ref{fig:buddies}, along with the T-to-P leakage functions.

For PA1 and PA2, we re-use the sidelobe models from DR3, constructed using observations of Saturn from 2014. For PA3, which was not part of DR3, we use Saturn observations from 2015 to characterize the sidelobes for DR4 in the same way as was done for DR3.

Looking more closely at how these sidelobe models are constructed, we begin with a series of observations of Saturn to which we apply the same data selection criteria as we did to Uranus, as described in \S\ref{sec:obs}. We then map the chosen observations with the \verb|moby2|\footnote{GitHub repository: \\ \hspace*{2em} \url{https://github.com/ACTCollaboration/moby2}} map-maker \citep[the same map-maker that was used for the beam analysis in][]{louis_2017} and coadd the maps together to produce one map of Saturn for each detector array (PA1, PA2, PA3).
The maps are coadded with weights based on an estimate of their white noise level determined outside the planet region.

We then visually identify regions in the maps containing compact polarized sidelobes. As can be seen in Figure~\ref{fig:buddies}, the pattern of the sidelobes in each map has approximate four-fold symmetry, with four groups of sidelobes appearing at roughly equal distances from the beam centroid. For each sidelobe we choose how many copies of the main beam should be used to model it. Stronger, elongated sidelobes are modelled by two copies of the main beam, to capture the elongation; weaker sidelobes are modelled with a single copy of the beam. Then for each of the four groups of sidelobes we fit the position and amplitude of the copies of the main beam. In theory this fit could be done with any of the signals, but currently we perform this fit to a map of $P = \sqrt{Q^2+U^2}$ because that is a bright, clear signal. With the model positions fixed, we then fit the amplitudes for each signal ($T$, $Q_r$, $U_r$) independently. The result is a base model for the sidelobes, but this model does not yet contain per-detector detail.

We next account for the fact that the sidelobes do not always appear in all the detectors. This is relevant to the extent that the per-detector weights in the planet maps are different from the per-detector weights in the survey maps (which are used for the CMB analysis, for example). This could possibly be a significant (up to tens of percent) effect, since the \verb|moby2| map-maker used to make the planet maps doesn't weight the detectors by their noise, whereas the \verb|enki| map-maker used for the survey maps does. The sidelobes occur primarily in the detectors at the periphery of the array, which is also where the noise tends to be higher, so this is a correlated effect.

To deproject the sidelobes from the per-detector time-ordered data, we need to estimate whether or not the sidelobes are seen by individual detectors, based on their position on the focal plane. The parameter we want to estimate is the radius of the \enquote{aperture}, or circle, around a detector such that if Saturn falls within the circle, the beam sidelobes are visible to that detector. To estimate this radius, we first divide the detectors into subsets.  For PA3, the subsets are the three hexagonal wafers of detectors, described in \cite{thornton_2016}. For each detector subset, we re-make maps of Saturn, and measure the total sidelobe flux in the resulting coadded map. We then compare these measurements to a model for the sidelobe flux as a function of radius, which scales with the fraction of detectors in each subset that would see each sidelobe.

As for the main beam analysis in \S\ref{subsec:add_corr}, small corrections are made to the model to account for systematic effects.

The sidelobe removal described here is not perfect, so there is still residual T-to-P leakage in the ACT data due to the sidelobes. The residual $TE$ and $TB$ leakages, shown in Figure~\ref{fig:buddies}, are estimated by making new maps of Saturn after our sidelobe removal.
Since the sidelobes for PA3 are already significantly weaker than PA1 and PA2, we do not compute residuals for PA3.
We add the residuals for PA1 and PA2 to the main beam leakage for use in the power spectrum analysis, as described in the next section. 
 
\begin{figure}
    \centering
    \includegraphics[width=\linewidth]{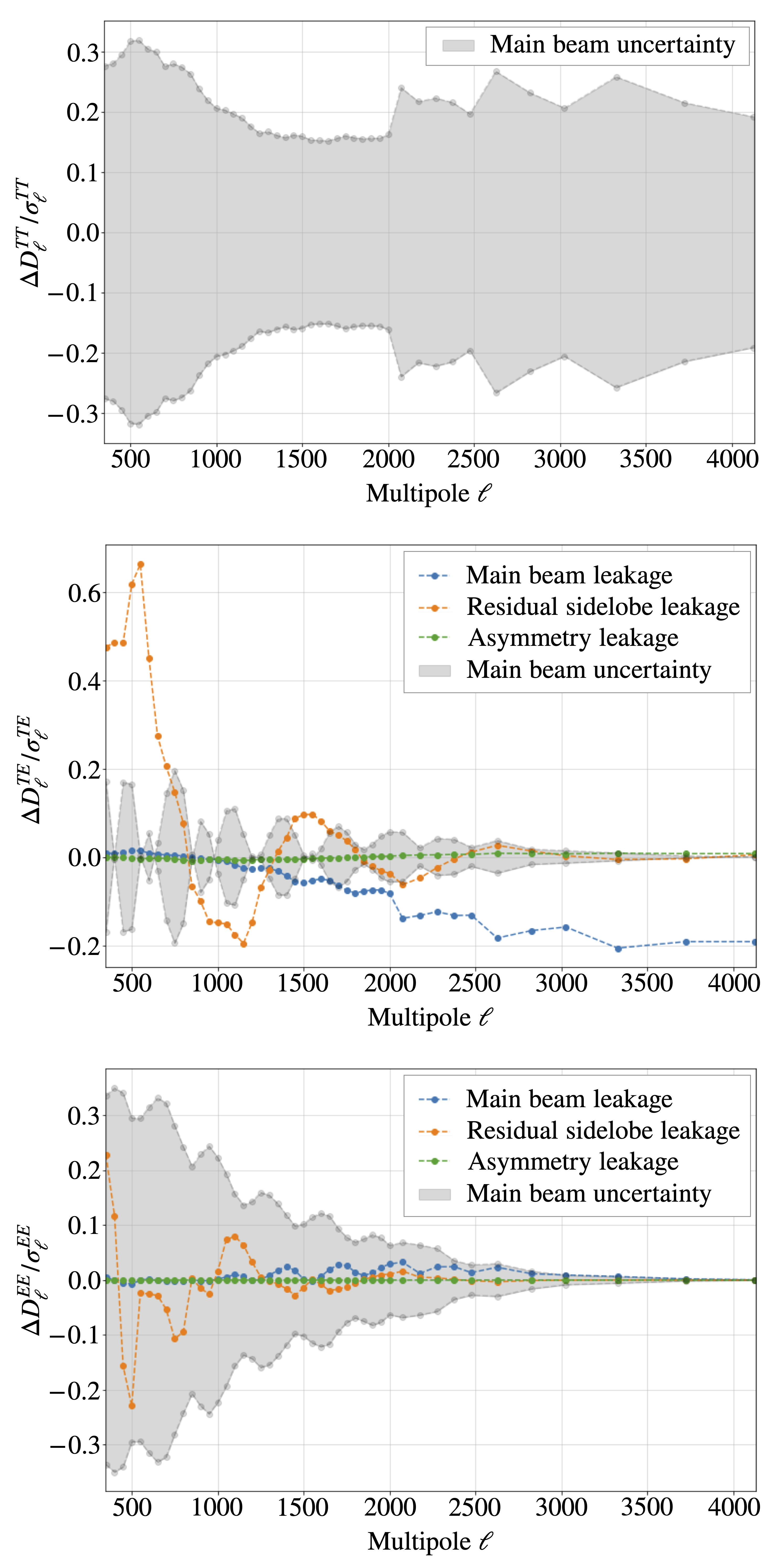}
    \caption{An example of the size of various effects on the inferred spectra, compared to our final uncertainties. Here we used the leakage in the main beam for S15 PA2 at 150 GHz, the residual sidelobe leakage for PA2 at 150 GHz, simulated leakage due to asymmetry for S15 PA2 at 150 GHz, and the uncertainty in the main beam for S15 PA2 deep56 at 150 GHz. The best fit $\Lambda$CDM plus foreground model for the deep 150 $\times$ 150 GHz spectra for ACT only from \cite{choi_2020} was used to compute the resulting change in $D^{TT}_{\ell}$, $D^{TE}_{\ell}$, and $D^{EE}_{\ell}$ due to these elements. These changes are then divided by the corresponding uncertainties $\sigma^{TT}_{\ell}$, $\sigma^{TE}_{\ell}$, and $\sigma^{EE}_{\ell}$ on the final binned deep spectra. 
    At larger multipoles, eventually the uncertainty on the main beam would come to dominate.
    Note that we correct for the main beam leakage and the residual sidelobe leakage in the power spectrum likelihood, as explained in \S\ref{subsec:pol_corr}.
    The asymmetry leakage (simulated using an all-sky beam convolution code, as mentioned in \S\ref{subsec:asym}) is negligible for all detector arrays.
    Sudden changes around $\ell=2000$, as can be seen in the top panel, for example, are due to a change in the bin width at this scale.}
    \label{fig:leakage_effects}
    \vspace{1em}
\end{figure}

The residual sidelobe leakage in Figure~\ref{fig:buddies} can be compared to the leakage in the main beam shown in Figure~\ref{fig:leak}.
The effect of each of these components on the final spectra (if there is no leakage correction as in \S\ref{subsec:pol_corr}) is shown in Figure \ref{fig:leakage_effects}.
At low $\ell$, the residual sidelobe leakage dominates, whereas the main beam leakage grows larger by $\ell \sim 2000$, and dominates at high $\ell$.

\subsection{Leakage Correction}
\label{subsec:pol_corr}

The main beam leakage and the residual leakage from the sidelobes are included in the power spectrum likelihood \citep[see Section 12 of][]{choi_2020} by making use of a leakage-corrected model for the $TE$ and $EE$ theory spectra, computed each time the likelihood is estimated. The corrected model spectra, $T_iE_j'$ and $E_iE_j'$, are related to the input theory spectra, $T_iT_j$, $T_iE_j$ and $E_iE_j$, via
\begin{subequations}
\begin{align}
    T_iE_j' &= T_iE_j + T_iT_j \gamma_j \\ 
    E_iE_j' &= E_iE_j + T_iE_j \gamma_i + T_jE_i \gamma_j + T_iT_j \gamma_i \gamma_j \;.
\end{align}
\end{subequations}

\begin{figure}
    \centering
    \includegraphics[width=\linewidth]{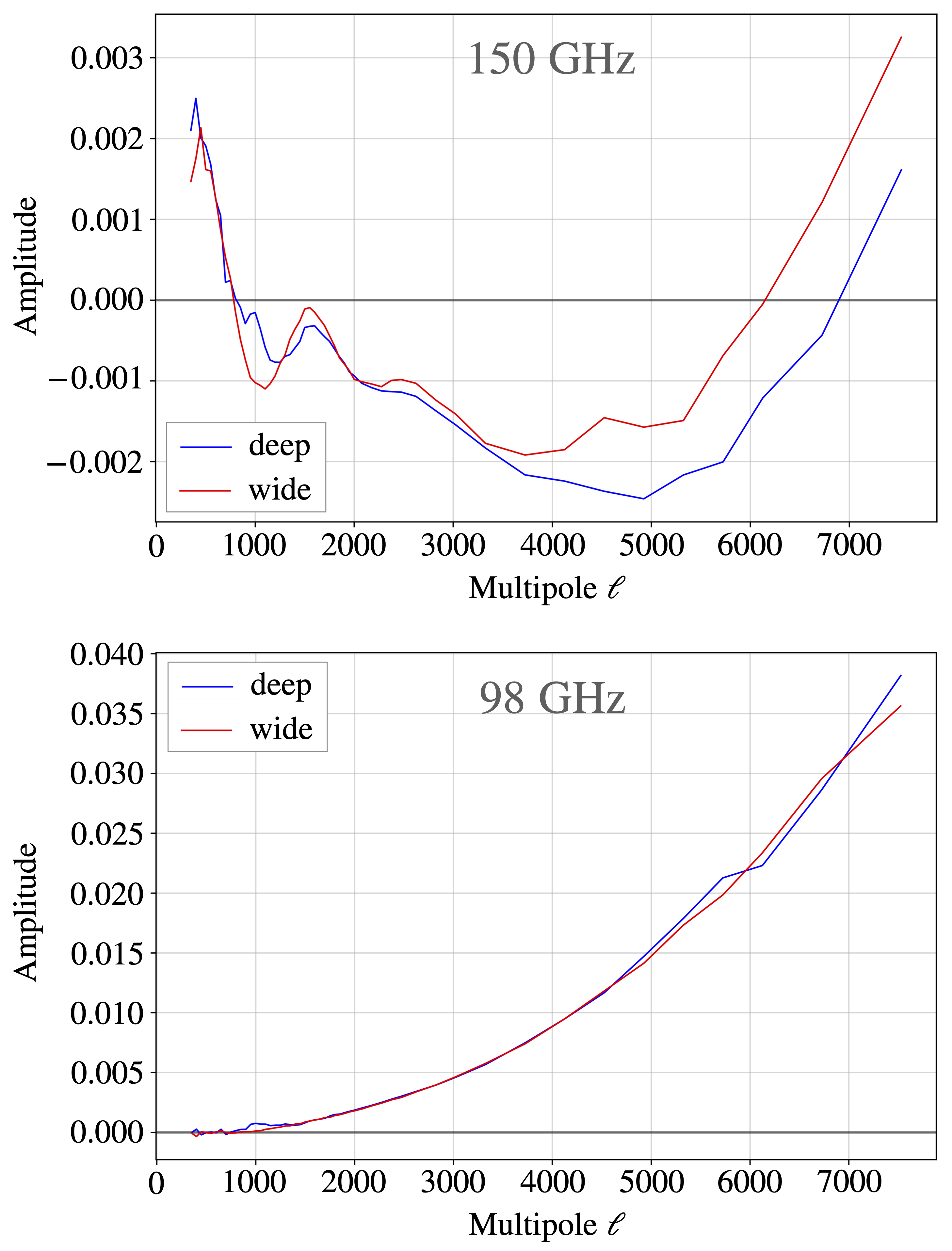}
    \caption{The $\gamma$ factors encoding the $\ell$-dependent T-to-P leakage, used in the power spectrum likelihood for both the deep and wide spectra at 150 GHz (top) and 98 GHz (bottom). The $\gamma$ factors for 150 GHz include both the main beam leakage from \S\ref{subsec:pol_main} and the residual sidelobe leakage from \S\ref{subsec:sidel}. Since no sidelobes are seen at 98 GHz, those $\gamma$ factors consist of only the main beam leakage.}
    \label{fig:gamma}
    \vspace{1em}
\end{figure}

Here the $i$ and $j$ subscripts denote different spectra and the $\gamma$ factors encode the $\ell$-dependent leakage (the sum of the main beam leakage from \S\ref{subsec:pol_main} and the residual sidelobe leakage from \S\ref{subsec:sidel}) and are shown in Figure \ref{fig:gamma}. At 150 (98) GHz, the amplitude of $\gamma$ is never greater than 0.0035 (0.038). 
Around $\ell \sim 1000$--$3000$, this leakage correction is roughly a few-percent ($\sim 1$--$4 \%$) effect for $TE$ and a sub-percent ($\sim 0$--$0.4 \%$) effect for $EE$.

As explained in \cite{choi_2020}, in the power spectrum analysis we first compute a power spectrum for each season, sky region, detector array, and frequency. We then coadd over seasons and arrays to obtain one spectrum per region and frequency. Finally, the regions are divided into two groups based on the detection thresholds for point sources: deep (Deep1, Deep5, Deep6, and Deep8) and wide (BOSS and AdvACT). The spectra for the regions in these two groups are coadded, resulting in a single deep and a single wide power spectrum at each frequency.

To obtain the $\gamma$ factors, we coadd the leakage beams for individual seasons and detector arrays using the same weights used to coadd the spectra, giving an effective leakage beam for both the deep and wide coadded spectra at each frequency.
The uncertainties in the $\gamma$ factors are incorporated in the data covariance matrix by using the main leakage beam uncertainties and the sidelobe residuals to compute another covariance matrix (similar to the main temperature beam covariance matrix) which is then added (in quadrature) to the data covariance matrix.

As described in \cite{aiola_2020,choi_2020}, including this leakage correction in the likelihood reduces the residuals compared to the best-fitting $\Lambda$CDM model but does not have a significant effect on the inferred cosmology. 
\cite{choi_2020} (in Section 12.3) also carried out a test by fitting for two scaling factors (one at 150\;GHz and one at 98 GHz), that multiply the nominal values of the $\gamma$ factors. The data support the baseline model, where the scaling factors are unity.

In the DR3 analysis in \cite{louis_2017} the residual leakage from the sidelobes was similarly treated as a systematic uncertainty in the cosmological power spectrum analysis, but the correction for the main beam polarized leakage is new to DR4.\footnote{As described in \cite{choi_2020}, for DR4 the ACT team adopted a blinding strategy in an attempt to prevent confirmation bias on the cosmological parameters. The analysis pre-unblinding used maps that deprojected the polarized sidelobes, but did not include the additional polarized leakage correction from the main beam or the sidelobe residuals. The post-unblinding analysis revealed some features in the $TE$ residuals to $\Lambda$CDM that led us to a more thorough search for possible sources of systematic uncertainty in the $TE$ spectrum. As a result, it was at this point that we added the correction in the likelihood for the T-to-P leakage from both the polarized sidelobe residuals, and the main beam.}

\section{Discussion}
\label{sec:disc}

\subsection{Beam Products}
\label{subsec:prods}

As part of DR4, several ACT data products were made publicly available on the NASA Legacy Archive for Microwave Background Data Analysis\footnote{\url{lambda.gsfc.nasa.gov/product/act/actpol_prod_table.cfm}} (LAMBDA) and at the National Energy Research Scientific Computing Center (NERSC). Details of all these data products are given in \cite{mallabykay_2021}. 

The beams described in this paper and used for the analysis in \cite{choi_2020} and \cite{aiola_2020} are included in the \enquote{ancillary products} section of the data release. For each season/region/array/frequency combination, there are multiple beam files. Both the real-space radial beam profiles and their harmonic-space transforms are provided, both for the instantaneous beams and for the beams with the jitter corrections included, and in each case are available both with and without the Rayleigh-Jeans-to-CMB spectral correction. 

Note that the instantaneous beams are not region-dependent, so only the jitter-corrected beams contain region flags in their filenames. The instantaneous beams are suitable for time-domain analysis, but they should not be used when analyzing maps. The maps released as part of DR4 have not been corrected for the instrument beam. When working with the maps, one should use the jitter-corrected beams where the pointing variance and its uncertainty have been accounted for. The beam harmonic-space transforms can be used to correct for the beam effects in harmonic space \citep{bond_1987}.

The $TE$ and $TB$ harmonic-space leakage beams for the main beam and the sidelobes and their residuals are also included in the ancillary products for DR4.

Since DR4 includes the data from DR3 as a subset, it is possible to compare some of the DR4 beams with the beams for the same season, sky region, detector array, and frequency released as part of DR3. As shown in Appendix \ref{appen:dr3_dr4}, despite the differences in the analyses, the beams from DR3 and DR4 are consistent.

\subsection{Beam Asymmetry}
\label{subsec:asym}

The asymmetry of the ACT main beams is relatively well described as elliptical, with aspect ratios varying from 1\% to 20\% across the different arrays, as shown in Table 1 of \cite{choi_2020}. Here we give a brief summary of an investigation into this azimuthal asymmetry and the associated spurious signal.

Beam asymmetry is, to leading order, responsible for two effects. The asymmetric convolution creates anisotropy in the sky maps that distorts the shape of point sources and introduces statistical anisotropy in the inferred CMB. The magnitude of the effect is reduced by increased cross-linking: as the telescope observes a position on the sky using approximately orthogonal scan directions, the spurious signal from one scan roughly cancels with that of the other scan. The cross-linking in the ACT maps is sufficient to reduce the contamination to the power spectrum from this effect to an insignificant amount.

The second effect of beam asymmetry is to introduce T-to-P leakage. The beam asymmetry introduces a dependency in the time-ordered data on the position angle of the instrument: observations with different scan directions yield systematically different data. 
For asymmetric beams with a quadrupole shape, which is the dominant asymmetric azimuthal mode of our approximately elliptical beams, the dependence on the position angle is interpreted by the map-maker as a linearly polarized sky component \citep{hu_2003}. In contrast to the first effect, the T-to-P leakage is not averaged down by cross-linking. Instead, the leakage is reduced by the instrument's orthogonally polarized co-pointing detectors: the leakage picked up by one detector approximately cancels with the leakage picked up by its partner.\footnote{This cancellation is not perfect for detector pairs with incorrect relative gain or pointing or for pairs with slightly different beams, but these effects are small compared to the leakage from detectors without a partner, which see no reduction.} As mentioned in \cite{choi_2020} the residual leakage causes an additive bias to the $TE$ and $TB$ power spectra that is roughly constant with multipole with an amplitude that is less than 0.2$\sigma$ away from zero. No attempt to remove this leakage has been made. 

For the investigation described here, both effects were simulated in the time domain using an all-sky beam convolution code similar to \cite{wandelt_2001, reinecke_2006, prezeau_2010, duivenvoorden_2019}.

\subsection{Conclusion}
\label{subsec:concl}

In this paper, we have presented the analysis of the ACT beams for DR4, which includes data from 2013--16. Improvements to the beam pipeline include: better atmosphere subtraction for the Uranus maps, a new scattering term in the model that is fitted to the main beams, a better estimate of the uncertainty in these fits, and residual T-to-P leakage terms that are included in the ACT power spectrum likelihood. Considerable effort was spent developing a realistic model of the beams (including optical effects) and the mapping process, in order to study all elements of the analysis, including the propagation of systematic uncertainties. 

The DR4 beams presented here were also used for the 2013--16 data that were part of the ACT DR5 maps \citep{naess_dr5}. Finally, the pipeline presented here was used to obtain preliminary beams for the 2017--18 Advanced ACT data included in DR5.
Looking forward, as we collect and analyze more ACT data and approach the cosmic variance limit, we will become increasingly sensitive to details of the instrument beams. While some details may change\footnote{Future changes likely will be informed in part by the work currently being done to model the beams in 2D.}, we expect to adopt similar methods as described here for analysis of the post-2016 ACT data.

\section*{Acknowledgements}
Support for ACT was through the U.S. National Science Foundation through awards AST-0408698, AST-0965625, and AST-1440226 for the ACT project, as well as awards PHY-0355328, PHY-0855887 and PHY-1214379. Funding was also provided by Princeton University, the University of Pennsylvania, and a Canada Foundation for Innovation (CFI) award to UBC. ACT operates in the Parque Astron\'omico Atacama in northern Chile under the auspices of the Agencia Nacional de Investigaci\'on y Desarrollo (ANID). The development of multichroic detectors and lenses was supported by NASA grants NNX13AE56G and NNX14AB58G. Detector research at NIST was supported by the NIST Innovations in Measurement Science program. 
Computations were performed on Tiger and Della as part of Princeton Research Computing resources at Princeton University, on Feynman at Princeton University, and on the Niagara supercomputer at the SciNet HPC Consortium. SciNet is funded by the CFI under the auspices of Compute Canada, the Government of Ontario, the Ontario Research Fund---Research Excellence, and the University of Toronto.

Research at Perimeter Institute is supported in part by the Government of Canada through the Department of Innovation, Science and Industry Canada and by the Province of Ontario through the Ministry of Colleges and Universities.

ML was supported by a Dicke Fellowship. 
ES and JD are supported through NSF grant AST-1814971 and AST-2108126. 
EC acknowledges support from the STFC Ernest Rutherford Fellowship ST/M004856/2, STFC Consolidated Grant ST/S00033X/ and from the Horizon 2020 ERC Starting Grant (Grant agreement No 849169).
SKC acknowledges support from NSF award AST-2001866. 
JCH acknowledges support from NSF grant AST-2108536.  
ADH acknowledges support from the Sutton Family Chair in Science, Christianity and Cultures and from the Faculty of Arts and Science, University of Toronto. 
KM acknowledges support from the National Research Foundation of South Africa.
LP gratefully acknowledges support from the Mishrahi and Wilkinson funds.
ZX is supported by the Gordon and Betty Moore Foundation through grant GBMF5215 to the Massachusetts Institute of Technology.

We gratefully acknowledge the publicly available software packages that were used for parts of this analysis. They include 
\texttt{healpy}~\citep{Healpix1}, \texttt{HEALPix}~\citep{Healpix2}, \texttt{libsharp}~\citep{reinecke/2013}, and 
\texttt{pixell}\footnote{https://github.com/simonsobs/pixell}. This research made use of \texttt{Astropy}\footnote{http://www.astropy.org}, a community-developed core Python package for Astronomy \citep{astropy:2013, astropy:2018}. We also acknowledge use of the \texttt{matplotlib}~\citep{Hunter:2007} package and the Python Image Library for producing plots in this paper.

\bibliographystyle{yahapj}
\bibliography{main}

\newpage

\appendix
\section{A. SHRINKING ALGORITHM FOR RADIAL PROFILE COVARIANCE MATRIX}
\label{appen:shrink}

In general, shrinkage is useful when estimating a covariance matrix (or any matrix, really) from a limited number of data points \citep{schafer_2005}. This technique works by combining an empirical estimate of the covariance matrix (a high-dimensional estimate of the underlying covariance with little or no bias) with a model (a low-dimensional estimate which may be biased but has much smaller variance) to minimize the total mean squared error (sum of bias squared and variance) with respect to the true underlying covariance. This is useful when the off-diagonal elements of the covariance matrix are excessively noisy. One can analytically calculate the optimal combination of the low and high dimensional estimates, parametrized by the shrinkage intensity. A review of covariance matrix shrinkage with an example of application to cosmological analysis is given in \cite{pope_2008}. Shrinkage can result in a much better estimation of the covariance matrix when few measurements are available, and it does not adversely affect the covariance in the case of a large number of measurements. Since the method is computationally very simple, \cite{pope_2008} suggest it should always be employed.

The method works as follows. Suppose we have $n$ sets of observations and each observation yields a data vector $\mathbf{x}$ of length $p$. 
In the case of the beams analysis for DR4, for a given season, detector array, and frequency, the number of observations $n$ is between 6 and 51 (as listed in Table \ref{tab:mapsel}) and in each case the resulting data vector $\mathbf{x}$ (radial profile) is of length $p = 55$.
Let $x_i^{(k)}$ then represent the $i^{\mathrm{th}}$ element of the vector for the $k^{\mathrm{th}}$ observation. The estimated empirical mean of the $i^{\mathrm{th}}$ element across all observations is then $\bar{x_i} = ({1}/{n})\sum_{k=1}^{n}x_i^{(k)}$. 

If we define
\begin{equation}
    W_{ij}^{(k)} = (x_i^{(k)}-\overline{x}_i)(x_j^{(k)}-\overline{x}_j),
\end{equation}
and
\begin{equation}
    \overline{W}_{ij} = \frac{1}{n}\sum_{k=1}^{n}W_{ij}^{(k)},
\end{equation}
then an unbiased, empirical estimate of the covariance, $\mathbf{S}$, of the data, is

\begin{subequations}
\begin{align}
    S_{ij} & = \widehat{\mathrm{Cov}}(x_i,x_j) \\ & = \frac{n}{n-1} \overline{W}_{ij} \\ & = \frac{1}{n-1}\sum_{k=1}^{n}(x_i^{(k)}-\overline{x}_i)(x_j^{(k)}-\overline{x}_j) \; .
\end{align}
\end{subequations}

We can also compute the covariance of the elements of this covariance matrix,
\begin{equation}
    \widehat{\mathrm{Cov}}(S_{ij},S_{lm}) = \frac{n}{(n-1)^3}\sum_{k=1}^{n}(W_{ij}^{(k)}-\overline{W}_{ij})(W_{lm}^{(k)}-\bar{W}_{lm}) \; ,
\end{equation}
writing the variance of an individual covariance matrix entry as $\widehat{\mathrm{Var}}(S_{ij}) = \widehat{\mathrm{Cov}}(S_{ij},S_{ij})$. 

Let $\mathbf{T}$ be the target matrix, our model with fewer (or no) free parameters. An example of such a matrix could be the identity matrix times a constant, or the target we use, which is described later, in Equation \ref{eq:target}. Then we can similarly write an equation for $\widehat{\mathrm{Cov}}(T_{ij},S_{ij})$. The optimal shrinkage intensity, ${\lambda}^{*}$,  and resulting final estimate of the covariance matrix $\mathbf{C}$ are then given by
\begin{equation}
\label{eq:lambda_star_general}
    \hat{\lambda}^{*} = \frac{\sum_{i,j}\widehat{\mathrm{Var}}(S_{ij})-\widehat{\mathrm{Cov}}(T_{ij},S_{ij})}{\sum_{i,j}(T_{ij}-S_{ij})^2}\; ,
\end{equation}
\begin{equation}
\label{eq:shrink_cov}
    \mathbf{C} = \widehat{\lambda}^{*}\mathbf{T} + (1-\widehat{\lambda}^{*})\mathbf{S} \; .
\end{equation}

The expression for $\hat{\lambda}^{*}$ above is the practical estimator suggested by \cite{schafer_2005} in order to estimate the optimal shrinkage intensity. This is based on the analytic solution for the optimal shrinkage intensity, $\lambda^*$, introduced by \cite{ledoit_2003}, which is identical to the expression above except the unbiased sample estimates $\widehat{\mathrm{Var}}$ and $\widehat{\mathrm{Cov}}$ are replaced by the true underlying $\mathrm{Var}$ and $\mathrm{Cov}$. Before this analytic solution, matrix shrinkage was much less practical, since complex and computationally intensive methods were needed to find the optimal shrinkage intensity.

Looking more closely at Equation \ref{eq:lambda_star_general}, the term $\widehat{\mathrm{Var}}(S_{ij})$ in the numerator means that as the variances of the elements of the empirical covariance matrix decrease (so, as $n$ becomes much larger than $p$) the shrinkage intensity decreases and our final estimate $\mathbf{C}$ of the covariance matrix approaches the empirical estimate $\mathbf{S}$. In the case of our radial beam profile, $n \ll p$, which is precisely when shrinkage is the most important. The term $\widehat{\mathrm{Cov}}(T_{ij},S_{ij})$ in the numerator accounts for the fact that $\mathbf{S}$ and $\mathbf{T}$ are estimated from the same data, so if some elements of $\mathbf{T}$ are equal to elements of $\mathbf{S}$ then the two terms in the numerator cancel for those elements, and they do not affect the estimate of $\hat{\lambda}^{*}$. The denominator in Equation \ref{eq:lambda_star_general} ensures that if our choice of $\mathbf{T}$ is very different from $\mathbf{S}$, then $\hat{\lambda}^{*}$ will be small and $\mathbf{C}$ will be close to $\mathbf{S}$. A poor choice of target matrix should not therefore negatively affect $\mathbf{C}$.

If Equation \ref{eq:lambda_star_general} leads to a value of $\hat{\lambda}^*$ that is greater than one, then we set $\hat{\lambda}^*=1$, which means our final estimate of the covariance matrix is composed of only the target matrix. On the other hand, if the equation leads to a value of $\hat{\lambda}^*$ that is less than zero, we set $\hat{\lambda}^*=0$, and our final estimate of the covariance matrix is equal to the empirical covariance matrix.

Several common choices for the target matrix are listed in Table 2 of \cite{schafer_2005}, along with simplified expressions for the associated shrinkage intensities $\hat{\lambda}^*$. In this analysis we use a \enquote{diagonal, unequal variance} target matrix, that is, a matrix whose diagonal elements are equal to the diagonal elements of our empirical estimate $\mathbf{S}$ of the covariance matrix and whose off-diagonal elements are zero:
\begin{equation}
\label{eq:target}
    T_{ij} = 
    \begin{cases}
    S_{ii} & \text{if $i=j$}\\
    0 & \text{if $i \neq j$} \; .
    \end{cases}
\end{equation}
In this case the expression for the shrinkage intensity simplifies to
\begin{equation}
    \hat{\lambda}^* = \frac{\sum_{i\neq j}\widehat{\mathrm{Var}}(S_{ij})}{\sum_{i\neq j} S_{ij}^2}\; .
\end{equation}

\newpage

\section{B. COMPUTING THE HARMONIC TRANSFORM OF THE BEAM}
\label{appen:trans}

The spherical harmonics are functions defined on the sphere as
\begin{equation}
    Y_l^m(\theta,\phi) = P_l^m(\cos\theta)\exp{i m\phi} \; ,
\end{equation}
where $P_l^m$ are the Legendre polynomials normalized for spherical harmonics. These are defined as 
\begin{equation}
    P_l^m(x) = (-1)^m\sqrt{\frac{2l+1}{4\pi}}\sqrt{\frac{(l-\abs{m})!}{(l+\abs{m})!}}(1-x^2)^{\abs{m}/2}\frac{d^{\abs{m}}}{dx^{\abs{m}}}P_l(x) \; ,
\end{equation}
where the Legendre polynomials $P_l(x)$ are defined via the recurrence relation

\begin{subequations}
\begin{align}
        P_0(x) & = 1\\
        P_1(x) & = x\\
        nP_n(x) & = (2n-1)xP_{n-1}(x)-(n-1)P_{n-2}(x) \; .
\end{align}
\end{subequations}

The spherical harmonics are orthonormal, so they obey the relation
\begin{equation}
    \int_0^{2\pi}\int_0^{\pi} Y_l^m(\theta,\phi)\overline{Y_{l'}^{m'}}(\theta,\phi) \sin\theta \; d\theta\; d\phi = \delta_{ll'}\delta_{mm'} \; ,
\end{equation}
where $\overline{z}$ is the complex conjugate of $z$ and $\delta_{ij}$ is the Kronecker symbol.

In a spherical harmonic transform, we compute the coefficients $f_{l}^{m}$ used to express a function $f (\theta,\phi)$ as
\begin{equation}
    f(\theta,\phi) = \sum_{l=0}^{\infty}\sum_{m=-l}^{l}f_{l}^{m} Y_l^m(\theta,\phi) \; .
\end{equation} 
The coefficients can be computed using the equation 
\begin{subequations}
\begin{align}
    f_{l}^{m} &= \int_{0}^{2\pi}\int_{0}^{\pi} f(\theta,\phi) \overline{Y_l^m}(\theta,\phi)\sin\theta \; d\theta\; d\phi   \\
    & = \int_{0}^{2\pi}\int_{0}^{\pi} f(\theta,\phi) P_{l}^{m}(\cos\theta)\exp{-im\phi} \sin\theta\; d\theta\; d\phi \; .
\end{align}
\end{subequations}

If $f (\theta,\phi)$ is independent of $\phi$ (as is the case for our beam), then we can write $f (\theta,\phi)$ = $f (\theta)$ and the equation above becomes
\begin{equation}
    f_{l}^{m}= \int_{0}^{2\pi}\exp{-im\phi}d\phi\int_{0}^{\pi} f(\theta) P_{l}^{m}(\cos\theta) \sin\theta\; d\theta \; .
\end{equation}
The integral over $\phi$ then simplifies to 
\begin{equation}
    \int_0^{2\pi} e^{-im\phi}\; d\phi = 2\pi \delta_{m0}\; .
\end{equation}
So $f_l^m$ is only non-zero for $m=0$, in which case we have 

\begin{subequations}
\begin{align}
    f_l^0 & =  2\pi\int_0^{\pi}f(\theta)P_l(\cos\theta)\sin\theta \; d\theta\\
    & = 2\pi \int_{-1}^{1}f(\theta)P_l (\cos\theta) \; d\cos \theta\; .
\end{align}
\end{subequations}

This is the equation for the Legendre polynomial transform, presented as a means of converting the radial beam profile $B(\theta)$ to the harmonic transform $B_{\ell}$. However, this can be time-consuming to compute. For small beams such as ours, it is not necessary to work in the curved sky regime. We instead perform a 2D Fourier transform, which effectively becomes a Hankel transform, as shown below. The difference between the Hankel and Legendre polynomial transforms is less than a factor of $4\times 10^{-5}$ between $\ell=0$ and $\ell=10,000$ and the Hankel transform is much faster to compute. 

Now let's consider the 2D Fourier transform of a function $f(x,y)$, 
\begin{equation}
    F(k_x,k_y) = \int_{-\infty}^{\infty}\int_{-\infty}^{\infty}f(x,y)\exp{-i(xk_x+ yk_y)}\;dx\; dy \; .
\end{equation}
Introducing the polar coordinates
\begin{equation*}
    \begin{split}
    x = \theta\cos\phi \hspace{0.5cm}& \hspace{0.5cm} y = \theta\sin\phi\\
    k_x = k\cos\psi \hspace{0.5cm}& \hspace{0.5cm} k_y = k\sin\psi
    \end{split}
\end{equation*}
where $\theta$ and $\phi$ here correspond to the $\theta$ and $\phi$ in spherical coordinates used throughout the paper, we then have, in the flat sky approximation,
\begin{equation}
        F(k\cos\psi,k\sin\psi) \equiv \mathcal{F}(k,\psi) = \int_{0}^{\infty}\int_{0}^{2\pi}f(\theta,\phi)\exp{-i\theta k(\cos\phi\cos\psi + \sin\phi\sin\psi)} \theta\; d\theta\; d\phi \; .
\end{equation}

If our function is circularly symmetric, so independent of $\phi$ (as is the case for our beam model), we have ${f (x,y) = \mathit{f} (\theta,\phi) = {f} (\theta)}$ and the equation above becomes

\begin{subequations}
\begin{align}
        \mathcal{F}(k,\psi) & = \int_0^{\infty}\theta f(\theta) \int_0^{2\pi}\exp{-i\theta k (\cos\phi\cos\psi + \sin\phi\sin\psi)}\; d\theta\; d\phi\\ 
        & = \int_0^{\infty}\theta f(\theta) \int_0^{2\pi}\exp{-i\theta k \cos(\phi-\psi)}\; d\theta\; d\phi \\
        & = \int_0^{\infty}\theta f(\theta) \int_0^{2\pi}\exp{-i\theta k \cos\alpha}\; d\theta\; d\alpha\\
        & = \int_0^{\infty}\theta f(\theta)\; 2 \int_0^{\pi}\exp{-i\theta k \cos\alpha}\; d\theta\; d\alpha \; .
\end{align}
\end{subequations}

Using the integral representation 
\begin{equation}
    J_n(z) = \frac{(-i)^n}{\pi}\int_0^{\pi} \exp{i z \cos \varphi}\cos(n\varphi)\;d\varphi
\end{equation}
for the Bessel functions $J_n$ of the first kind, we have
\begin{equation}
    J_0(z) = \frac{1}{\pi} \int_0^{\pi} \exp{i z \cos\varphi}\; d\varphi \; ,
\end{equation}
and so the final expression for the 2D Fourier transform of a circularly symmetric function $f(\theta)$ may be written as
\begin{subequations}
\begin{align}
    \mathcal{F}(k) &= 2\pi \int_0^{\infty}\theta f(\theta) J_0(-\theta k)\; d\theta\\
    & = 2\pi \int_0^{\infty}\theta f(\theta) J_0(\theta k)\; d\theta \; ,
\end{align}
\end{subequations}
which is a Hankel transform of order zero, and where the last line follows from the identity $J_n(-z) = J_n(z)$ for integer $n$.

In order to compute the harmonic transform of our beam profile, we evaluate the expression above separately for the three main terms in our beam profile fit: the core term (composed of the sum of basis functions), the scattering term, and the $1/\theta^3$ asymptotic term. The integrals for the core and scattering terms are computed numerically, but we derive an analytic expression for the integral of the $1/\theta^3$ term, shown below.

Given a fit amplitude $\alpha$, the Hankel transform for the $1/\theta^3$ term may be written as
\begin{equation}
    \mathcal{F}_{1/\theta^3}(k) = \alpha \int_0^{\infty} \theta\Big(\frac{1}{\theta^3} \Big)  J_0(\theta\ell)\; d\theta\; = \alpha \int_0^{\infty} \frac{J_0(\theta\ell)}{\theta^2}  \; d\theta\; .
\end{equation}
The analytic expression we use for this integral is 
\begin{equation}
    \int \frac{J_0(\theta\ell)}{\theta^2}  \; d\theta  = \ell \Big[ J_1(\theta \ell) - J_0(\theta \ell)\Big(\frac{\theta^2\ell^2+1}{\theta\ell} \Big)- \frac{\pi\theta\ell}{2}\Big(H_0(\theta\ell)J_1(\theta\ell)-H_1(\theta\ell)J_0(\theta\ell) \Big)\Big] \; ,
\end{equation}
where $H_n(x)$ is the Struve function.

\newpage

\section{C. BEAM SOLID ANGLES FOR DIFFERENT EFFECTIVE FREQUENCIES }
\label{appen:sas}

To indicate the effect of the passbands on the beams, we tabulate the solid angles for select effective frequencies. The effective frequencies of the band centers are currently uncertain to approximately 2.4 GHz. However, since this is in part due to systematic errors in the measurements, the relative uncertainties are smaller.

Similar to the correction we make to the main beams for use with the CMB, in each case we take the beam to be $B'(\ell) = B(\ell \nu_{\mathrm{RJ}}/\nu_{\mathrm{S}})$, where $\nu_{\mathrm{RJ}}$ is the effective frequency for radiation with a Rayleigh-Jeans spectrum and $\nu_{\mathrm{S}}$ is the effective frequency for the source of interest (either CMB, synchrotron emission, dust, or the thermal Sunyaev-Zel’dovich (tSZ) effect).

As described in \S\ref{subsec:add_corr}, for the beam analysis for DR4, the effective frequencies from \cite{thornton_2016} were used for the RJ-to-CMB beam spectral correction. Subsequently, the effective frequencies were re-computed for the foreground modeling for DR4, using improved passband data and upgraded code, as detailed in Appendix D of \cite{choi_2020}. These updated frequencies are shown here in Table \ref{tab:eff_freq} and the corresponding beam solid angles are shown in the following tables. Considering the uncertainties on the passbands, the frequencies from \cite{thornton_2016} and \cite{choi_2020} are consistent. In addition, given that the uncertainty on the beams is subdominant in the power spectrum analysis, which of these effective frequencies one uses for the beam spectral correction does not have a significant effect on the results. Still, the solid angles for the CMB in Table \ref{tab:app_sa_cmb} below are slightly different from those in Table \ref{tab:eff_sa}.

For any particular season/region/detector array/frequency, the uncertainty on the solid angles is 2--5\%. This includes both the instantaneous beam uncertainty and the uncertainty due to the region-dependent jitter correction. As is apparent here, the derived solid angle and passband are intimately connected. 

In addition, the beam and the passband are coupled, an effect which we considered for the first time in detail in \cite{madhavacheril_2020}, and described in Appendix A of said paper. In short, the beam shape evolves as a function of frequency across the passband, and so our customary assumption of separability of these two components does not hold to high precision. This beam-bandpass coupling was modeled as part of a systematic check of the DR4 results in \cite{choi_2020} and was found to not have a significant effect on the inferred cosmological parameters. 

\vspace{2em}
{\renewcommand{\arraystretch}{1.8}
\begin{table*}[h]
\caption{Effective frequencies [GHz].}
\centering
\begin{tabular}{|c|c|c|c|c|c|c|}
\hline
Array & Band & RJ & CMB & tSZ & dust & sync \\
\hline
PA1 & 150 GHz & 150.8 & 149.6 & 150.0 & 151.2 & 146.9 \\
\hline
PA2 & 150 GHz & 151.2 & 149.9 & 150.4 & 151.6 & 147.3 \\
\hline
PA3 & 150 GHz & 148.4 & 147.6 & 147.9 & 148.7 & 145.8 \\
\hline
PA3 & 98 GHz & 98.7 & 97.9 & 98.4 & 98.8 & 95.5 \\
\hline
\end{tabular}
\label{tab:eff_freq}
\end{table*}
}

\begin{table*}[htbp]
\caption{Effective beam solid angles (nsr) for the CMB.}
\centering
\begin{tabular}{|c|c|c|C{4em}|C{4em}|C{4em}|C{4em}|C{4em}|C{4em}|C{4em}|}
\hline
Array & Band & Season  & Deep1 & Deep5 & Deep6 & Deep56 & Deep8 & BOSS & AdvACT \\
\hline
\multirow{3}{*}{PA1} & \multirow{3}{*}{150 GHz} & S13 &  213  & 209  &  210  & - & - & - & - \\
 &  & S14 & - & - & - &  204  & - & - & - \\
 &  & S15 & - & - & - &  206  &  199  &  205  & - \\
\hline
\multirow{3}{*}{PA2} & \multirow{3}{*}{150 GHz} & S14 & - & - & - &  190  & - & - & - \\
 & & S15 & - & - & - &  196  &  192  &  193  & - \\
 & & S16  & - & - & - & - & - & - &  194  \\
\hline
\multirow{2}{*}{PA3} & \multirow{2}{*}{150 GHz} & S15 &  - & - & - & 274  &  268  &  285  & - \\
 & & S16 &  - & - & - & - & - & - &  242  \\
\hline
\multirow{2}{*}{PA3} & \multirow{2}{*}{98 GHz} & S15 &  - & - & - &  546  &  541  & 556 & - \\
 & & S16 & - & - & - & - & - & - &  524  \\
\hline
\end{tabular}
\label{tab:app_sa_cmb}
\end{table*}

\begin{table*}[htbp]
\caption{Effective beam solid angles (nsr) for synchrotron emission.}
\centering
\begin{tabular}{|c|c|c|C{4em}|C{4em}|C{4em}|C{4em}|C{4em}|C{4em}|C{4em}|}
\hline
Array & Band & Season  & Deep1 & Deep5 & Deep6 & Deep56 & Deep8 & BOSS & AdvACT \\
\hline
\multirow{3}{*}{PA1} & \multirow{3}{*}{150 GHz} & S13 &  221  & 217  &  218  & - & - & - & - \\
 &  & S14 & - & - & - &  211  & - & - & - \\
 &  & S15 & - & - & - &  213  &  207  &  212  & - \\
\hline
\multirow{3}{*}{PA2} & \multirow{3}{*}{150 GHz} & S14 & - & - & - &  196  & - & - & - \\
 & & S15 & - & - & - &  203  &  198  &  200  & - \\
 & & S16  & - & - & - & - & - & - &  201  \\
\hline
\multirow{2}{*}{PA3} & \multirow{2}{*}{150 GHz} & S15 &  - & - & - & 281  &  275  &  292  & - \\
 & & S16 &  - & - & - & - & - & - &  248  \\
\hline
\multirow{2}{*}{PA3} & \multirow{2}{*}{98 GHz} & S15 &  - & - & - &  575  &  569  & 585 & - \\
 & & S16 & - & - & - & - & - & - &  552  \\
\hline
\end{tabular}
\end{table*}

\begin{table*}[htbp]
\caption{Effective beam solid angles (nsr) for dusty sources.}
\centering
\begin{tabular}{|c|c|c|C{4em}|C{4em}|C{4em}|C{4em}|C{4em}|C{4em}|C{4em}|}
\hline
Array & Band & Season  & Deep1 & Deep5 & Deep6 & Deep56 & Deep8 & BOSS & AdvACT \\
\hline
\multirow{3}{*}{PA1} & \multirow{3}{*}{150 GHz} & S13 &  208  & 205  &  205  & - & - & - & - \\
 &  & S14 & - & - & - &  199  & - & - & - \\
 &  & S15 & - & - & - &  201  &  195  &  200  & - \\
\hline
\multirow{3}{*}{PA2} & \multirow{3}{*}{150 GHz} & S14 & - & - & - &  186  & - & - & - \\
 & & S15 & - & - & - &  192  &  187  &  189  & - \\
 & & S16  & - & - & - & - & - & - &  190  \\
\hline
\multirow{2}{*}{PA3} & \multirow{2}{*}{150 GHz} & S15 &  - & - & - & 270  &  264  &  281  & - \\
 & & S16 &  - & - & - & - & - & - &  239  \\
\hline
\multirow{2}{*}{PA3} & \multirow{2}{*}{98 GHz} & S15 &  - & - & - &  536  &  531  & 546 & - \\
 & & S16 & - & - & - & - & - & - &  515  \\
\hline
\end{tabular}
\end{table*}

\begin{table*}[htbp]
\caption{Effective beam solid angles (nsr) for the tSZ effect.}
\centering
\begin{tabular}{|c|c|c|C{4em}|C{4em}|C{4em}|C{4em}|C{4em}|C{4em}|C{4em}|}
\hline
Array & Band & Season  & Deep1 & Deep5 & Deep6 & Deep56 & Deep8 & BOSS & AdvACT \\
\hline
\multirow{3}{*}{PA1} & \multirow{3}{*}{150 GHz} & S13 &  212  & 208  &  209  & - & - & - & - \\
 &  & S14 & - & - & - &  202  & - & - & - \\
 &  & S15 & - & - & - &  204  &  198  &  204  & - \\
\hline
\multirow{3}{*}{PA2} & \multirow{3}{*}{150 GHz} & S14 & - & - & - &  189  & - & - & - \\
 & & S15 & - & - & - &  195  &  190  &  192  & - \\
 & & S16  & - & - & - & - & - & - &  193  \\
\hline
\multirow{2}{*}{PA3} & \multirow{2}{*}{150 GHz} & S15 &  - & - & - & 273  &  267  &  284  & - \\
 & & S16 &  - & - & - & - & - & - &  241  \\
\hline
\multirow{2}{*}{PA3} & \multirow{2}{*}{98 GHz} & S15 &  - & - & - &  541  &  536  & 551 & - \\
 & & S16 & - & - & - & - & - & - &  520  \\
\hline
\end{tabular}
\vspace{18em}
\end{table*}

\clearpage

\section{ D. TRANSFORMING FROM $\{Q_{\lowercase{r}},U_{\lowercase{r}}\}$ TO $\{E,B\}$}
\label{appen:qr_ur}

Instead of estimating the polarized $\ell$-space beams directly from non-local $E$ and $B$ transform maps, we opt to use a locally defined map-space polarization basis for this purpose. As we show below, the fields $Q_r$ and $U_r$ turn out to be a convenient choice; in the flat-sky limit, they are defined in terms of local linear combinations of the usual $Q$ and $U$ maps as follows:
\begin{equation} \label{eq:q_r}
    Q_r(\boldsymbol{\theta}) = Q(\boldsymbol{\theta}) \cos 2\phi_{\theta} + U(\boldsymbol{\theta}) \sin 2\phi_{\theta}
\end{equation}

\begin{equation} \label{eq:u_r}
    U_r(\boldsymbol{\theta}) = U(\boldsymbol{\theta}) \cos 2\phi_{\theta} - Q(\boldsymbol{\theta}) \sin 2\phi_{\theta}
\end{equation}
where $\boldsymbol{\theta} \equiv (\theta,\phi_\theta)$ are standard polar coordinates with the beam centroid as their origin and $\phi_{\theta}$ increasing clockwise from the positive $y$-axis (assuming that one uses the convention in which $+x$ points to the right and $+y$ points upward). Conversely, we may also define these fields in terms of their local contributions to both $Q$ and $U$ in the same coordinate system:
\begin{equation} \label{eq:q}
    Q(\boldsymbol{\theta}) = Q_r(\boldsymbol{\theta}) \cos 2\phi_{\theta} - U_r(\boldsymbol{\theta}) \sin 2\phi_{\theta}
\end{equation}

\begin{equation} \label{eq:u}
    U(\boldsymbol{\theta}) = U_r(\boldsymbol{\theta}) \cos 2\phi_{\theta} + Q_r(\boldsymbol{\theta}) \sin 2\phi_{\theta} \; .
\end{equation}

Since we are ultimately interested in how leakage manifests itself in the usual angular power spectra, we need to translate any polarized beam models of $Q_r$ and $U_r$ to an $\ell$-space representation of $E$ and $B$. As it turns out, there exists a simple relation between the azimuthally averaged versions of these components, which we derive here in the flat-sky limit.
We begin with the Fourier-space expressions for $E$ and $B$:
\begin{equation} \label{eq:e}
    E(\boldsymbol{\ell}) = \hat{Q}(\boldsymbol{\ell}) \cos 2\phi_{\ell} + \hat{U}(\boldsymbol{\ell}) \sin 2\phi_{\ell}
\end{equation}
\begin{equation} \label{eq:b}
    B(\boldsymbol{\ell}) = \hat{U}(\boldsymbol{\ell}) \cos 2\phi_{\ell} - \hat{Q}(\boldsymbol{\ell}) \sin 2\phi_{\ell}
\end{equation}
where $\boldsymbol{\ell} \equiv (\ell,\phi_\ell)$ is the Fourier conjugate of $\boldsymbol{\theta}$, and $\{\hat{Q},\hat{U}\}$ are just standard Fourier transforms of $\{Q,U\}$:
\begin{equation} \label{eq:q_hat}
    \hat{Q}(\boldsymbol{\ell}) = \int Q(\boldsymbol{\theta}) e^{i\boldsymbol{\ell}\cdot\boldsymbol{\theta}} d\boldsymbol{\theta} = \int Q(\theta,\phi_{\theta}) e^{i\ell\theta\cos(\phi_{\theta} - \phi_\ell)} \theta\; d\theta\; d\phi_{\theta}
\end{equation}
\begin{equation} \label{eq:u_hat}
    \hat{U}(\boldsymbol{\ell}) = \int U(\boldsymbol{\theta}) e^{i\boldsymbol{\ell}\cdot\boldsymbol{\theta}} d\boldsymbol{\theta} = \int U(\theta,\phi_{\theta}) e^{i\ell\theta\cos(\phi_{\theta} - \phi_{\ell})} \theta\; d\theta\; d\phi_{\theta} \; .
\end{equation}

Taking the azimuthal average of Equations \ref{eq:e} and \ref{eq:b}, we get the one-dimensional transforms $\tilde{E}$ and $\tilde{B}$:
\begin{equation} \label{eq:tilde_e1}
    \tilde{E}(\ell) = \frac{1}{2\pi} \int \hat{Q}(\boldsymbol{\ell}) \cos 2\phi_{\ell}\;d\phi_{\ell} + \frac{1}{2\pi} \int \hat{U}(\boldsymbol{\ell}) \sin 2\phi_{\ell}\;d\phi_{\ell}
\end{equation}
\begin{equation} \label{eq:tilde_b1}
    \tilde{B}(\ell) = \frac{1}{2\pi} \int \hat{U}(\boldsymbol{\ell}) \cos 2\phi_{\ell}\;d\phi_{\ell} - \frac{1}{2\pi} \int \hat{Q}(\boldsymbol{\ell}) \sin 2\phi_{\ell}\;d\phi_{\ell} \; .
\end{equation}

The expression for $\tilde{E}$ in Equation \ref{eq:tilde_e1} may be rewritten in terms of map-space $Q$ and $U$ with the help of Equations \ref{eq:q_hat} and \ref{eq:u_hat}:
\begin{equation} \label{eq:tilde_e2}
    \tilde{E}(\ell) = \frac{1}{2\pi} \int Q(\theta,\phi_{\theta}) e^{i\ell\theta\cos(\phi_{\theta} - \phi_{\ell})} \theta\;d\theta\;d\phi_{\theta} \cos 2\phi_{\ell}\;d\phi_{\ell} 
    + \frac{1}{2\pi} \int U(\theta,\phi_{\theta}) e^{i\ell\theta\cos(\phi_{\theta} - \phi_{\ell})} \theta\;d\theta\;d\phi_{\theta} \sin 2\phi_{\ell}\;d\phi_{\ell} \; .
\end{equation}

Substituting for $Q(\boldsymbol{\theta})$ and $U(\boldsymbol{\theta})$ in Equation \ref{eq:tilde_e2} using Equations \ref{eq:q} and \ref{eq:u}, we obtain a relation between $\tilde{E}$ and $\{Q_r,U_r\}$:
\begin{equation} \label{eq:tilde_e3}
\begin{split}
    \tilde{E}(\ell) = & \frac{1}{2\pi} \int \big(Q_r(\theta,\phi_{\theta})\cos 2\phi_{\theta} - U_r(\theta,\phi_{\theta})\sin 2\phi_{\theta}\big) e^{i\ell\theta\cos(\phi_{\theta} - \phi_{\ell})} \theta\;d\theta\;d\phi_{\theta} \cos 2\phi_{\ell}\;d\phi_{\ell} \\ & 
    + \frac{1}{2\pi} \int \big(U_r(\theta,\phi_{\theta})\cos 2\phi_{\theta} + Q_r(\theta,\phi_{\theta})\sin 2\phi_{\theta}\big) e^{i\ell\theta\cos(\phi_{\theta} - \phi_{\ell})} \theta\;d\theta\;d\phi_{\theta} \sin 2\phi_{\ell}\;d\phi_{\ell}\; .
\end{split}
\end{equation}

Grouping together the terms with $Q_r(\boldsymbol{\theta})$ and $U_r(\boldsymbol{\theta})$, the equation above becomes:
\begin{equation} \label{eq:tilde_e4}
\begin{split}
    \tilde{E}(\ell) = & \frac{1}{2\pi} \int Q_r(\theta,\phi_{\theta})\big(\cos 2\phi_{\theta} \cos 2\phi_{\ell} + \sin 2\phi_{\theta} \sin 2\phi_{\ell}\big) e^{i\ell\theta\cos(\phi_{\theta} - \phi_{\ell})} \theta\;d\theta\;d\phi_{\theta}\;d\phi_{\ell} \\ &
    + \frac{1}{2\pi} \int U_r(\theta,\phi_{\theta})\big(\cos 2\phi_{\theta} \sin 2\phi_{\ell} - \sin 2\phi_{\theta} \cos 2\phi_{\ell}\big) e^{i\ell\theta\cos(\phi_{\theta} - \phi_{\ell})} \theta\;d\theta\;d\phi_{\theta}\;d\phi_{\ell}\; .
\end{split}
\end{equation}

Then, making use of a simple trigonometric identity, Equation \ref{eq:tilde_e4} may be written as:
\begin{equation} \label{eq:tilde_e5}
\begin{split}
    \tilde{E}(\ell) = & \frac{1}{2\pi} \int Q_r(\theta,\phi_{\theta})\cos 2(\phi_{\theta}-\phi_{\ell})\;e^{i\ell\theta\cos(\phi_{\theta} - \phi_{\ell})} \theta\;d\theta\;d\phi_{\theta}\;d\phi_{\ell} \\ &
    - \frac{1}{2\pi} \int U_r(\theta,\phi_{\theta})\sin 2(\phi_{\theta}-\phi_{\ell})\;e^{i\ell\theta\cos(\phi_{\theta} - \phi_{\ell})} \theta\;d\theta\;d\phi_{\theta}\;d\phi_{\ell} \; .
\end{split}
\end{equation}
And making the substitution $\phi_{\rho} \equiv \phi_{\theta} - \phi_{\ell}$, Equation \ref{eq:tilde_e5} may be expressed as:
\begin{equation} \label{eq:tilde_e6}
\begin{split}
    \tilde{E}(\ell) = & \frac{1}{2\pi} \int Q_r(\theta,\phi_{\theta})\cos 2\phi_{\rho}\;e^{i\ell\theta\cos\phi_{\rho}}\;\theta\;d\theta\;d\phi_{\theta}\;d\phi_{\rho}\\ &
    - \frac{1}{2\pi} \int U_r(\theta,\phi_{\theta})\sin 2\phi_{\rho}\;e^{i\ell\theta\cos\phi_{\rho}}\;\theta\;d\theta\;d\phi_{\theta}\;d\phi_{\rho}\; .
\end{split}
\end{equation}

We are now able to write each of the two terms in the expression for $\tilde{E}$ as three separate integrals:
\begin{equation} \label{eq:sep_int}
\begin{split}
    \tilde{E}(\ell) = & \int \big(\frac{1}{2\pi} \int Q_r(\theta,\phi_{\theta})\,d\phi_{\theta}\big) \big(\int \cos 2\phi_{\rho}\;e^{i\ell\theta\cos\phi_{\rho}}\;d\phi_{\rho}\big) \theta\;d\theta \\ & - \int \big(\frac{1}{2\pi} \int U_r(\theta,\phi_{\theta})\;d\phi_{\theta}\big) \big(\int \sin 2\phi_{\rho}\;e^{i\ell\theta\cos\phi_{\rho}}\;d\phi_{\rho}\big) \theta\;d\theta \; .
\end{split}
\end{equation}
Note that the integrals of $Q_r(\boldsymbol{\theta})$ and $U_r(\boldsymbol{\theta})$ over $\phi_{\theta}$ --- the first set of parentheses --- are simply the azimuthal averages $\tilde{Q}_r$ and $\tilde{U}_r$, while the integrals over $\phi_{\rho}$ - the second set of parentheses - turn out to have simple analytic counterparts:
\begin{equation}
    \int \cos 2\phi_{\rho}\;e^{i\ell\theta\cos\phi_{\rho}}\;d\phi_{\rho} = -2\pi J_2(\ell\theta)
\end{equation}
\begin{equation}
    \int \sin 2\phi_{\rho}\;e^{i\ell\theta\cos\phi_{\rho}}\;d\phi_{\rho} = 0 \; .
\end{equation}

So the azimuthally averaged $\ell$-space $E$ beam is simply the second-order Hankel transform of the azimuthally averaged map-space $Q_r$ beam:
\begin{equation} \label{eq:final1}
    \tilde{E}(\ell) = -2\pi\int \tilde{Q}_r(\theta) J_2(\ell\theta)\;\theta\;d\theta \; .
\end{equation}
One can similarly show that the same relation exists between the azimuthally averaged $\ell$-space $B$ and map-space $U_r$ beams:
\begin{equation} \label{eq:final2}
    \tilde{B}(\ell) = -2\pi\int \tilde{U}_r(\theta) J_2(\ell\theta)\;\theta\;d\theta \; .
\end{equation}
With Equations \ref{eq:final1} and \ref{eq:final2} in hand, we have a complete formalism for transforming the polarized beams using the $\{Q_r,U_r\}$ basis.

\clearpage

\section{ E. DR3 vs DR4 BEAMS}
\label{appen:dr3_dr4}

\begin{figure}[h]
    \centering
    \includegraphics[width=0.49\linewidth]{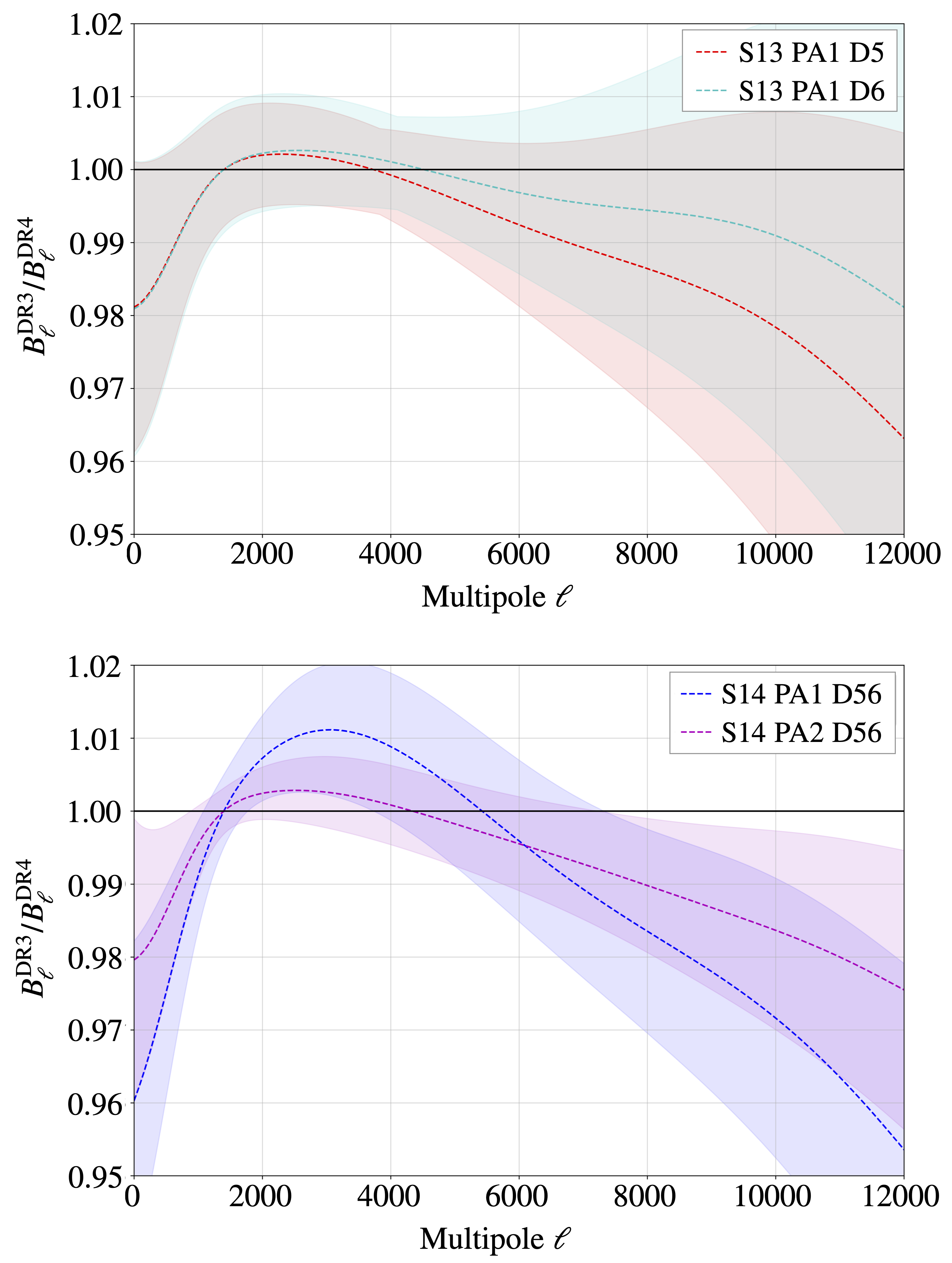}
    \caption{Ratio of the beam transforms for DR3 and DR4 at 150 GHz, for S13 (top) and S14 (bottom). The shaded bands indicate the 1$\sigma$ uncertainty bounds, determined using the maximum uncertainty of the two transforms being compared. For these plots, the transforms have been normalized at $\ell=1400$, which corresponds roughly to the effective calibration scale.}
    \label{fig:dr3_dr4}
    \vspace{1em}
\end{figure}

The beam transforms made publicly available for the DR3 and DR4 releases have been compared for each season, sky region, detector array, and frequency in common, as shown in Figure~\ref{fig:dr3_dr4}. Despite the changes in the analyses, the beam transforms are consistent.

\end{document}